 \documentclass[final,5p,times,twocolumn,authoryear]{elsarticle}

\usepackage{amssymb}
\usepackage{lipsum}
\usepackage{amsmath}
\setcitestyle{square,numbers,sort&compress}
\setcitestyle{citesep={,}}
\biboptions{numbers,sort&compress}
\usepackage{xcolor}
\makeatletter
\def\ps@pprintTitle{%
	\let\@oddhead\@empty
	\let\@evenhead\@empty
	\let\@oddfoot\@empty
	\let\@evenfoot\@oddfoot
}
\makeatother

\usepackage{caption}
\usepackage{float}
\usepackage{graphicx}
\usepackage{subcaption}
\begin{document}

\begin{frontmatter}



\title{Holographic Einstein Ring of Deformed AdS-Schwarzschild Black Holes
}


  \author[first]{Jin-Yu Gui\fnref{1}} 
\fntext[1]{Email: guijinyu0619@qq.com}

\affiliation[first]{organization={Department of Mechanics, Chongqing Jiaotong University},
            city={Chongqing},
            postcode={400074}, 
            %% state={Nanan},
            country={China}}           
\author[second]{Xiao-Xiong Zeng\fnref{2}}
\fntext[2]{Email: xxzengphysics@163.com (Corresponding author)}
\affiliation[second]{organization={College of Physics and Electronic Engineering, Chongqing Normal University},
            city={Chongqing},
            postcode={401331}, 
            %% state={Nanan},
            country={China}} 

\author[first]{Ke-Jian He\fnref{3}}
\fntext[3]{Email: kjhe94@163.com (Corresponding author)}

\author[third]{Huan Ye\fnref{4}}
\fntext[4]{Email: 1417640866@qq.com}
\affiliation[third]{organization={School of Material Science and Engineering, Chongqing Jiaotong University},
            city={Chongqing},
            postcode={400074}, 
            %% state={Nanan},
            country={China}} 
\begin{abstract}
In this work, the wave optics is employed to investigate the Einstein ring of a deformed AdS-Schwarzschild black hole (BH). When the source is fixed on the AdS boundary, one can obtain the corresponding response function generated on the antipodal side of the boundary. By utilizing a virtual optical system equipped with a convex lens, we are able to capture an image of the BH's holographic Einstein ring on the screen.  The influence of the relevant physical parameters and the observer's position on the characteristics of the Einstein ring is also investigated, revealing that variations in the observer's position result in a transition of the displayed image from an axisymmetric ring to an arc, ultimately converging into a solitary point of luminosity. In addition, variations in the relevant physical parameters naturally exert influences on the Einstein ring. The photon ring of the BH was also investigated from a geometric optics perspective, and the numerical results indicate that the incident angle of the photon ring aligns with that of the Einstein ring. In the context of modified gravity theories, the investigation of Einstein rings formed by deformed AdS-Schwarzschild BH is expected to not only contribute to advancing the development of gravitational theories but also facilitate a more comprehensive understanding of spacetime geometry and the physical properties of BHs, thereby distinguishing them from Schwarzschild BH.

\end{abstract}




\end{frontmatter}

%\tableofcontents



\section{Introduction}
\label{introduction}

\par
The AdS/CFT correspondence\cite{Maldacena:1997re}, also known as holographic duality or gauge/gravity duality, establishes a profound connection between classical gravitational theory in Anti-de Sitter (AdS) spacetime and quantum field theories (QFTs). Due to its particularity, it has been a focal point of research for numerous physicists over the past few decades and serves as a bridge spanning different disciplinary fields\cite{Natsuume:2014sfa}. The relevant research in the field of holographic duality not only serves to test the internal consistency of the theory, but also provides a novel platform for advancing the study of strongly coupled field theory. The concept of holographic duality has been extensively utilized in the investigation of holographic superconductivity\cite{Hartnoll:2008kx,Hartnoll:2008vx,Gubser:2008px}, Fermi and non-Fermi liquids\cite{Cubrovic:2009ye,Liu:2009dm,Faulkner:2009wj}, exotic metals\cite{Hartnoll:2009ns,Pal:2010sx,Kim:2010zq} and  Metal insulation phase transition in quantum phase transition\cite{Mefford:2014gia,Kiritsis:2015oxa,Donos:2014uba}.

The BH shadow, being an observable characteristic of a BH, has consistently remained a prominent subject in the realms of theoretical physics and astronomy. The Event Horizon Telescope(EHT) project successfully captured images of BH shadows located at the center of both the M87 galaxy and our Milky Way\cite{EventHorizonTelescope:2019dse,EventHorizonTelescope:2022wkp}, marking a significant milestone in astronomical observations. Currently, the study of BH shadows is extensive in various spacetime, including binary BH shadows\cite{Shipley:2016omi}, different configurations of BHs\cite{Bambi:2008jg,Atamurotov:2013dpa,He:2024qka, He:2022yse, He:2021htq,Li:2021ypw,He:2022aox}, dynamically evolving spacetimes \cite{Mishra:2019trb}, wormholes \cite{Nedkova:2013msa, Chen:2024wtw,He:2024yeg}, and naked singularities\cite{Shaikh:2018lcc}. The presence of a luminous ring positioned beyond the event horizon of BHs has been demonstrated through images and associated research. The ring structure, resulting from the extensive accumulation of high-energy matter surrounding the BH, is commonly referred to as the photon ring. Not only does the photon ring serve as a valuable tool for testing general relativity (GR), but it also provides crucial insights into the structural composition of spacetime.

The current research on BH shadows primarily relies on geometric optics, focusing on the behavior of photons in close proximity to BHs. Recently, Hashimoto $et~al.$ employed wave optics to investigate the shadows cast by BHs, presenting a captivating and innovative approach\cite{Hashimoto:2019jmw}. Especially in \cite{Hashimoto:2018okj}, a significant breakthrough was made concerning the existence of a sequence of luminous ring structures, known as Einstein rings, which encircling the AdS Schwarzschild BH based on the principle of holographic duality. It has been observed that holographic images of the BH, which result from its gravitational lensing effect, can be constructed using the response function for a thermal state on a two-dimensional sphere which is dual to the Schwarzschild-AdS BH. Furthermore, clear observation of Einstein rings is possible. This methodology has also been employed in holographic superconductivity models\cite{Kaku:2021xqp}, wherein researchers have unveiled a discontinuous alteration in the size of the photon ring. The charged BHs\cite{Liu:2022cev} have also been studied accordingly, and the result showed that the chemical potential has no influence on the radius of the Einstein ring, and the temperature dependence of the Einstein ring exhibits a unique characteristic. In the context of other modified gravity theory, the study of holographic Einstein rings of BHs has yielded some interesting results, which can be found in\cite{Zeng:2023zlf,Zeng:2023tjb,Zeng:2023ihy,Hu:2023mai,Li:2024mdd,He:2024bll,Luo:2024wih}.

\par
The AdS-Schwarzschild BH, being a classical solution in the AdS spacetime, has been extensively investigated due to its profound insights into the nature of BHs, gravity, and the universe's structure within the AdS framework\cite{ Khan:2019jkz,
Gentile:2012jm,Dunn:2015xya,Dey:2017xty,Banerjee:2020dww}. In the context of modified gravity theories, studies on AdS-Schwarzschild BHs, including investigations of deformed solutions, are also of paramount importance. In the work 
\cite{Khosravipoor:2023jsl}, the author obtained deformed AdS-Schwarzschild BH solutions using the gravity decoupling (GD) method, which extends known solutions of the standard gravitational action to additional sources and modified theories of gravitation. Furthermore, the effects of deformation parameters on the horizon structure, thermodynamics and Hawking-Page phase transition temperature are also studied. However, the shadow of the deformed AdS-Schwarzschild BH remains unexplored. 

However, the  holographic Einstein rings associated with deformed ADS-Schwarzschild BHs remains uncharted territory, constituting the focal point of this paper. In this study, we aim to investigate the Einstein rings of the deformed AdS-Schwarzschild BH using wave optics, based on the AdS/CFT correspondence, and explore the impact of relevant physical parameters on the Einstein rings.

The paper is organized as follows. In Section 2, we briefly introduce the deformed AdS-Schwarzschild BH and derive the lensing response function that describes the diffraction of a wave source by the BH in this spacetime background. In Section 3, we employ an optical system equipped with a convex lens to observe the Einstein ring formed on the screen. We pay particular attention to the influence of the relevant physical parameters and the observer’s position on the Einstein ring. Furthermore, we compare the relevant conclusions obtained through wave optics with those derived from geometric optics. Finally, in Section 4, we summarize the key findings and conclusions of our study.

\section{The deformed AdS-Schwarzschild BH and its response function}
We consider the BH solution of a deformed AdS-Schwarzschild BH with the following metric\cite{Khosravipoor:2023jsl}\begin{equation}
\mathrm{d}s^{2} =-F(r)\mathrm{d}t^{2}  +\frac{1}{F(r)} \mathrm{d}r^{2} +r^{2} (\mathrm{d}\theta ^{2}+\sin ^{2}\theta \mathrm{d}\varphi ^{2}  ),\label{jie}
\end{equation}
with
\begin{equation}
F(r)=1-\frac{2M}{r} +\frac{r^{2}}{\tilde{l ^{2}}}+\alpha \frac{\beta ^{2}+3 r^{2}+3\beta r }{3r(\beta +r)^{3} } .\label{jie1}
\end{equation}
The parameter $\alpha$, known as the deformation parameter, modulates the influence of geometric deformations on the background geometry by adjusting their intensity, while $M$ represents the ADM mass of the BH. In addition, the term $\beta$ is a control parameter with a length dimension that serves to prevent the occurrence of the central singularity. $\tilde{l}  =\sqrt{3/\left | \Lambda  \right |}$ is the AdS radius, where $\Lambda$ is the cosmological constant. In what follows, the value of $\tilde{l}$ is taken as  $\tilde{l} =1$ for the purpose of simplifying calculations. Substituting the metric function mentioned above with the variable $r=1/y$ and $G(y)=F(1/r)$, Eq.(\ref{jie}) can be written as \begin{equation}
\mathrm{d}s^{2} =\frac{1}{y^{2} }[-G(y)\mathrm{d}t ^{2}+\frac{\mathrm{d}y^{2} }{G(y)}+ \mathrm{d}\theta ^{2}+\sin ^{2}\theta \mathrm{d}\varphi ^{2}  ],\label{dugui}
\end{equation}
\begin{equation}
G(y)=1+y^{2} -2My^{3}+\frac{y^{3}\alpha (\frac{3}{y^{2}  }+\frac{3\beta }{y} +\beta ^{2} )}{3(\frac{1}{y}+\beta)^{3} } .
\end{equation}

At $y=\infty$, a space-time singularity emerges, while at $y=0$, the corresponding system resides on the AdS boundary. Note that the temperature of the BH is defined by the Hawking temperature, which is given by the relation $T=\frac{1}{4\pi} {G}'(y_{h} ) $, where $y=y_{h}$ represents the event horizon of the BH.

Next, we consider the dynamics of a massless scalar field based on the following Klein-Gordon equation\cite{Hashimoto:2018okj}\begin{equation}
D_{a}D ^{a}\tilde{\Psi}=0  .
\end{equation}

In order to solve the Klein-Gordon equation more conveniently, we use the incident Eddington-Finkelstein coordinate system\cite{Liu:2022cev}, which facilitates our observation of the BH's nature and ensures the continuity of physical quantities at the event horizon. The coordinates can be expressed as \begin{equation}
\upsilon \equiv t+y_{*} =t-\int \frac{\mathrm{d}y}{G(y)} ,
\end{equation} hence, the metric function in Eq.(\ref{dugui}) can be expressed as follows\begin{equation}
\mathrm{d}s^{2} =\frac{1}{y^{2} }\Big[-G(y)\mathrm{d}\upsilon  ^{2}-2\mathrm{d}y \mathrm{d}\upsilon + \mathrm{d}\theta ^{2}+\sin ^{2}\theta \mathrm{d}\varphi ^{2}  \Big].
\end{equation}

\begin{figure}
	\centering 
\includegraphics[width=0.4\textwidth, angle=0]{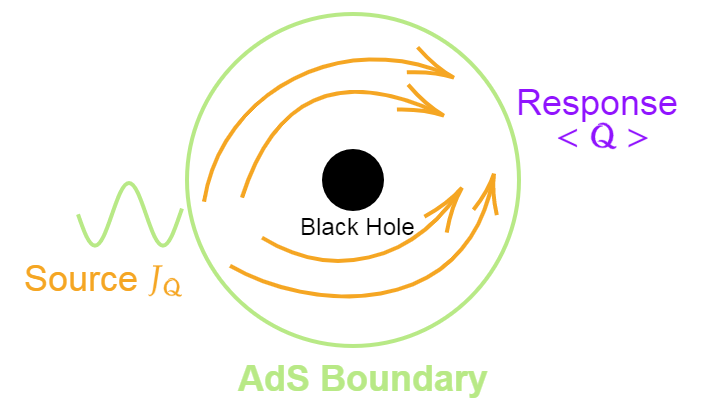}	
	\caption{Diagram of the response function and the source. } 
	\label{diagram-1}%
\end{figure}

Near the AdS boundary, the asymptotic solution of the scalar field is \begin{eqnarray}
\tilde{\Psi}(\upsilon ,y,\theta, \varphi  )&=&J_{\mathcal{Q} }  (\upsilon ,\theta, \varphi ) +y\partial_{\upsilon }  J_{\mathcal{Q} } (\upsilon ,\theta, \varphi )\nonumber\\&+&
\frac{1}{2} y^{2} D_{S}^{2} J_{\mathcal{Q} }  (\upsilon ,\theta, \varphi )+\left \langle \mathcal{Q}  \right \rangle y^{3} +Q(y^{4} ),
\end{eqnarray} where $D_{S}^{2}$ denotes the scalar Laplacian on the unit $S^{2}$. Based on the AdS/CFT framework, $J_{\mathcal{Q} } $ can be interpreted as the material source on the boundary\cite{Klebanov:1999tb}. We select a monochromatic and axisymmetric Gaussian wave packet as the wave source, located at the South pole $(\theta _{0} =\pi )$ of the AdS boundary,
as shown in Figure \ref{diagram-1}

\begin{eqnarray}
J_{\mathcal{Q} }  (\upsilon ,\theta )&=&e^{-i\omega \upsilon }(2\pi\eta ^{2} )^{-1} \mathrm{exp}[-\frac{(\pi -\theta )^{2} }{2\eta ^{2} }] \nonumber\\
&=&e^{-i\omega \upsilon }\sum_{l=0}^{\infty }C_{l0} X_{l0}(\theta )  ,
\end{eqnarray} where $\eta$ denotes the width of the wave  generated by the Gaussian source,  and $\eta \ll \pi $. The parameter $\omega$ represents the frequency of the incident wave. $X_{l0}(\theta )$  is the spherical harmonics function,  $C_{l0}$ is the coefficients of  $X_{l0}(\theta )$, which can be expanded as \begin{equation}
C_{l0}=(-1)^{l}\sqrt{\frac{l+1/2}{2\pi} }  \mathrm{exp}[-\frac{(l+1/2)^{2}\eta ^{2}  }{2} ].
\end{equation} Then, considering the symmetry of spacetime, the scalar field $\tilde{\Psi}(\upsilon ,y,\theta, \varphi  )$ can be decomposed into \begin{equation}
\tilde{\Psi}(\upsilon ,y,\theta, \varphi )=\sum_{l=0}^{\infty } \sum_{n=-l}^{l} e^{-i\omega \upsilon }C_{l0}Y_{l}(y)
X_{ln}(\theta ,\varphi )  .\label{biaoliang}
\end{equation} The corresponding response function  
$\left \langle \mathcal{Q} \right \rangle$ can be expanded as \begin{equation}
\left \langle \mathcal{Q} \right \rangle= \sum_{l=0}^{\infty } e^{-i\omega \upsilon }C_{l0}(\mathcal{Q})_{l} X_{l0} (\theta ).\label{xiangying}
\end{equation}

In Eq.(\ref{biaoliang}), 
the radial wave function $Y_{l}$  satisfies the equation of motion as follow \begin{eqnarray}
y^{2}G(y){Y}''  _{l} +(y^{2}{G}'(y)-2yG(y)+2i\omega y^{2}){Y}'_{l} \nonumber\\  +[-2i\omega y-y^{2}l(l+1)]{Y}_{l}=0.\label{yundong}
\end{eqnarray} According to the AdS/CFT correspondence, near the AdS boundary $Y_{l}$ can be expanded as \begin{equation}
\lim_{y \to 0} Y_{l} =1-i\omega y+\frac{y^{2}(-l(1+l)) }{2} +\left \langle \mathcal{Q}\right \rangle _{l} y^{3}
+Q (y^{4} ).
\end{equation} Obviously, the function $Y_{l}$ has two boundary conditions. One of which is at the event horizon $y=y_{h}$, and has the following form \begin{equation}
(y_{h}^{2} {G}' +2i\omega y_{h}^{2}){Y}'_{l}-[(2i\omega y_{h}+y_{h}^{2}l(l+1))]{Y}_{l}=0.
\end{equation}

\begin{figure}
	\centering 
\includegraphics[width=0.4\textwidth, angle=0]{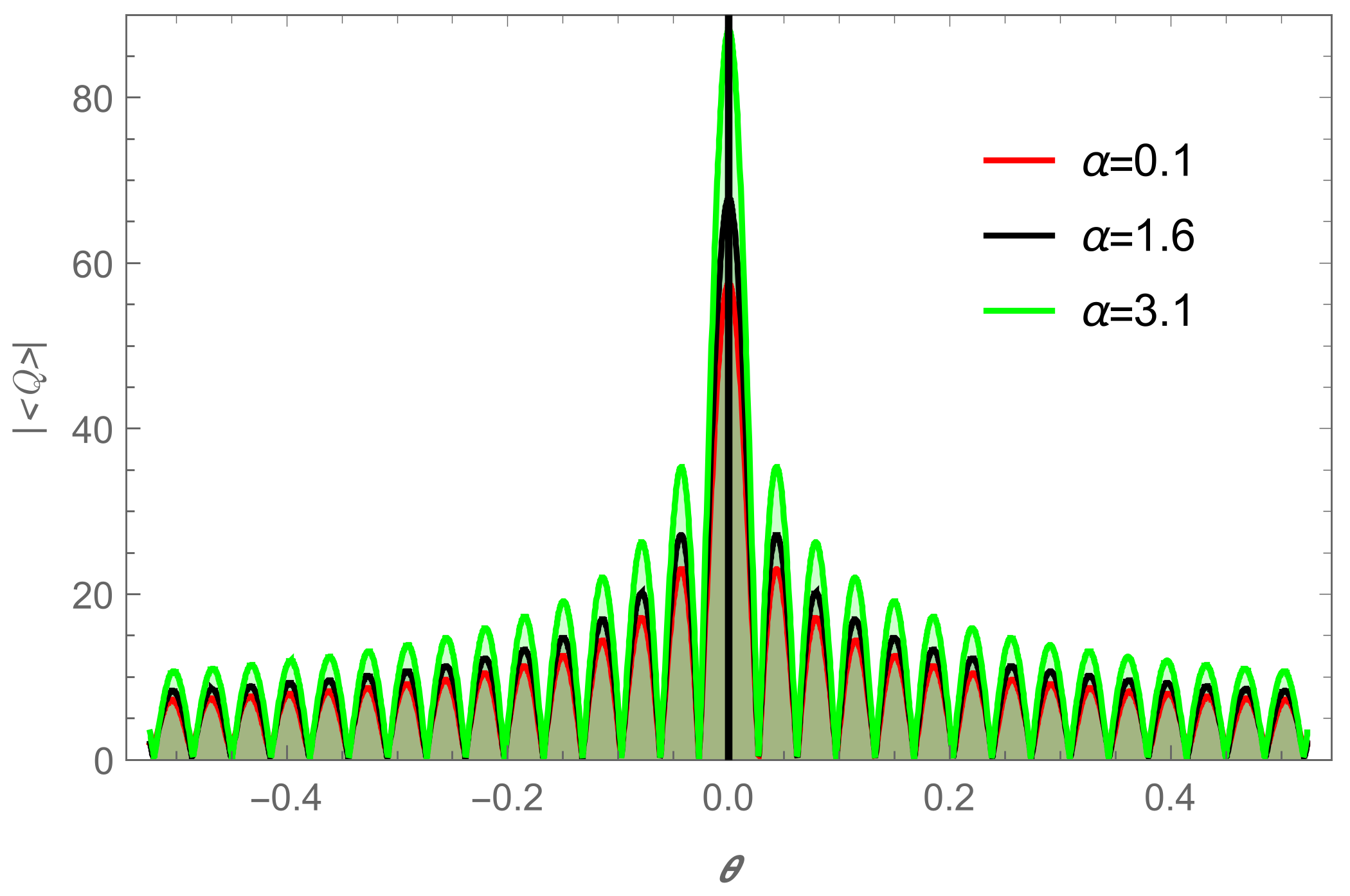}	
	\caption{Effect of different 
 $\alpha$  on the  response function, where $\beta=1$, $y_{h}=1$, $\omega =90$.}
	\label{2}%
\end{figure}

\begin{figure}
	\centering 
\includegraphics[width=0.4\textwidth, angle=0]{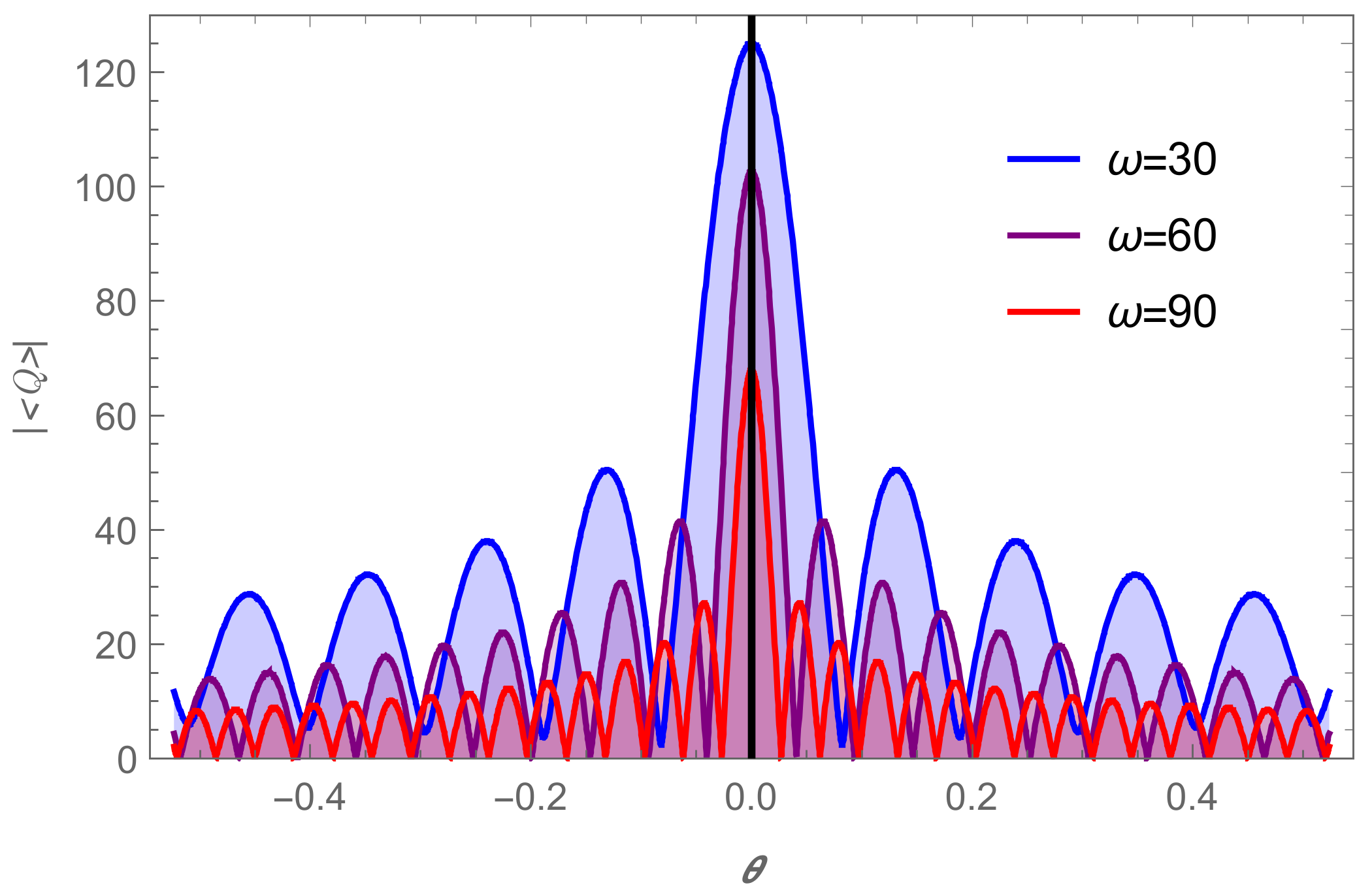}	
	\caption{Effect of different 
 $\omega$ on the response function, where $\alpha=1.6$, $y_{h}=1$, $\beta=1$.}
	\label{3}%
\end{figure}

\begin{figure}
	\centering 
\includegraphics[width=0.4\textwidth, angle=0]{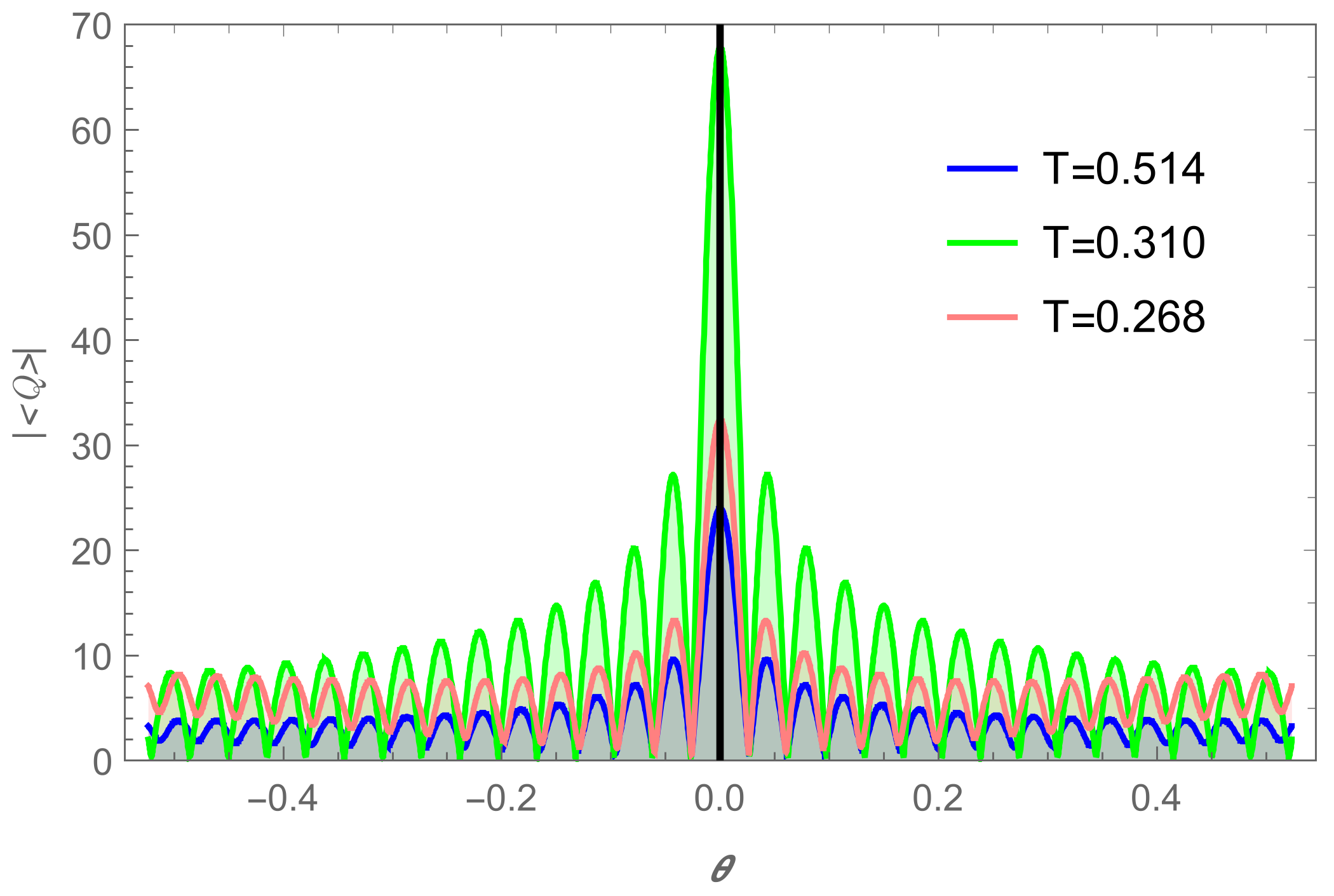}	
	\caption{Effect of different 
$\mathit{T}$  on the  response function, where $\alpha=1.6$, $\beta=1$, $\omega=90$.}
	\label{4}%
\end{figure}

The other is at $y=0$, namely the AdS boundary. In this case, the wave source  $J_{\mathcal{Q} }$ is the asymptotic form of the scalar field at infinity, and $Y_{l} (0)=1$, which can be seen from  Eq.(\ref{yundong}). We get the corresponding numerical solution of $Y_{l}$, and extract $\left \langle \mathcal{Q}\right \rangle _{l}$ by using the pseudo-spectral method\cite{Hashimoto:2019jmw,Hashimoto:2018okj}. Subsequently, using the Eq.(\ref{xiangying}) to get the value of the total response function. In Figure \ref{2} to Figure \ref{4}, fixing the control parameter $\beta$ at 1 and varying the deformation parameter $\alpha$ and other parameters to observe the amplitude of the response function. It can be observed from Figure \ref{2} that the amplitude of the response function increases with the increase of the deformation parameter $\alpha$. Meanwhile, Figure \ref{3} indicates that the period of the response function decreases with the increase of the wave source frequency $\omega$. As can be seen from Figure \ref{4}, the amplitude of the response function exhibits a notable variation with temperature, yet the relationship is not linear. For instance, at $\mathit{T} =0.310$, the amplitude of the response function attains its peak, and subsequently diminishes at $\mathit{T} =0.269$ and $\mathit{T} =0.514$.

\begin{figure}
	\centering 
\includegraphics[width=0.4\textwidth, angle=0]{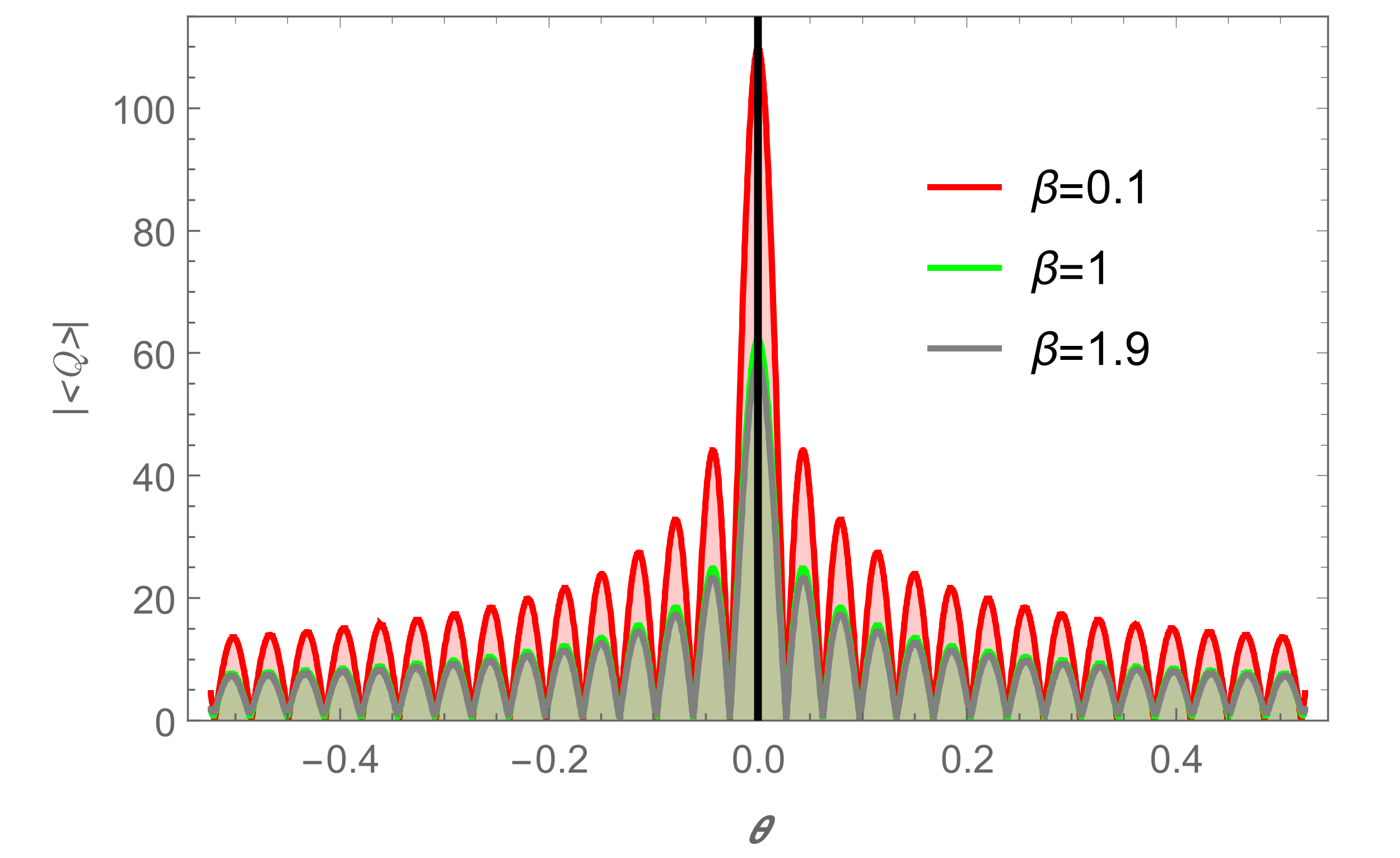}	
	\caption{Effect of different 
 $\beta$  on the  response function, where $\alpha=1$,  $y_{h}=1$, $\omega=90$.}
 \label{5}%
\end{figure}

\begin{figure}
	\centering 
\includegraphics[width=0.4\textwidth, angle=0]{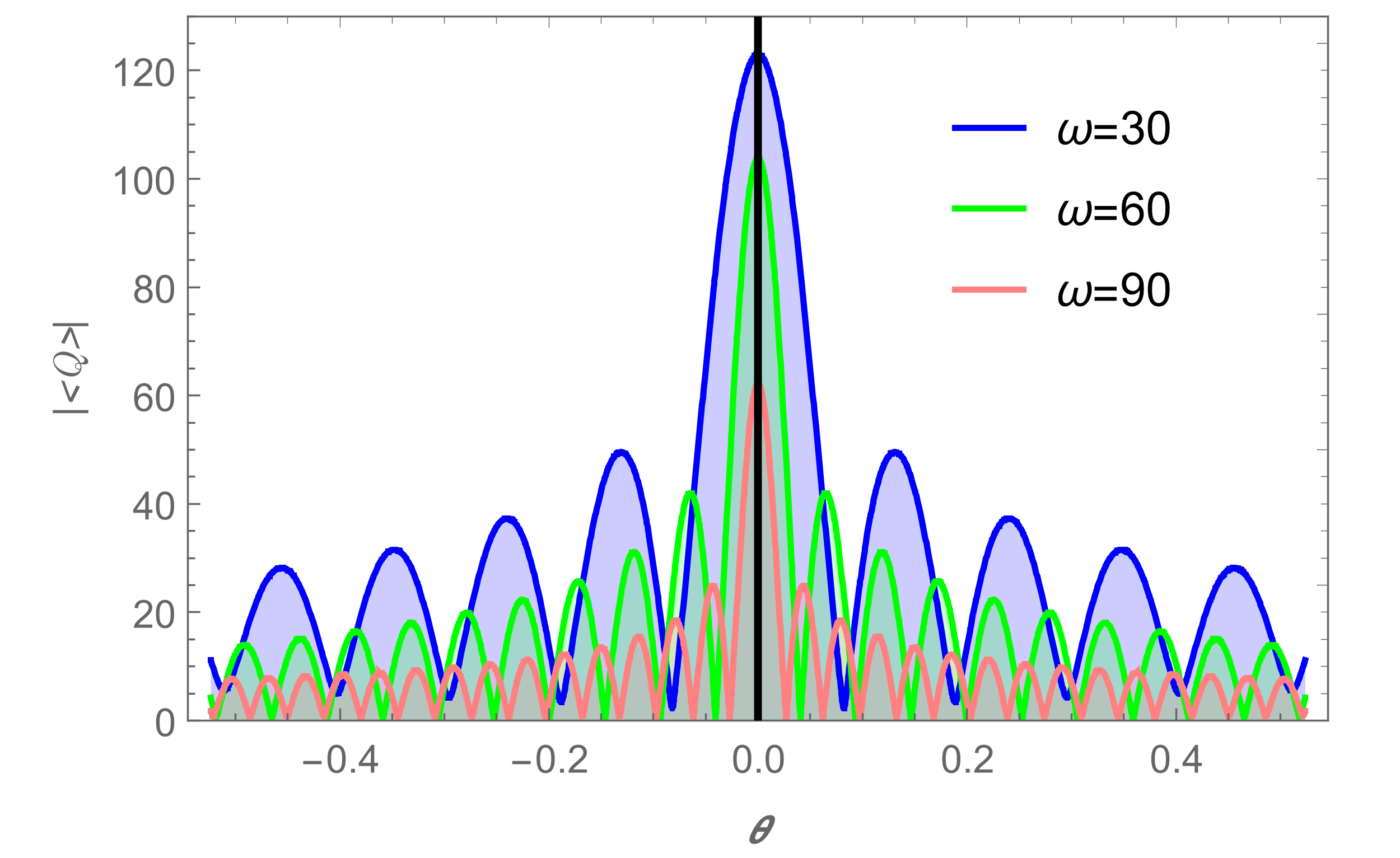}	
	\caption{Effect of different $\omega$ on the response function, where $\alpha=1$, $y_{h}=1$, $\beta=1$.} 
	\label{6}%
\end{figure}

\begin{figure}
	\centering 
\includegraphics[width=0.4\textwidth, angle=0]{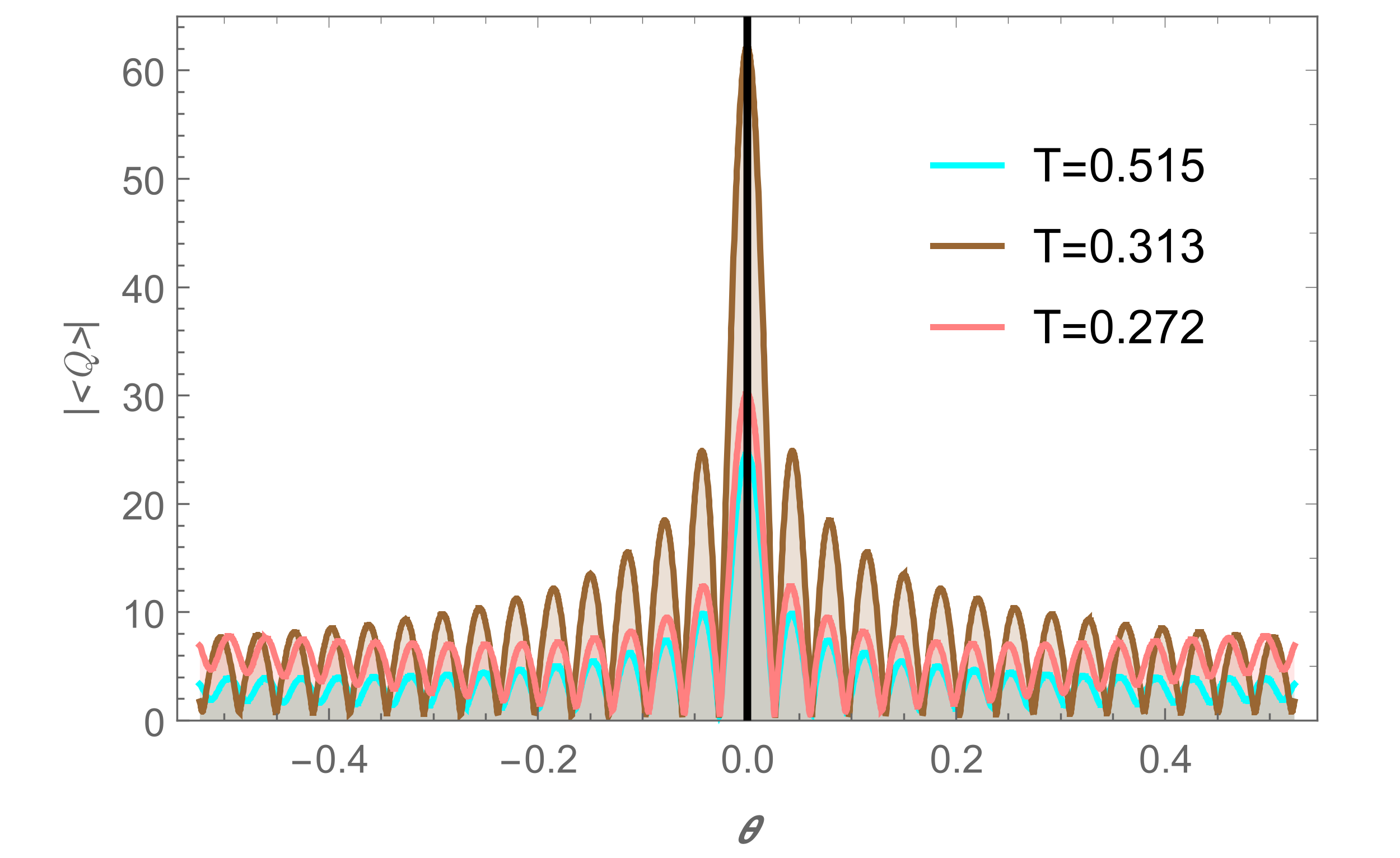}	
	\caption{Effect of different $\mathit{T}$ on the response function, where $\alpha=1$, $\beta=1$, $\omega =90$.}
	\label{7}%
\end{figure}

 For simplicity and without loss of generality, we fix the deformation parameter $\alpha=1$ and change the control parameter $\beta$ and other parameters, as shown in Figure \ref{5} to \ref{7}. The amplitude of the response function decreases with the increase of the control parameter $\beta$ and the wave source frequency $\omega$, as shown in Figure \ref{5} and \ref{6}, respectively. In Figure \ref{7}, the amplitude of the response function reached its maximum when $\mathit{T} =0.313$ and decreasing smoothly when $\mathit{T} =0.272$ and $\mathit{T} =0.515$.

\section{The formation of holographic ring}

Although the response function has been derived, it doesn't exhibit holographic image. To achieve this goal, it is necessary to incorporate an optical setup equipped with a convex lens. The convex lens converts a plane wave into a spherical wave, and the angle observed at the AdS boundary is $\theta _{obs}$. A new coordinate $(\theta ^{'} ,\varphi ^{'} )$ is obtained by rotating coordinate $(\theta  ,\varphi )$, and the following relation is satisfied\cite{Liu:2022cev} \begin{equation} \cos\varphi ^{'}  +i\cos\theta ^{'}  =e^{i\theta _{obs} }  (\sin\theta  \cos\varphi +i\cos\theta  ), \end{equation} where $\theta ^{'}=0 ,\varphi ^{'}=0$ correspond to the observational center. A Cartesian coordinate system $(x,y,z)$ is established such that  $(x,y)=(\theta^{'} \cos\varphi ^{'}  ,\theta^{'} \sin\varphi ^{'})$ at the boundary where the observer located, for a virtual optical system. The convex lens is adjusted on a two-dimensional plane $(x,y)$, in which the focal length of the lens and corresponding radius are denoted by $f$ and $d$, respectively. Further, the coordinates on the spherical screen are defined as $(x,y,z)=(x_{SC},y_{SC},z_{SC} )$, satisfying $x^{2} _{SC}+y^{2} _{SC}+z^{2} _{SC} =f^{2} $. The relationship between the incident wave $\Psi (\tilde{x})$ before passing through the convex lens and the outgoing wave  $\Psi _{T} (\tilde{x})$ after passing through the convex lens satisfies \begin{equation}
\Psi _{T} (\tilde{x} )=e^{-i\hat{\omega }\frac{\left |\tilde{x}  \right | }{f} } \Psi (\tilde{x} ).
\end{equation} The wave function on the screen can be represented as\cite{Liu:2022cev} \begin{eqnarray}
\Psi _{SC} (\tilde{x}_{SC}  )&=&\int_{\left |\tilde{x}  \right |\le d}d^{2} x\Psi _{T} (\tilde{x} )
e^{i\omega R}\nonumber\\&\propto &\int_{\left |\tilde{x}  \right |\le d}d^{2} x\Psi(\tilde{x} )e^{-i\frac{\omega}{2f}
  \tilde{x}\cdot \tilde{x}_{SC} }\nonumber\\&=&\int d^{2} x\Psi (\tilde{x} )\sigma (\tilde{x})e^{-i\frac{\omega}{2f},
  \tilde{x}\cdot \tilde{x}_{SC} },\label{bo}
\end{eqnarray} where $R$ is the distance from the point $(x,y,0)$ on the lens to the point $(x^{2} _{SC},y^{2} _{SC},z^{2} _{SC} )$ on the screen. In the formula, $\sigma (\tilde{x})$ is the window function, which is defined as \begin{equation}
\sigma (\tilde{x}): = 
\begin{cases}
  1,~~~~0\le\tilde{x} \le d;  \\
  0,~~~~~~~~~~~\tilde{x} \ge  d.
\end{cases} 
\end{equation}

From Eq.(\ref{bo}), it is evident  that the observed wave on the screen is associated with the incident wave through the Fourier transform. In this article, the response function is considered as the incident wave $\Psi (\tilde{x}$), and we can observe the holographic images on this screen, using Eq.(\ref{bo}). The effects of the deformation parameter $\alpha$, control parameter $\beta$, and Gaussian wave source on holographic images are investigated, with the source width  $\eta=0.02$  and radius of convex lens $d=0.6$.

\subsection{The effect of deformation
parameter $\alpha$ on the images}

\begin{figure*}[htbp]
\centering
\begin{subfigure}[b]{0.24\textwidth}
  \centering
  \includegraphics[width=\textwidth]{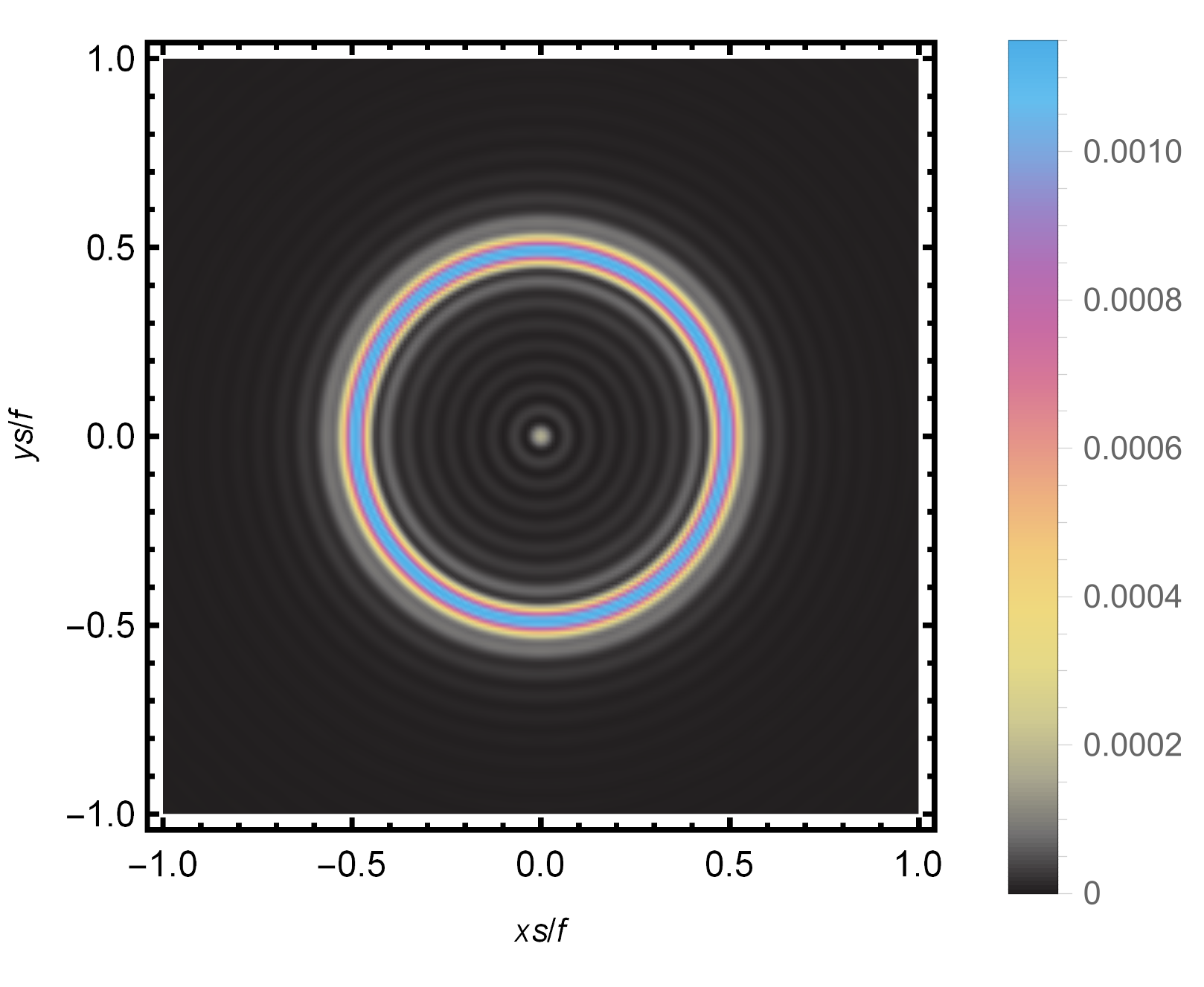} \caption{$\alpha=0.1$,$\theta _{obs}=0$}
\end{subfigure}
\hfill
\begin{subfigure}[b]{0.24\textwidth}
  \centering
  \includegraphics[width=\textwidth]{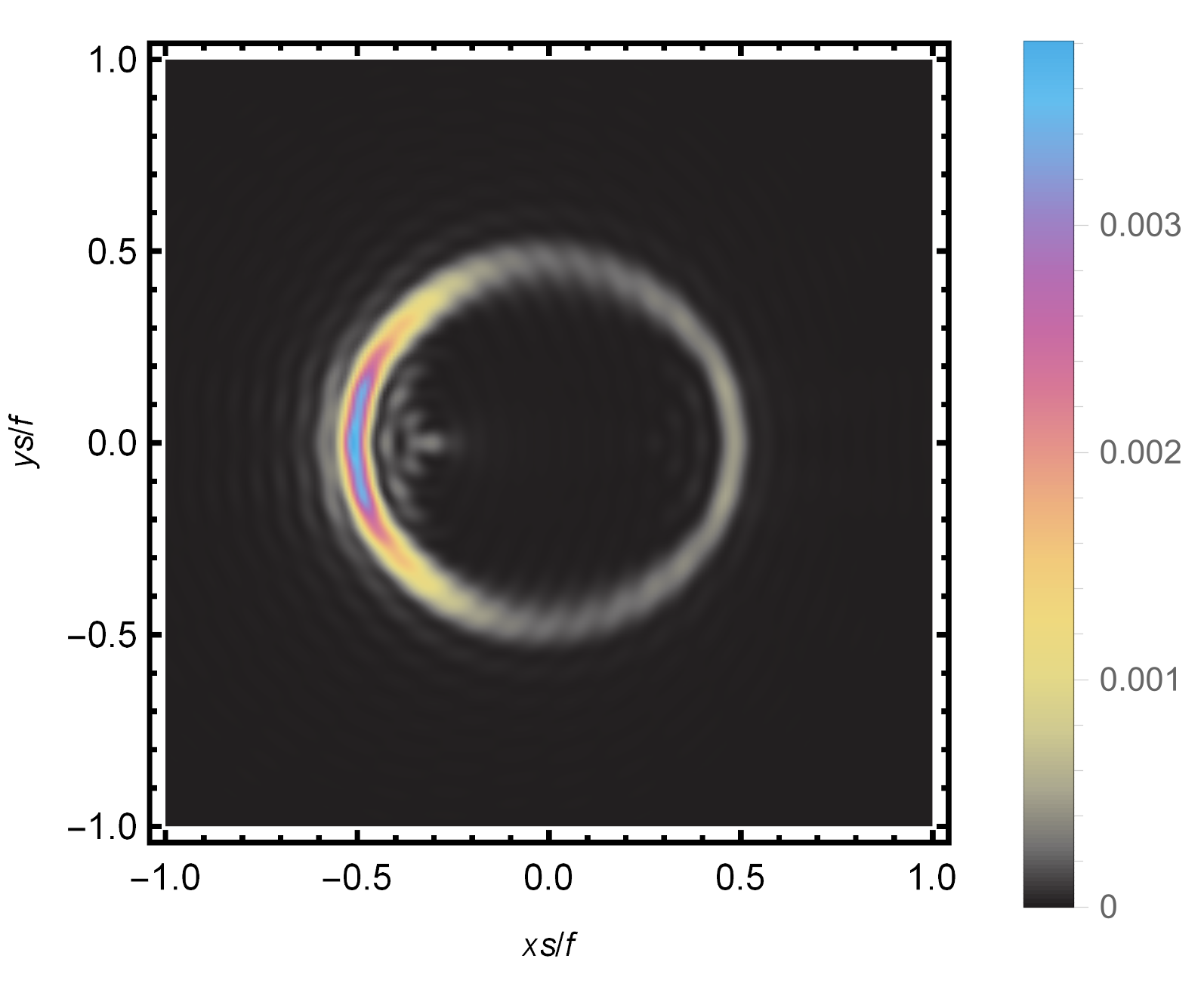}
  \caption{$\alpha=0.1$,$\theta _{obs}=\pi/6$}
\end{subfigure}
\hfill
\begin{subfigure}[b]{0.24\textwidth}
  \centering
  \includegraphics[width=\textwidth]{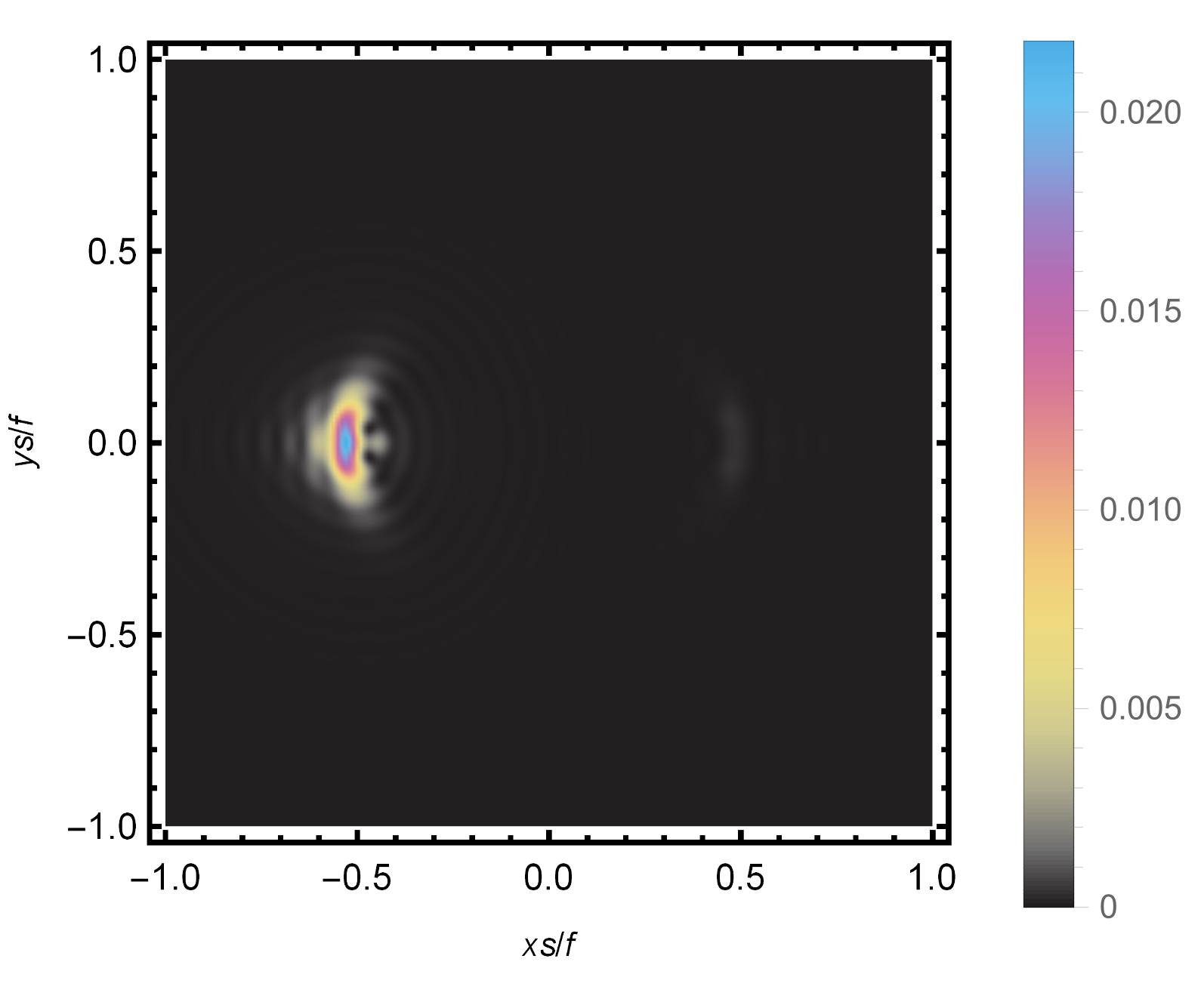}
  \caption{$\alpha=0.1$,$\theta _{obs}=\pi/3$}
\end{subfigure}
\hfill
\begin{subfigure}[b]{0.24\textwidth}
  \centering
  \includegraphics[width=\textwidth]{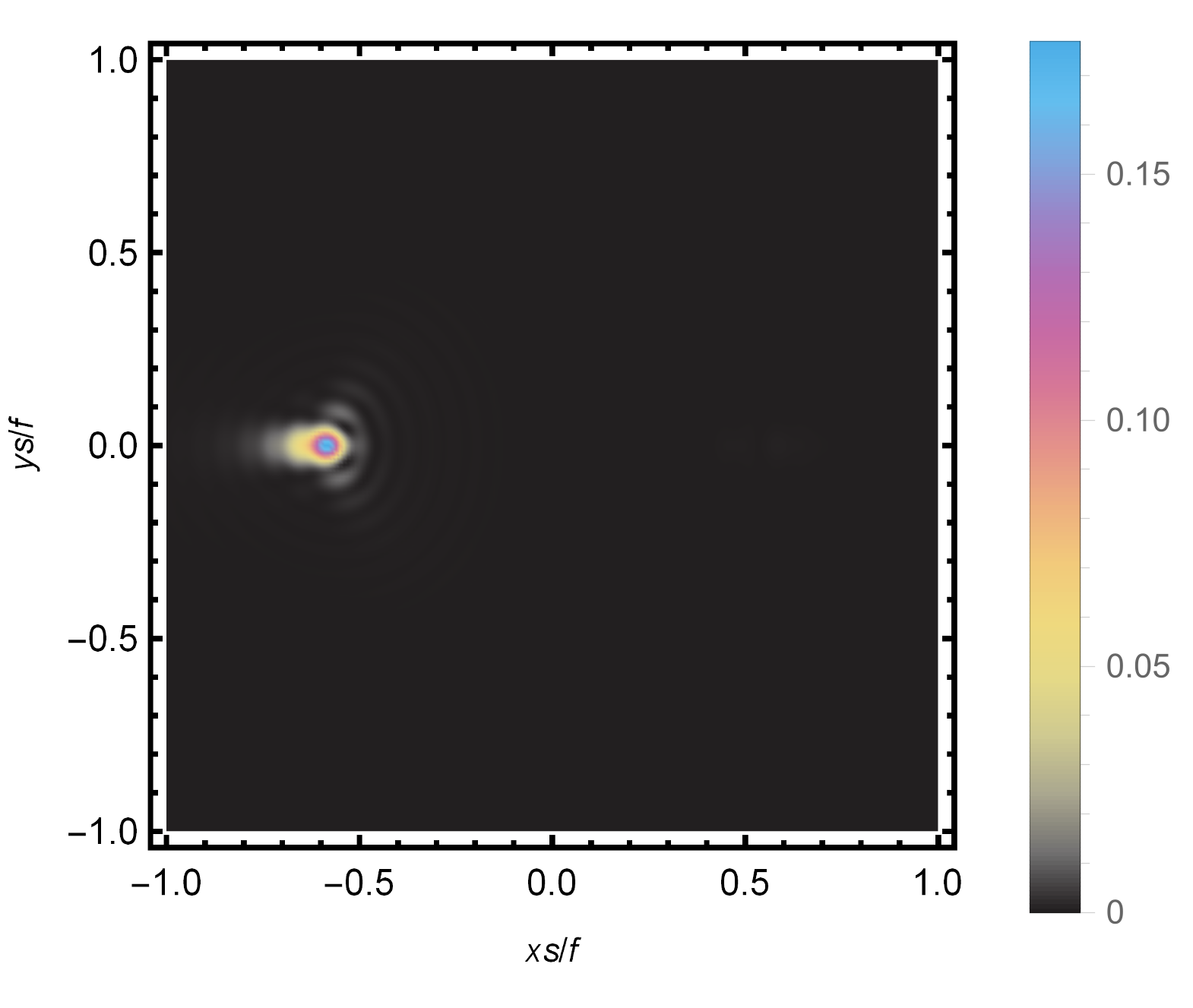}
  \caption{$\alpha=0.1$,$\theta _{obs}=\pi/2$}
\end{subfigure}
\hfill
\begin{subfigure}[b]{0.24\textwidth}
  \centering
\includegraphics[width=\textwidth]{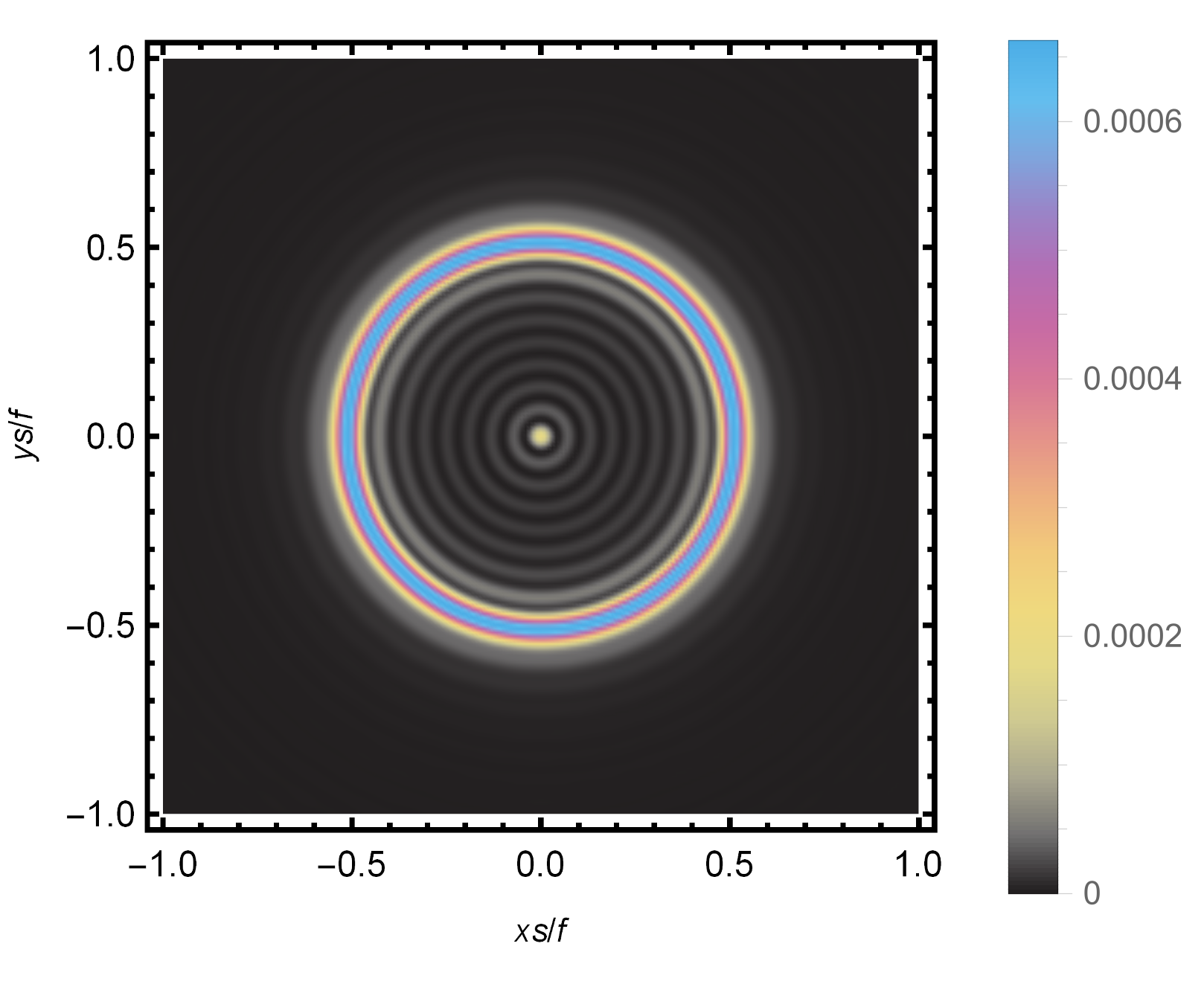}
  \caption{$\alpha=5.1$,$\theta _{obs}=0$}
\end{subfigure}
\hfill
\begin{subfigure}[b]{0.24\textwidth}
  \centering
  \includegraphics[width=\textwidth]{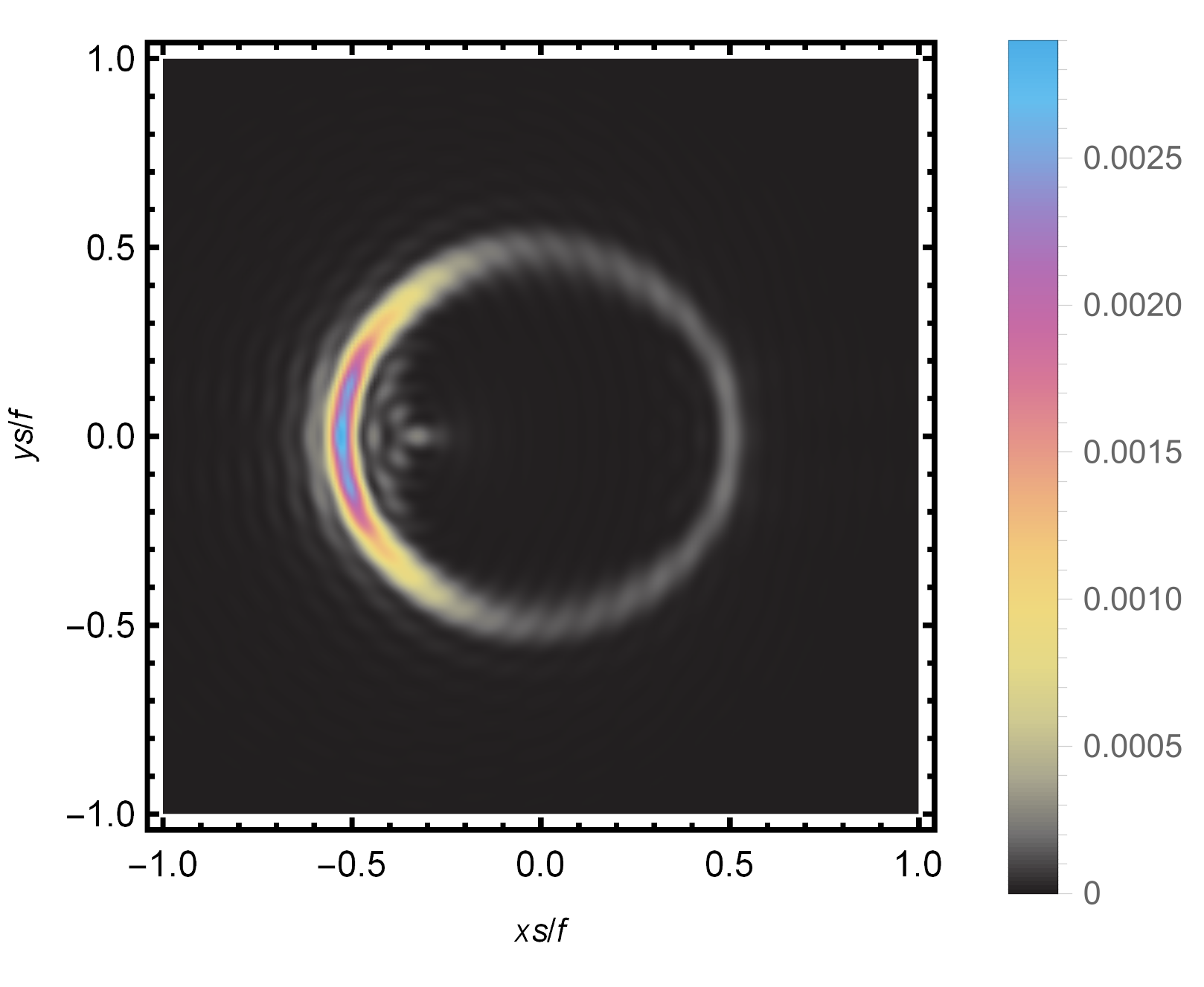}
  \caption{$\alpha=5.1$,$\theta _{obs}=\pi/6$}
\end{subfigure}
\hfill
\begin{subfigure}[b]{0.24\textwidth}
  \centering
  \includegraphics[width=\textwidth]{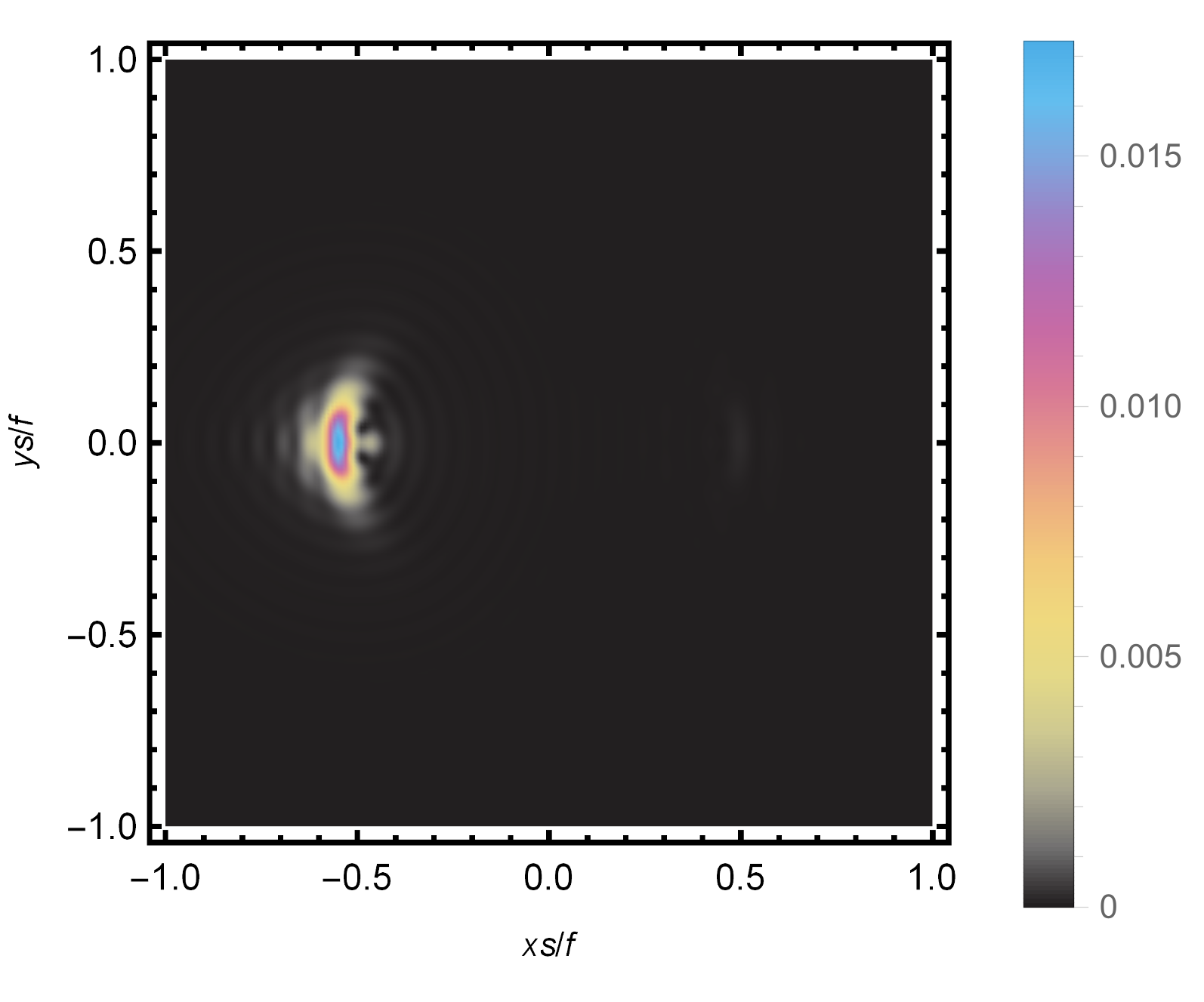}
  \caption{$\alpha=5.1$,$\theta _{obs}=\pi/3$}
\end{subfigure}
\hfill
\begin{subfigure}[b]{0.24\textwidth}
  \centering
  \includegraphics[width=\textwidth]{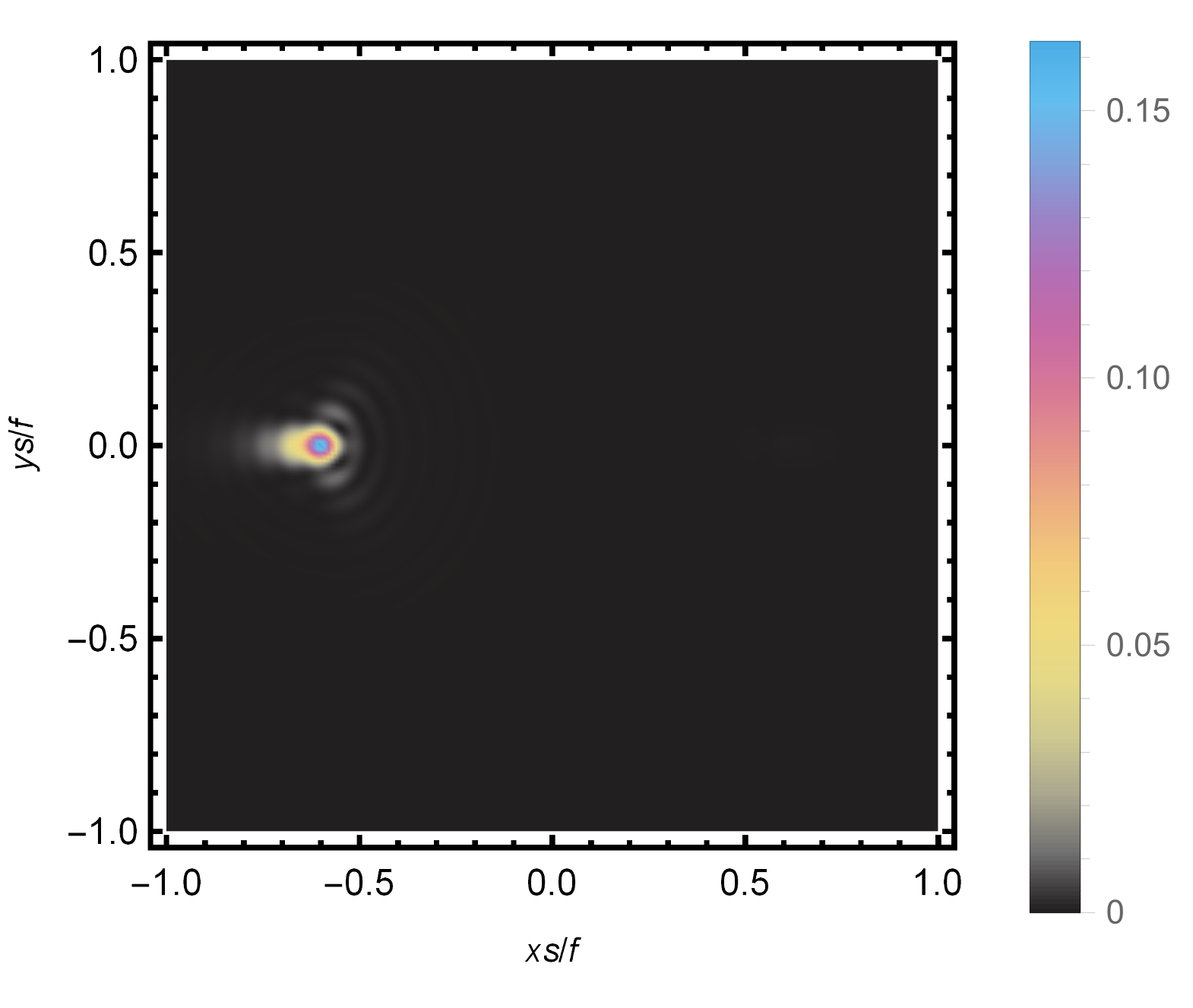}
  \caption{$\alpha=5.1$,$\theta _{obs}=\pi/2$}
\end{subfigure}
\hfill
\begin{subfigure}[b]{0.24\textwidth}
  \centering
\includegraphics[width=\textwidth]{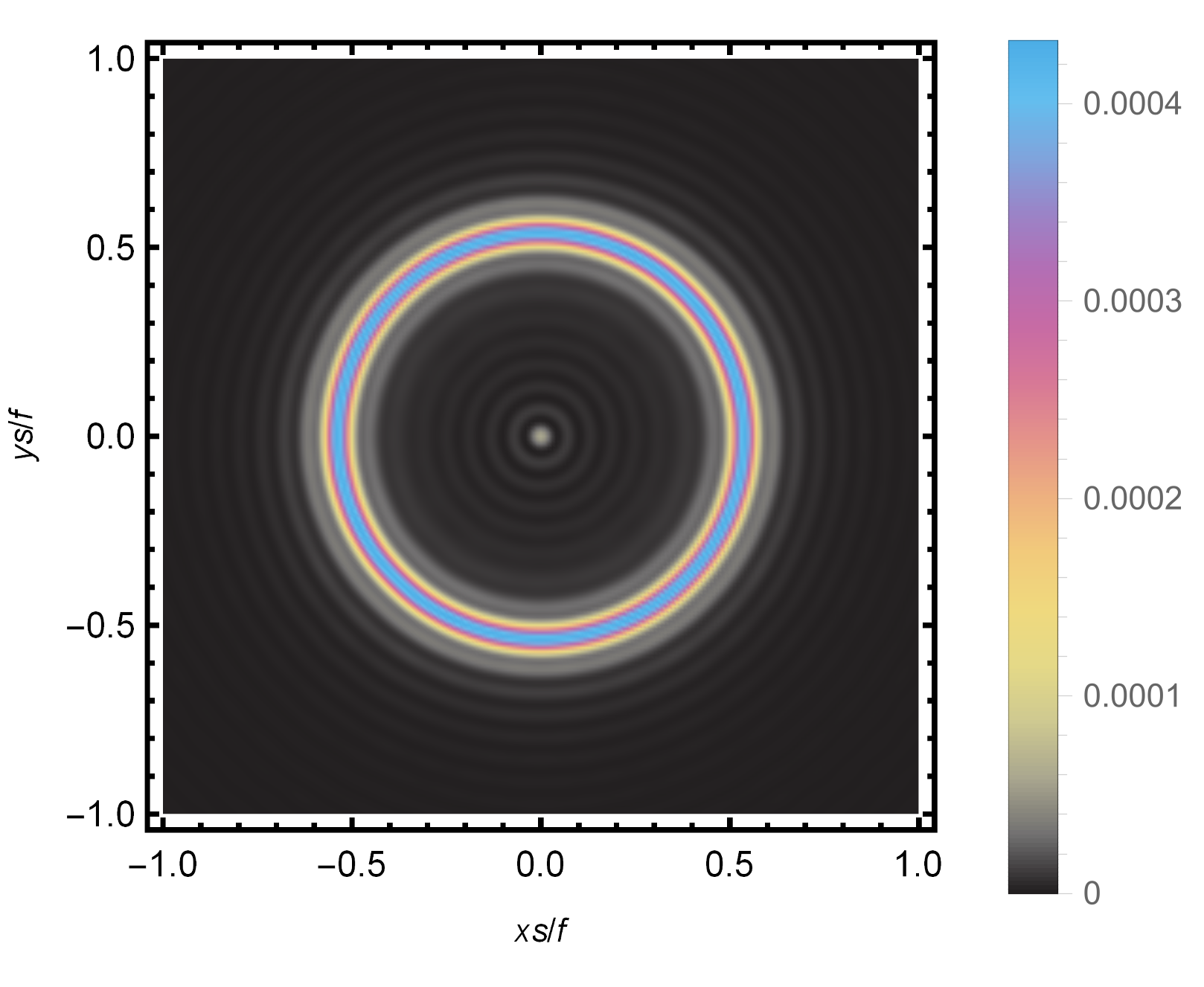} \caption{$\alpha=10.1$,$\theta _{obs}=0$}
\end{subfigure}
\hfill
\begin{subfigure}[b]{0.24\textwidth}
  \centering
  \includegraphics[width=\textwidth]{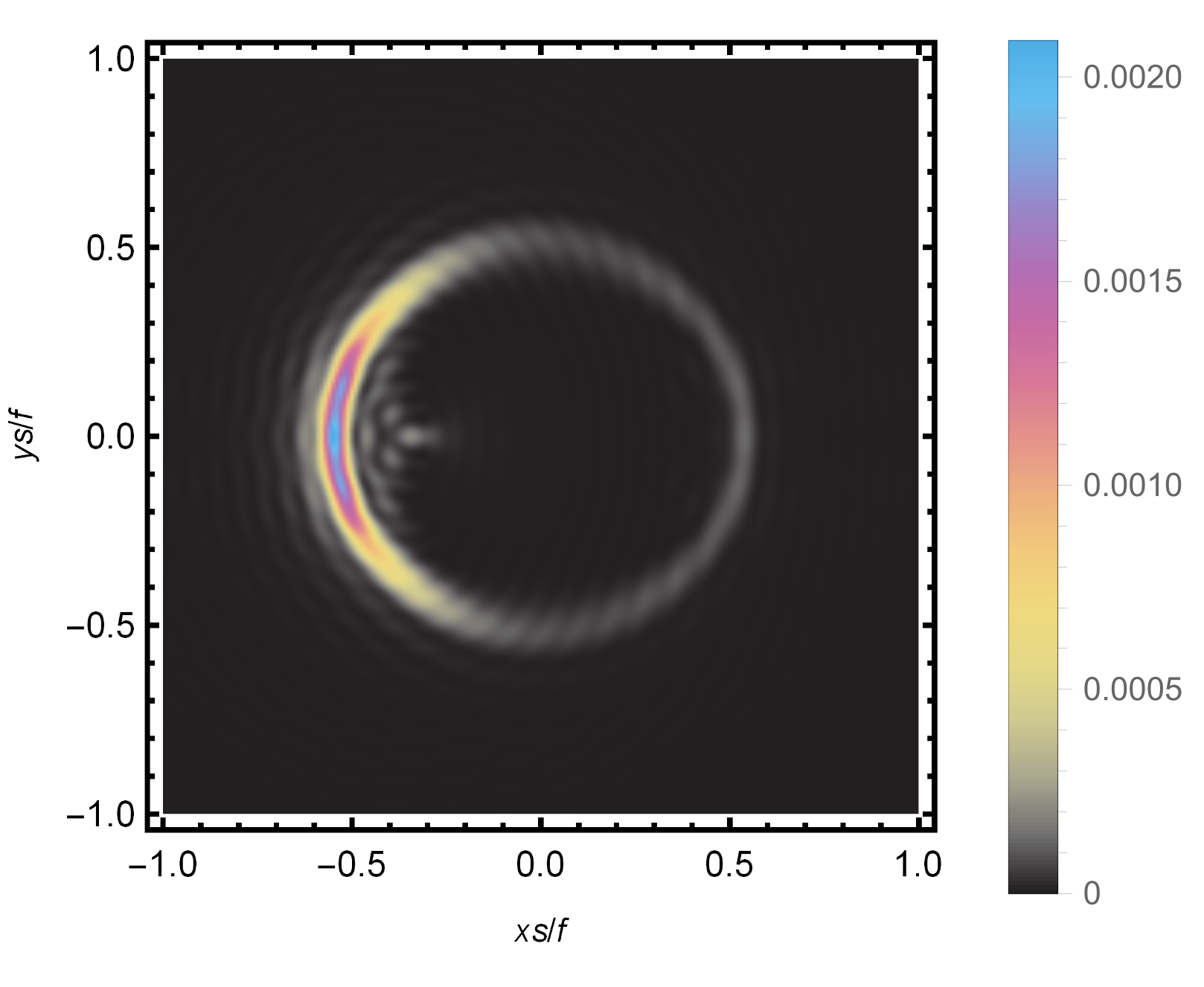}
  \caption{$\alpha=10.1$,$\theta _{obs}=\pi/6$}
\end{subfigure}
\hfill
\begin{subfigure}[b]{0.24\textwidth}
  \centering
  \includegraphics[width=\textwidth]{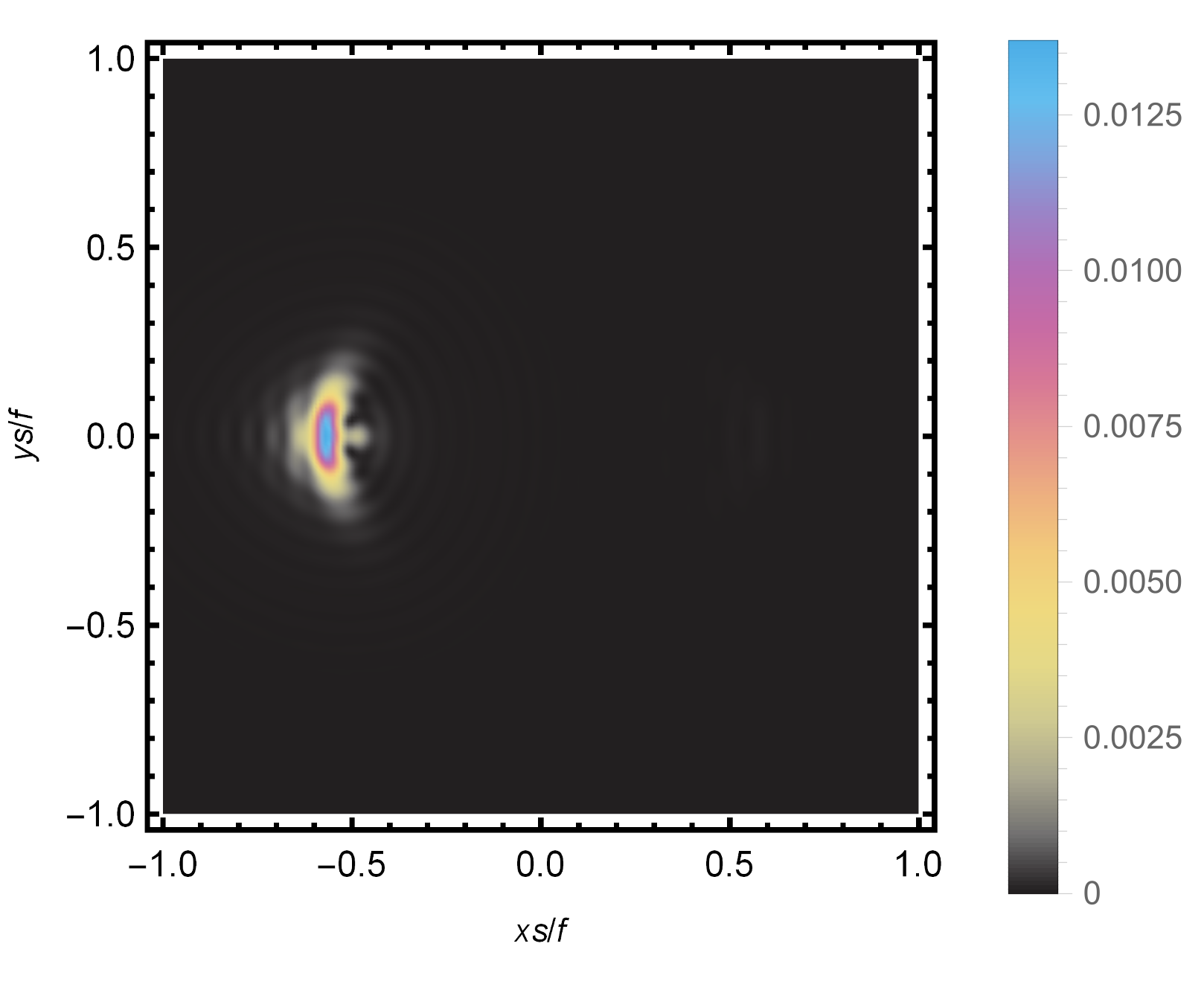}
  \caption{$\alpha=10.1$,$\theta _{obs}=\pi/3$}
\end{subfigure}
\hfill
\begin{subfigure}[b]{0.24\textwidth}
  \centering
  \includegraphics[width=\textwidth]{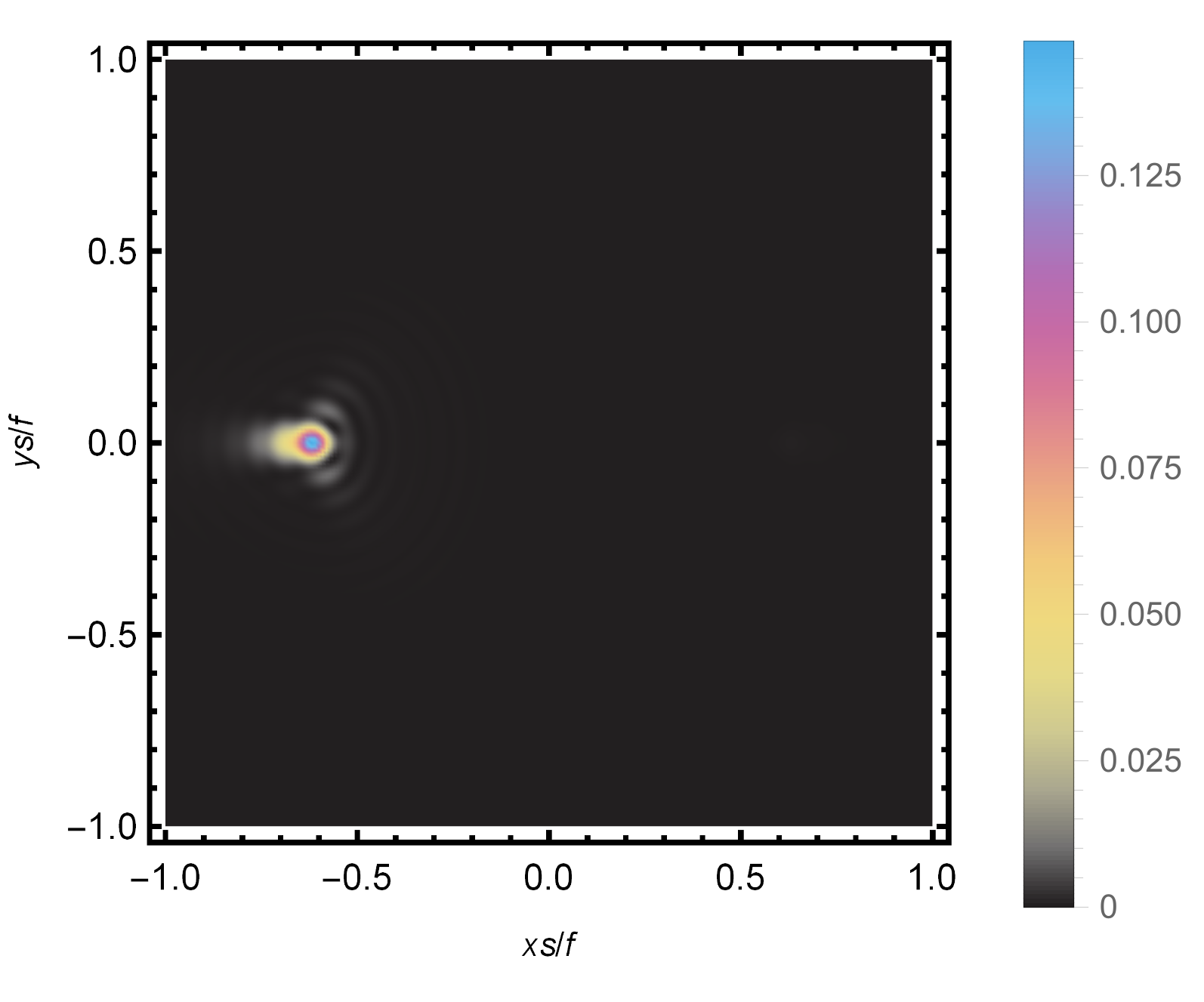}
  \caption{$\alpha=10.1$,$\theta _{obs}=\pi/2$}
\end{subfigure}
\hfill
\begin{subfigure}[b]{0.24\textwidth}
  \centering
\includegraphics[width=\textwidth]{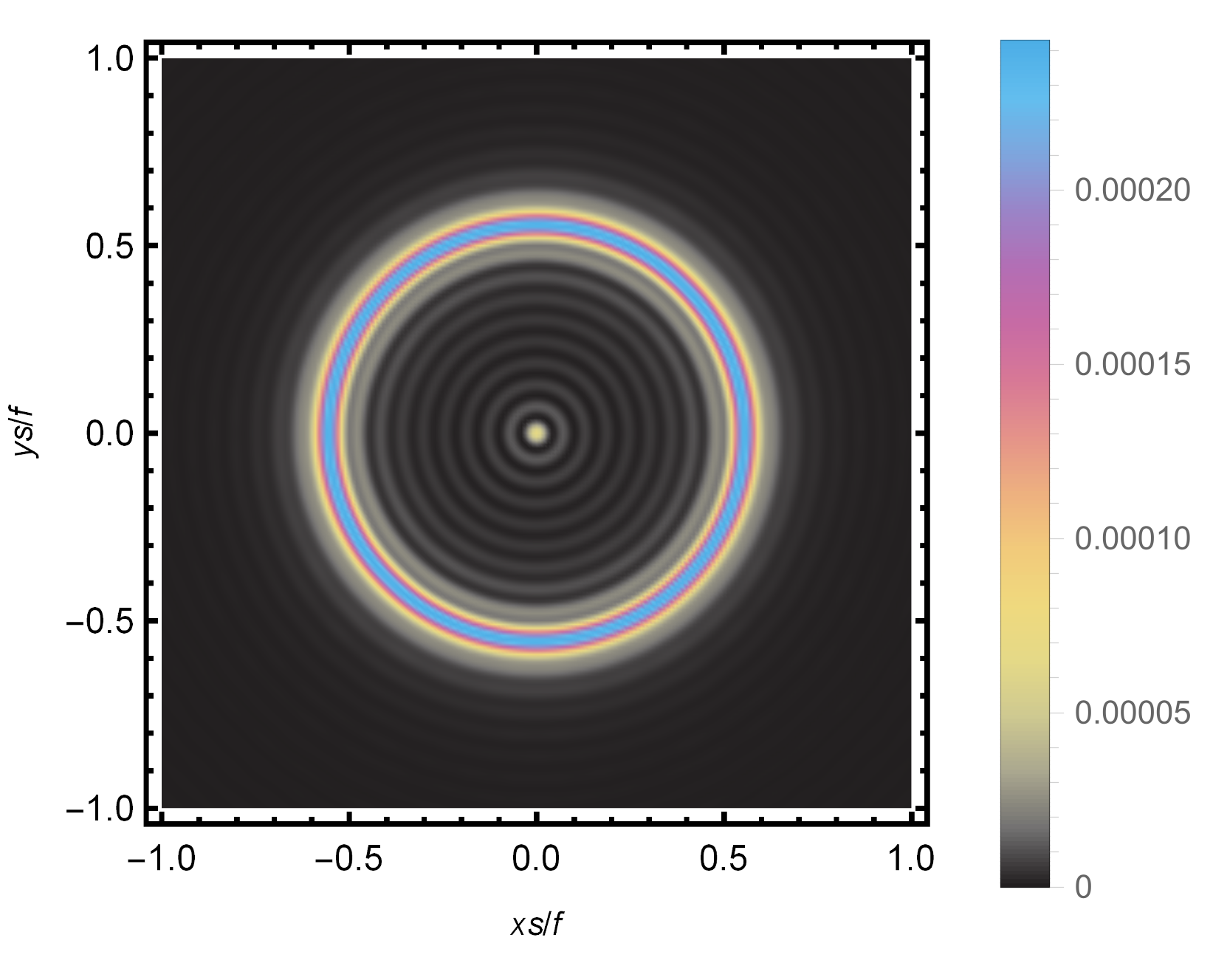}
  \caption{$\alpha=15.1$,$\theta _{obs}=0$}
\end{subfigure}
\hfill
\begin{subfigure}[b]{0.24\textwidth}
  \centering
  \includegraphics[width=\textwidth]{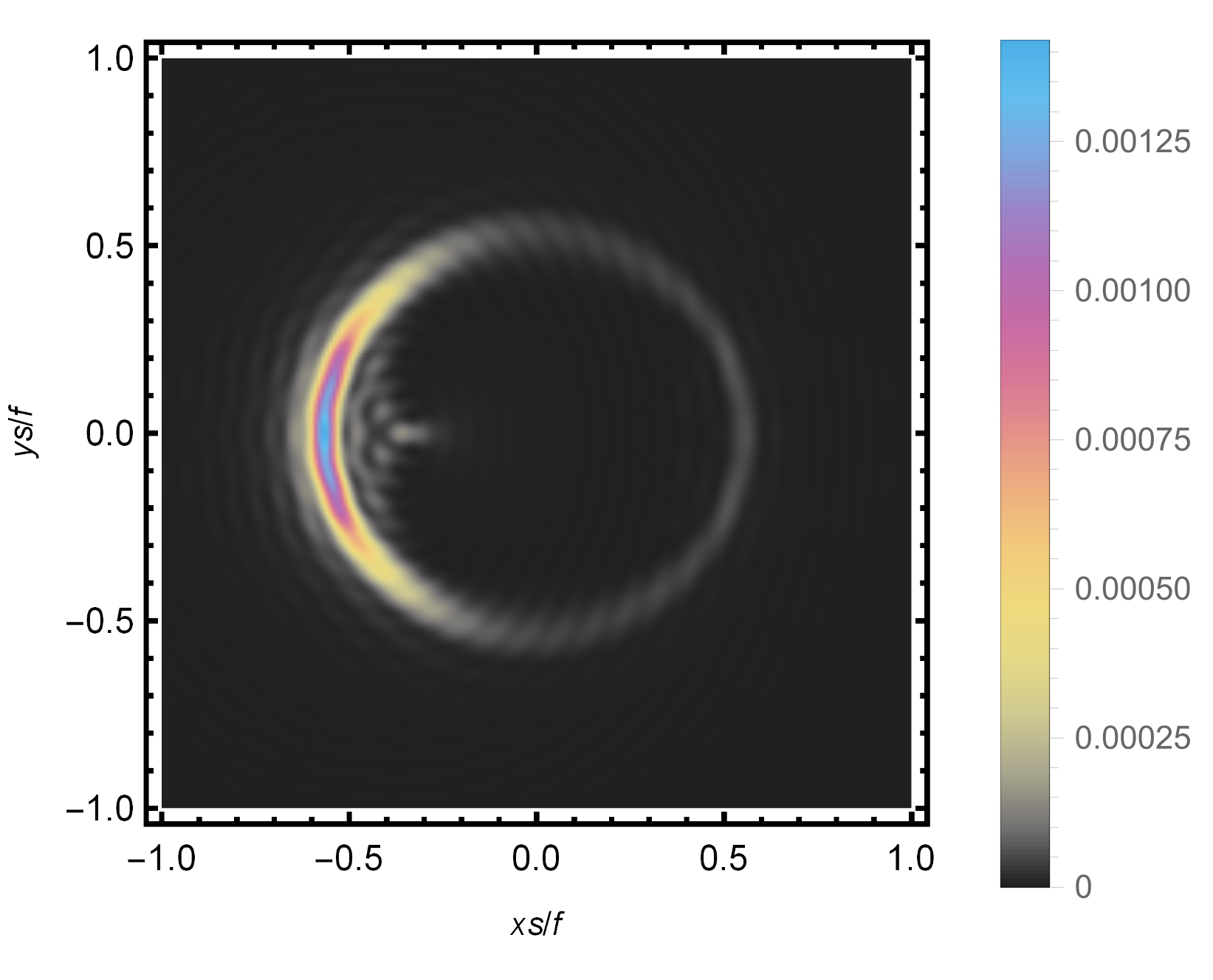}
  \caption{$\alpha=15.1$,$\theta _{obs}=\pi/6$}
\end{subfigure}
\hfill
\begin{subfigure}[b]{0.24\textwidth}
  \centering
  \includegraphics[width=\textwidth]{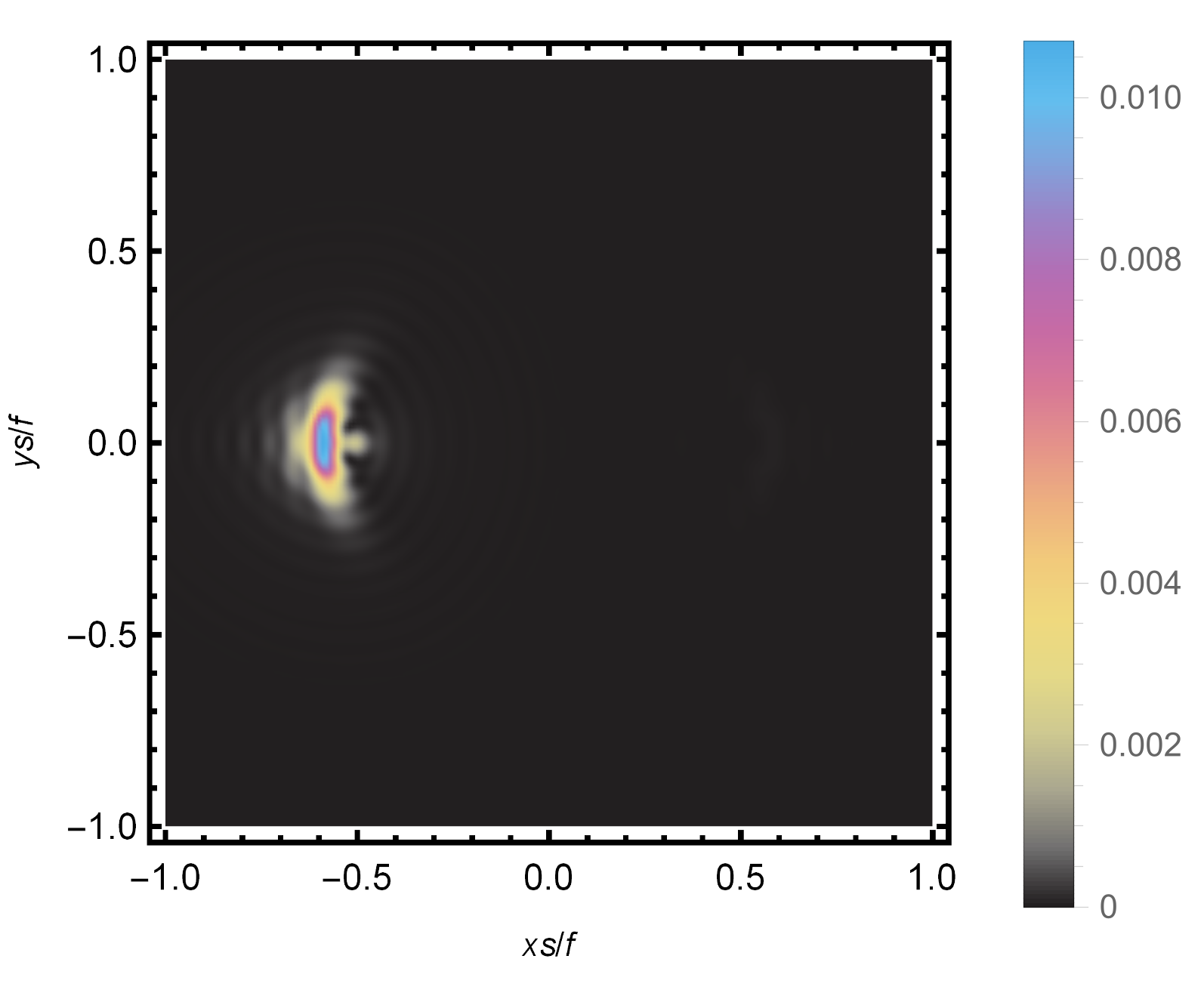}
  \caption{$\alpha=15.1$,$\theta _{obs}=\pi/3$}
\end{subfigure}
\hfill
\begin{subfigure}[b]{0.24\textwidth}
  \centering
  \includegraphics[width=\textwidth]{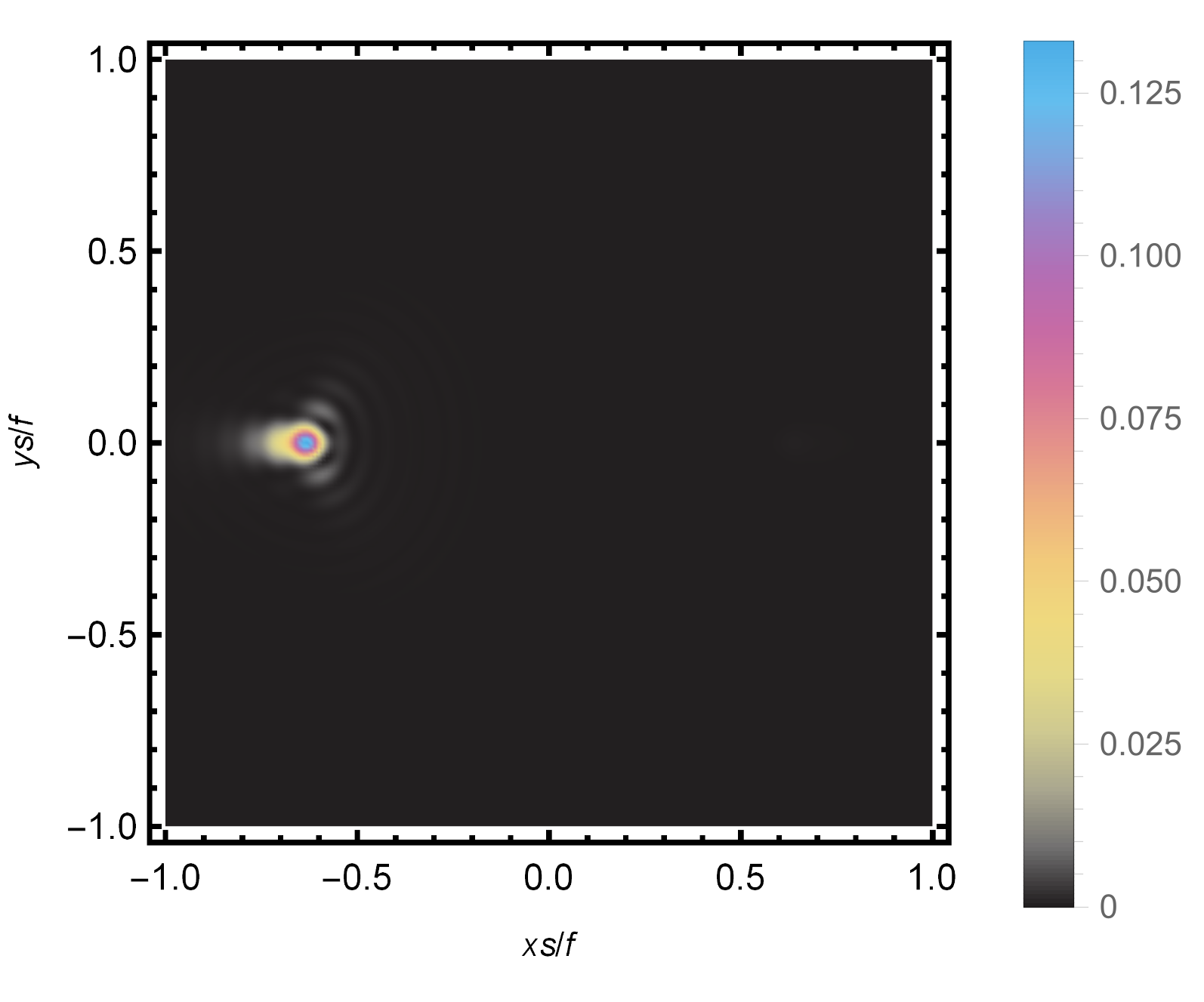}
  \caption{$\alpha=15.1$,$\theta _{obs}=\pi/2$}
\end{subfigure}
\hfill
\begin{subfigure}[b]{0.24\textwidth}
  \centering
\end{subfigure}
\caption{Observational appearance of the response on the screen for different $\alpha$ at various observation angles, where $y_{h}=5$, $\omega =90$.}
\label{8}%
\end{figure*}

\begin{figure}[htbp]
  \centering
  \begin{subfigure}[b]{0.45\columnwidth}
    \centering
    \includegraphics[width=\textwidth,height=0.9\textwidth]{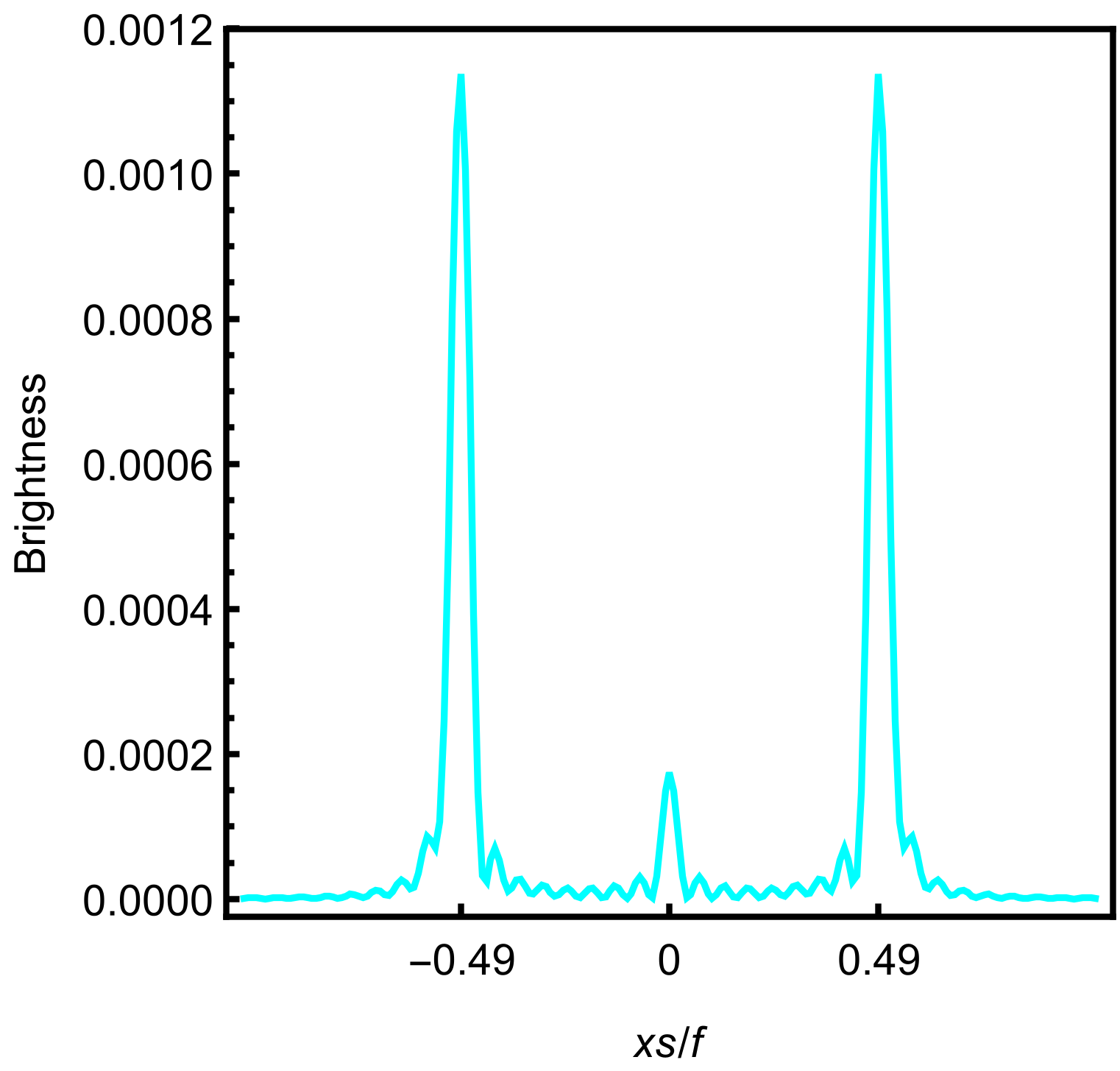}
    \caption{$\alpha=0.1$}
  \end{subfigure}
  \hfill
  \begin{subfigure}[b]{0.45\columnwidth}
    \centering
    \includegraphics[width=\textwidth,height=0.9\textwidth]{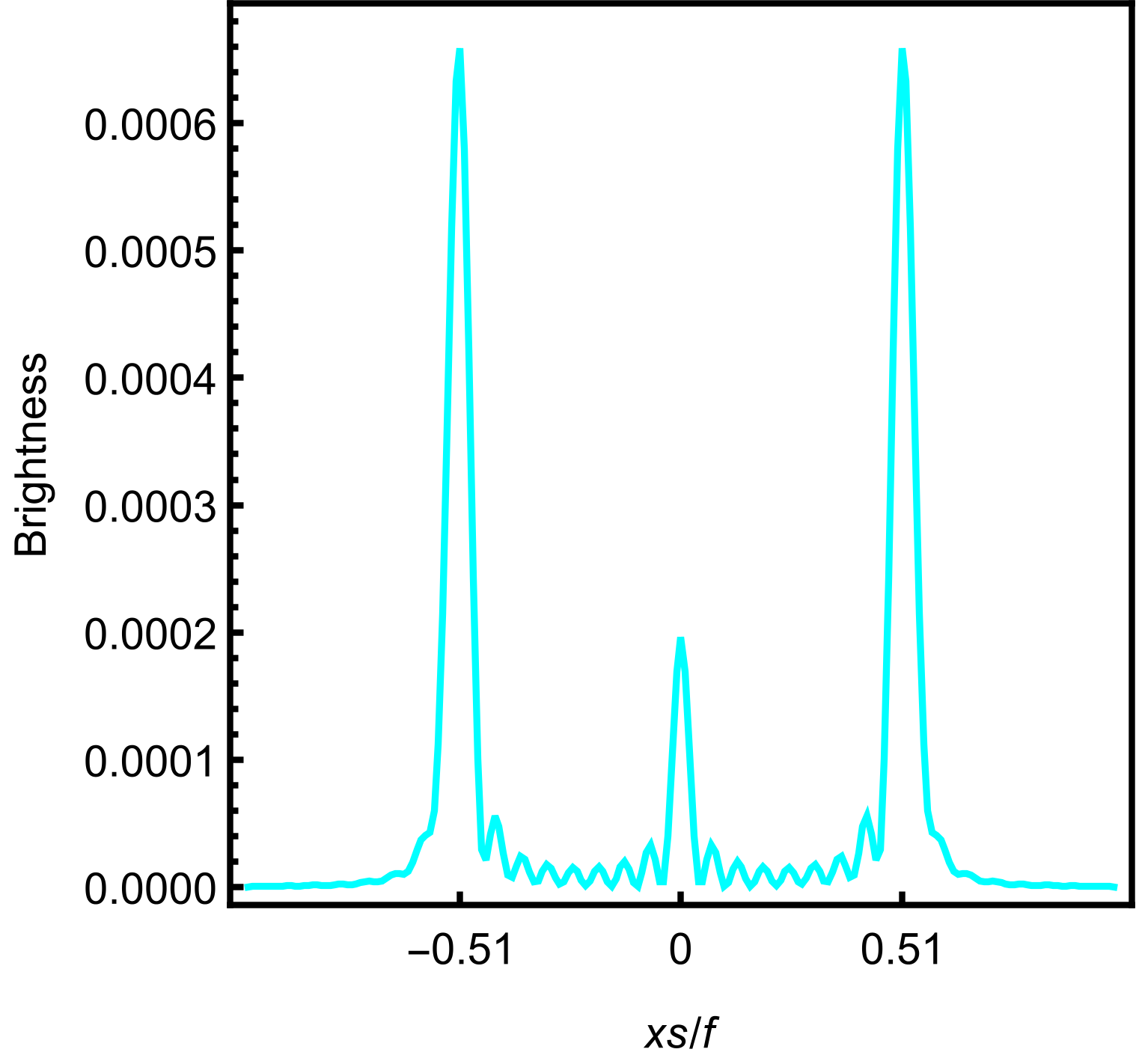}
    \caption{$\alpha=5.1$}
  \end{subfigure}
\begin{subfigure}[b]{0.45\columnwidth}
    \centering
    \includegraphics[width=\textwidth,height=0.9\textwidth]{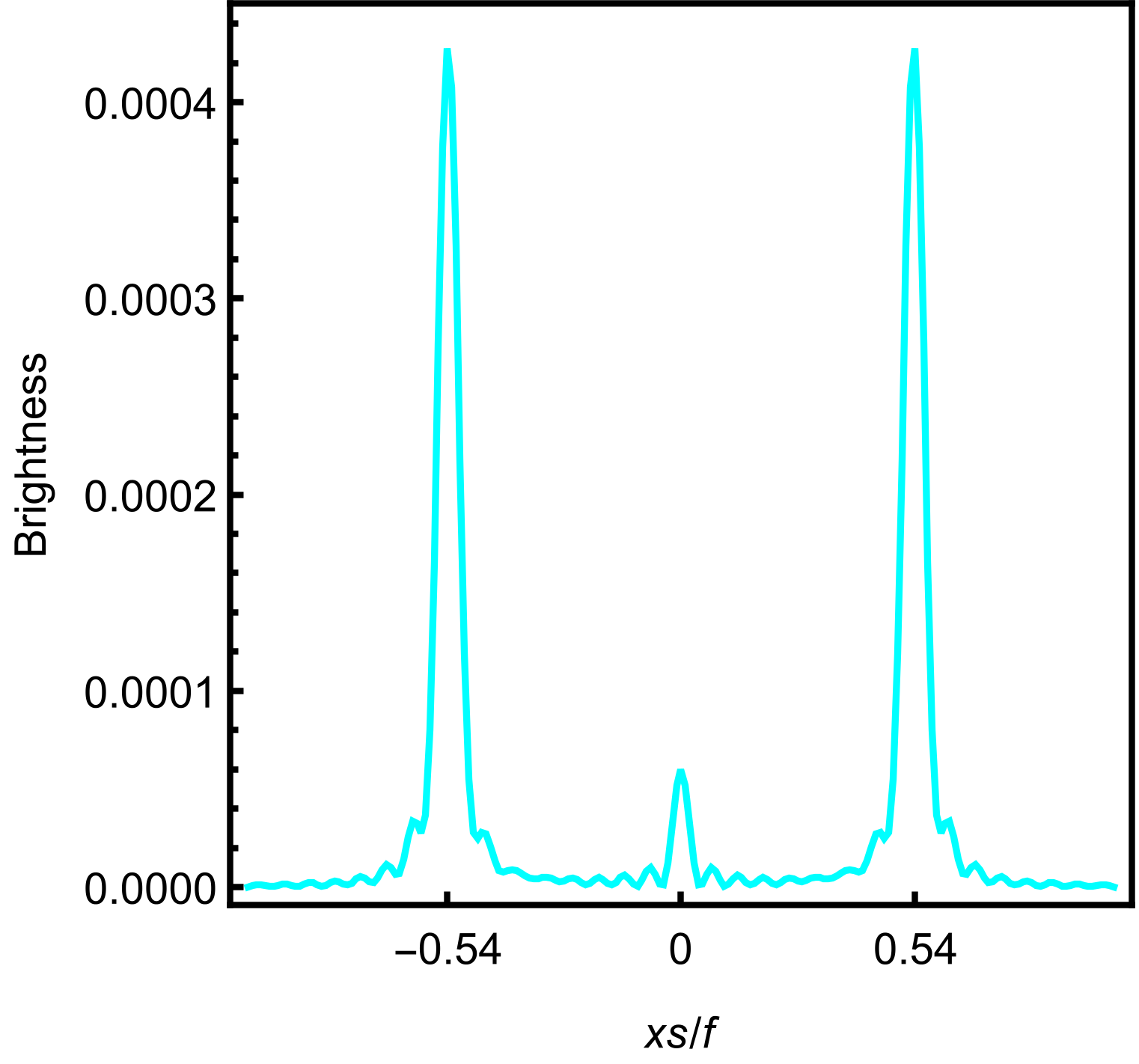}
    \caption{$\alpha=10.1$}
  \end{subfigure}
  \hfill
  \begin{subfigure}[b]{0.45\columnwidth}
    \centering
    \includegraphics[height=0.9\textwidth,width=\textwidth]{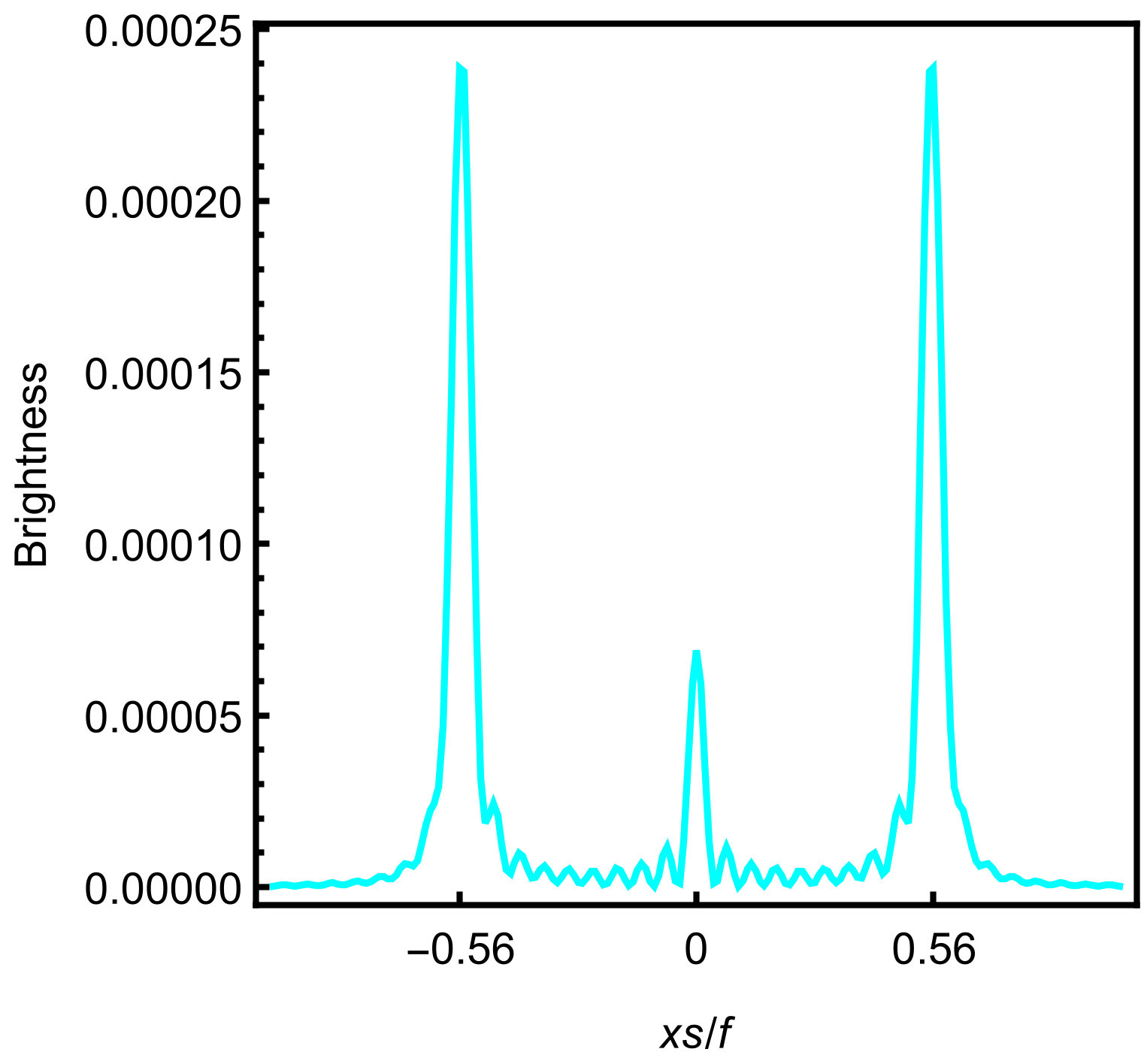}
    \caption{$\alpha=15.1$}
  \end{subfigure}
  \caption{Effect of $\alpha$ on the brightness, where $\theta _{obs}=0$, $y_{h}=5$, $\omega=90$.}
  \label{9}%
\end{figure}

Initially, holographic images obtained by varying the deformation parameter  $\alpha$ and the observer's position $\theta _{obs}$ are presented in Figure \ref{8}. When $\theta _{obs}=0$, it indicates  that the wave source is fixed at the south pole of the AdS boundary, while the observer is situated at the north pole, observing a series of axisymmetric concentric rings. As $\theta _{obs}=\pi/6$ (see the second column), the holographic ring transitions from strict spherical symmetry to axisymmetry, accompanied by a decrease in brightness. As it increases to $\theta _{obs}=\pi/3$ and $\theta _{obs}=\pi/2$, the holographic ring gradually disappears, especially at $\theta _{obs}=\pi/2$ where only bright spots are visible. This demonstrates that the shape of the holographic ring strongly depends on the observer's position. In the leftmost column of Figure \ref{8}, despite variations in the value of $\alpha$, a brightest circular ring persists, known as the Einstein ring. In addition, to further investigate the impact of the deformation parameter $\alpha$ on the Einstein ring, the brightness is plotted in Figure \ref{9}, in which the distance between the two peak trajectories corresponds to the diameter of the Einstein ring. The increase in $\alpha$ results in a gradual increase in the radius of the Einstein ring. When $\alpha=0.1 $, the radius of the ring is 0.49, and when $\alpha =15.1$, the radius of the ring is 0.56. Furthermore, the luminosity represented on the y-axis of Figure \ref{9} gradually decreases as $\alpha $ increases.

\begin{figure}[htbp]
  \centering
  \begin{subfigure}[b]{0.49\columnwidth}
    \centering
    \includegraphics[width=\textwidth,height=0.8\textwidth]{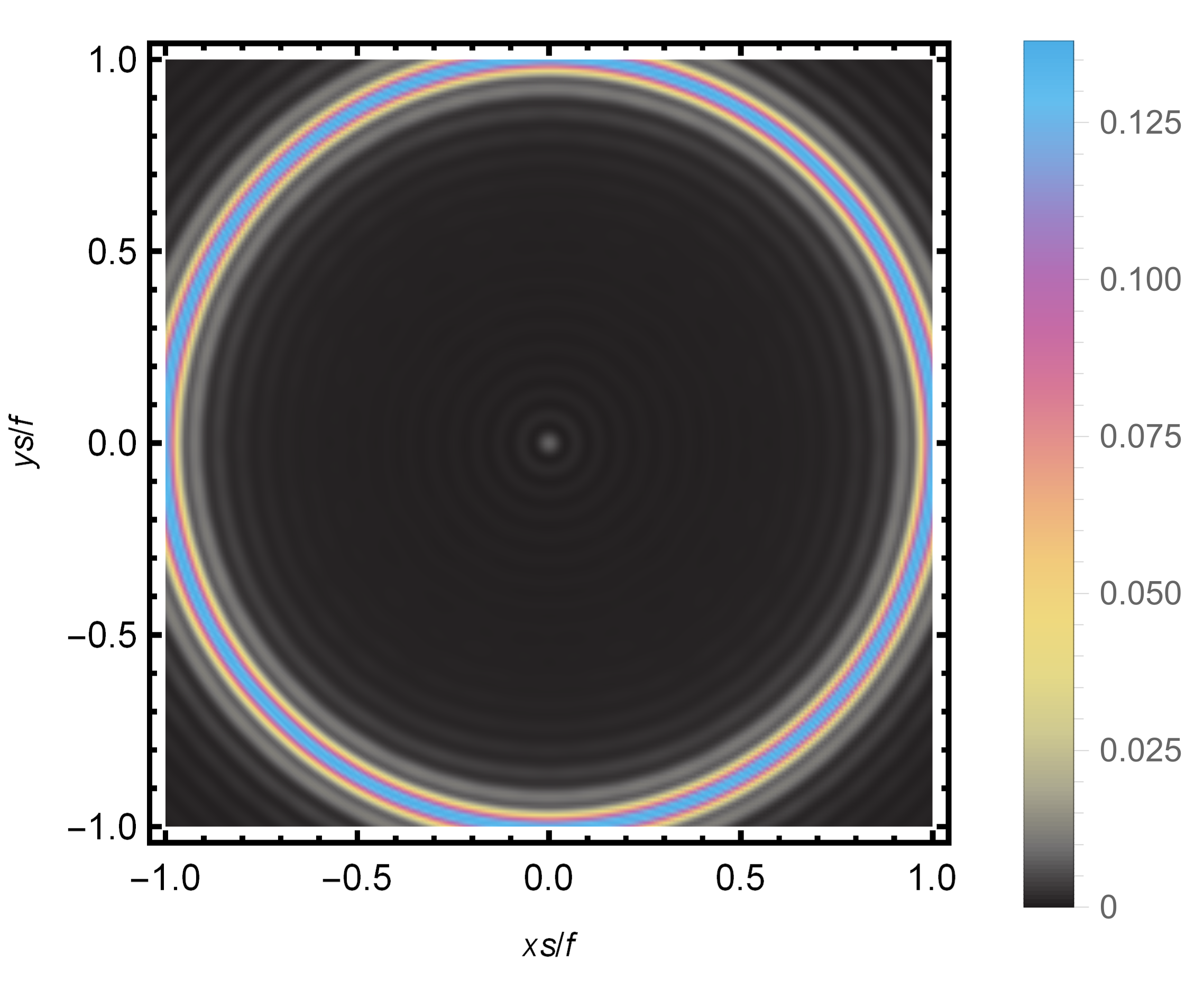}
    \caption{$y_{h}=0.5$($\mathit{T}=0.514$)}
  \end{subfigure}
  \hfill
  \begin{subfigure}[b]{0.49\columnwidth}
    \centering
    \includegraphics[width=\textwidth,height=0.8\textwidth]{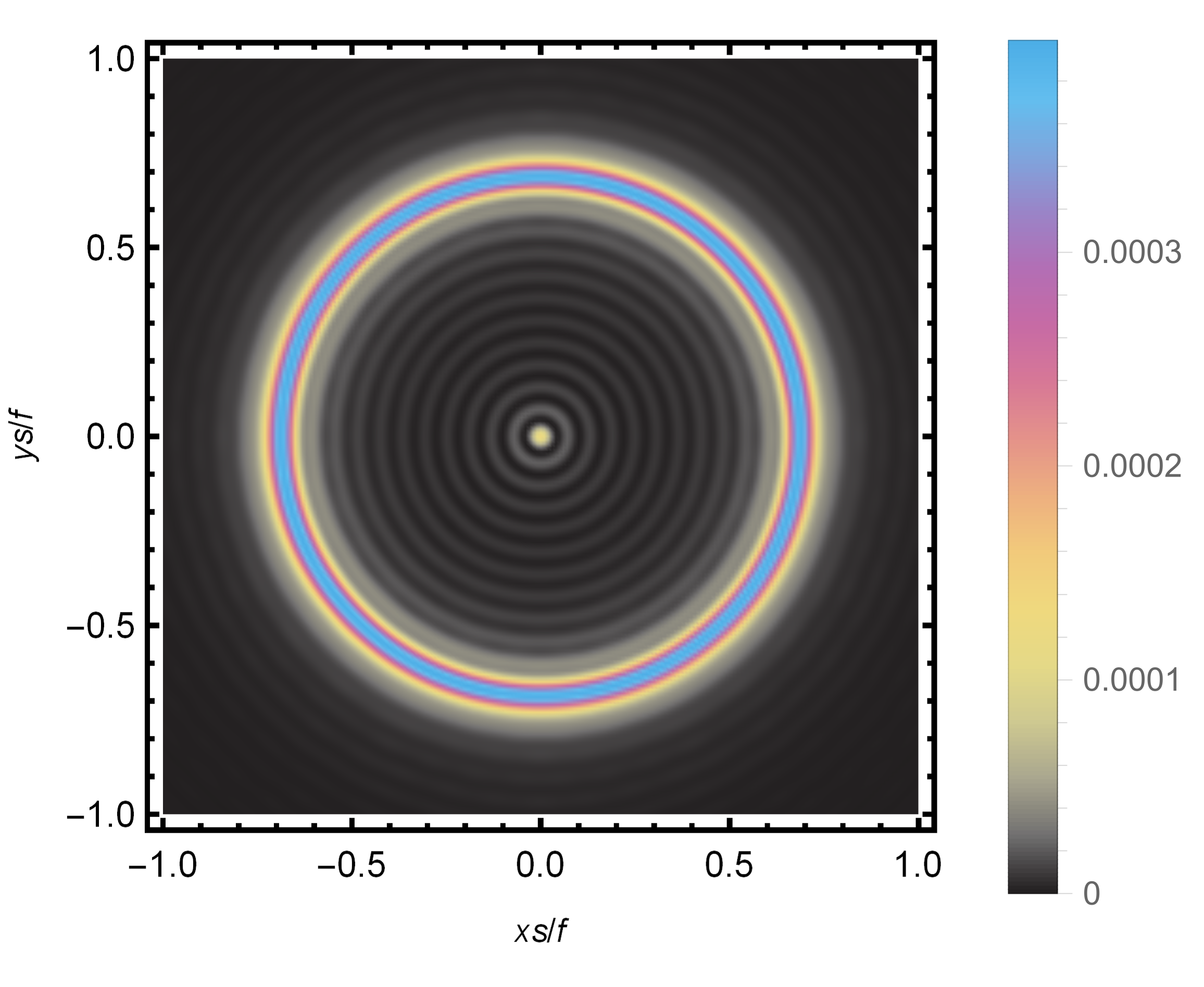}
    \caption{$y_{h}=3$($\mathit{T}=0.305$)}
  \end{subfigure}
\begin{subfigure}[b]{0.49\columnwidth}
    \centering
    \includegraphics[width=\textwidth,height=0.8\textwidth]{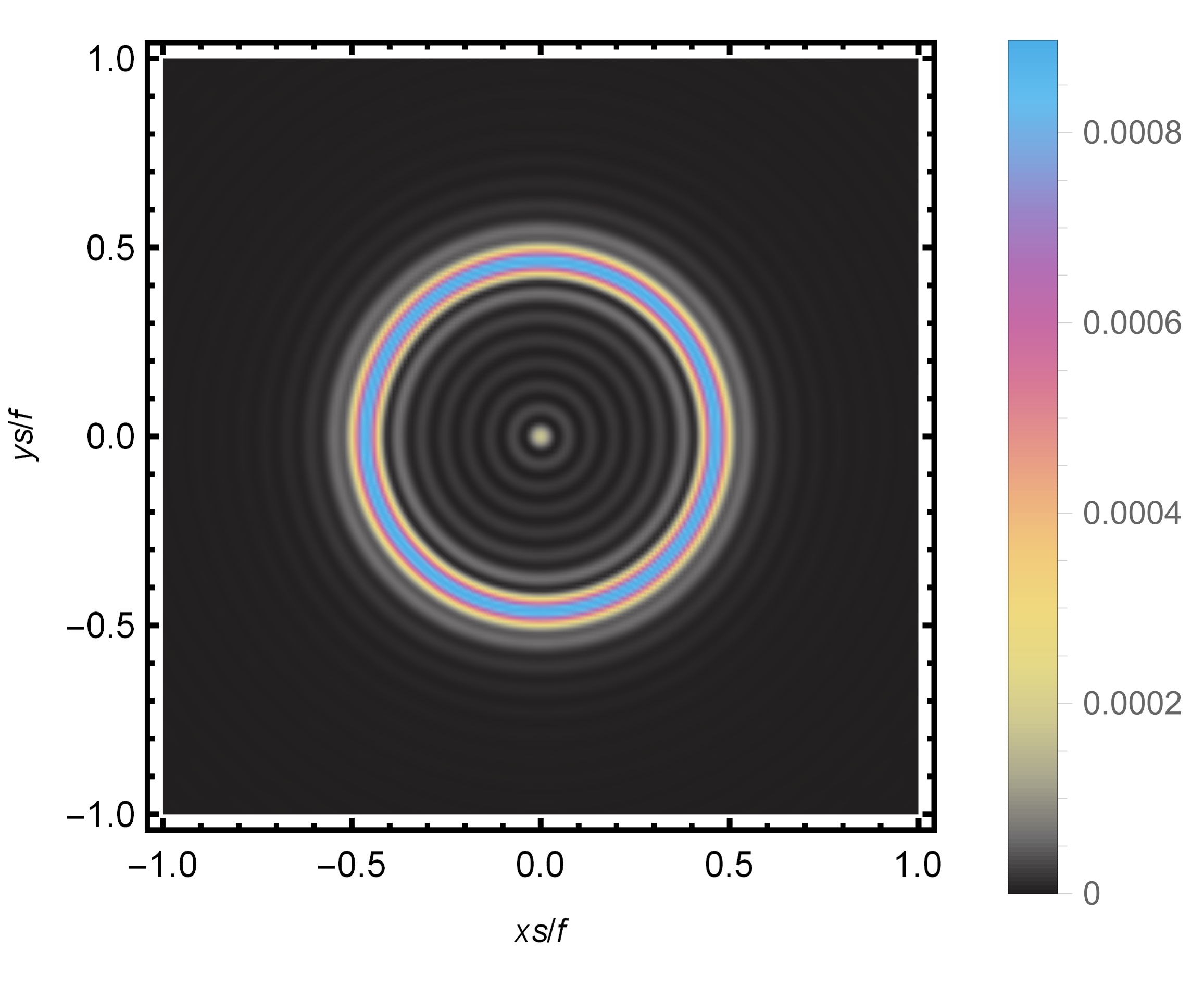}
    \caption{$y_{h}=5.5$($\mathit{T}=0.469$)}
  \end{subfigure}
  \hfill
  \begin{subfigure}[b]{0.49\columnwidth}
    \centering
    \includegraphics[width=\textwidth,height=0.8\textwidth]{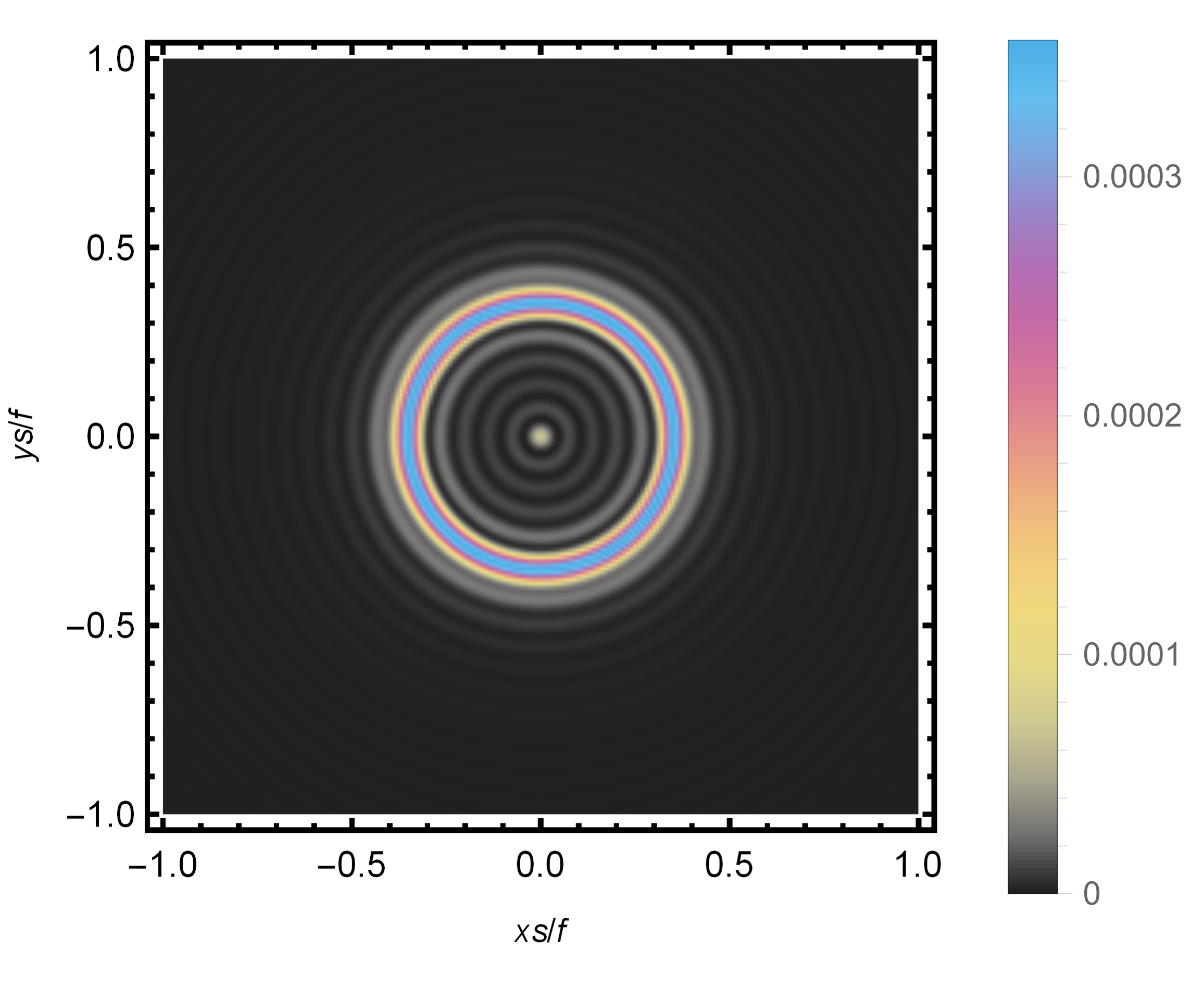}
    \caption{$y_{h}=8$($\mathit{T}=0.657$)}
  \end{subfigure}
  \caption{Effect of $y_{h}$ on the Einstein ring, where $\theta _{obs}=0$, $\alpha=1.6$, $\omega=90$.}
  \label{10}%
\end{figure}

\begin{figure}
	\centering 
\includegraphics[width=0.4\textwidth, angle=0]{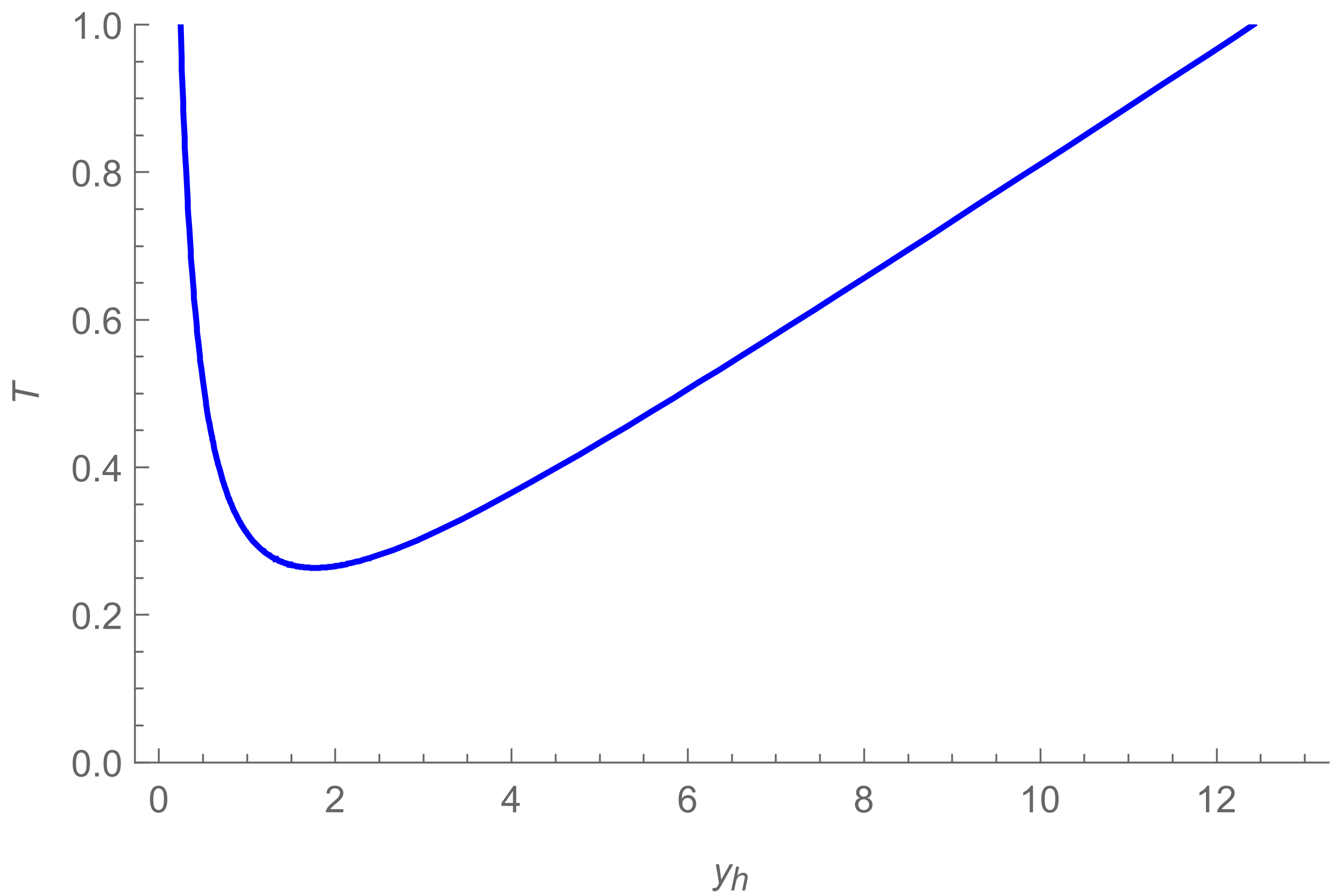}	
	\caption{
 The relationship between temperature and event horizon $y_{h}$, where 
$\alpha=1.6$, $\beta=1$.}
	\label{t17}%
\end{figure}

Next, the impact of the event horizon on the Einstein ring of BH is investigated. For simplicity, only the case where $\theta_{obs}=0$ is considered, as depicted in Figure \ref{10}. When $y_{h}=0.5$, a bright spherically symmetric ring appears on the screen. Additionally, as $y_{h}$ increases from 0.5 to higher values, such as $y_{h}=3$ and $y_{h}=5.5$, the ring gradually contracts. Notably, when $y_{h}=8$, the ring radius reaches its minimum. The observation indicates that as $y_{h}$ increases, the ring radius decreases. However, there is no apparent pattern of change in the ring radius with temperature. In literature \cite{Liu:2022cev}, the authors discussed the variation of ring radius with temperature. Here we argue that the variation of ring radius with the event horizon radius is more reasonable, as the relationship between the BH temperature and the event horizon is not monotonic, as depicted in Figure \ref{t17}.

\begin{figure}[htbp]
  \centering
  \begin{subfigure}[b]{0.45\columnwidth}
    \centering
    \includegraphics[width=\textwidth,height=0.9\textwidth]{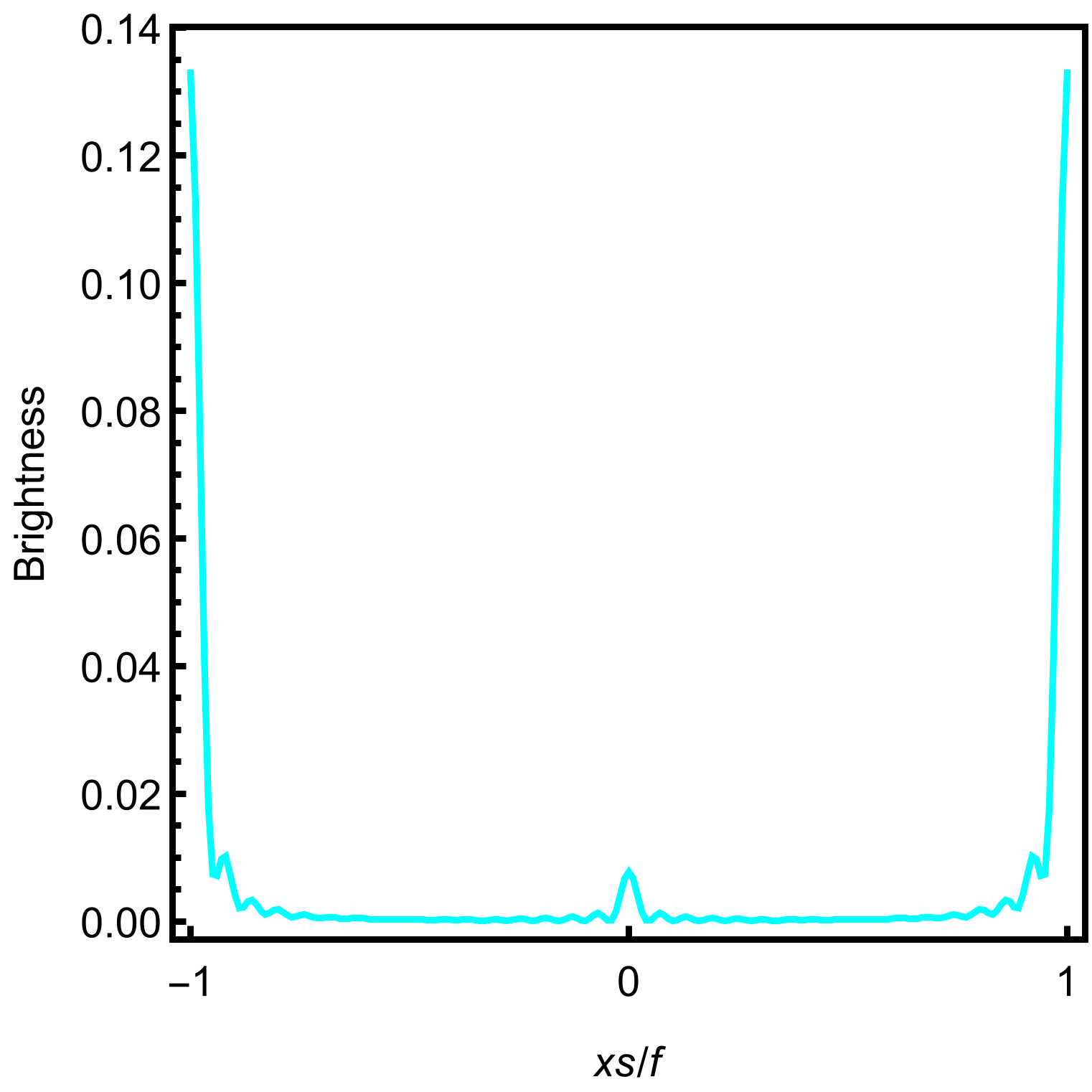}
    \caption{$y_{h}=0.5$($\mathit{T}=0.514$)}
  \end{subfigure}
  \hfill
  \begin{subfigure}[b]{0.45\columnwidth}
    \centering
    \includegraphics[width=\textwidth,height=0.9\textwidth]{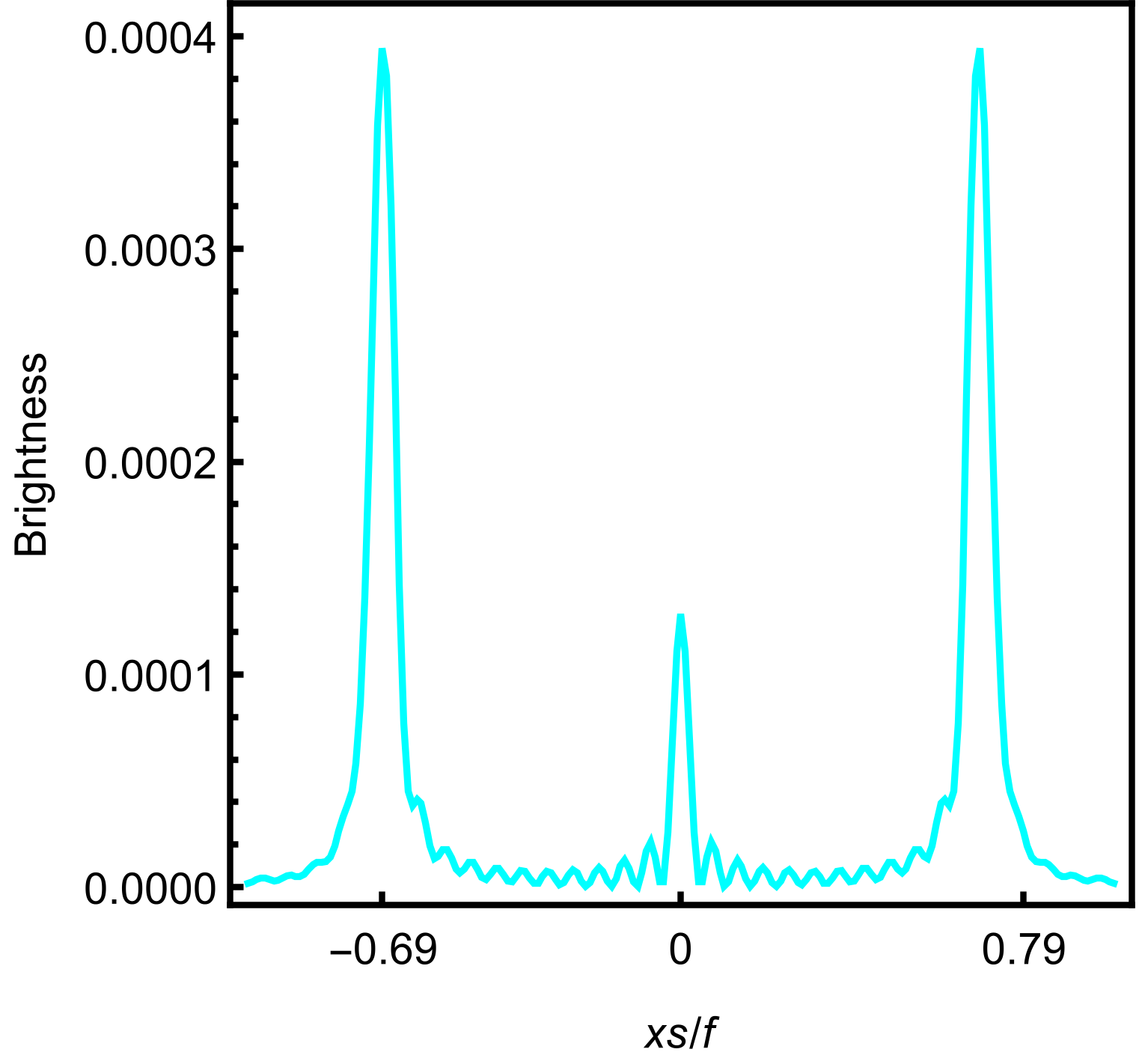}
    \caption{$y_{h}=3$($\mathit{T}=0.305$)}
  \end{subfigure}
\begin{subfigure}[b]{0.45\columnwidth}
    \centering
    \includegraphics[width=\textwidth,height=0.9\textwidth]{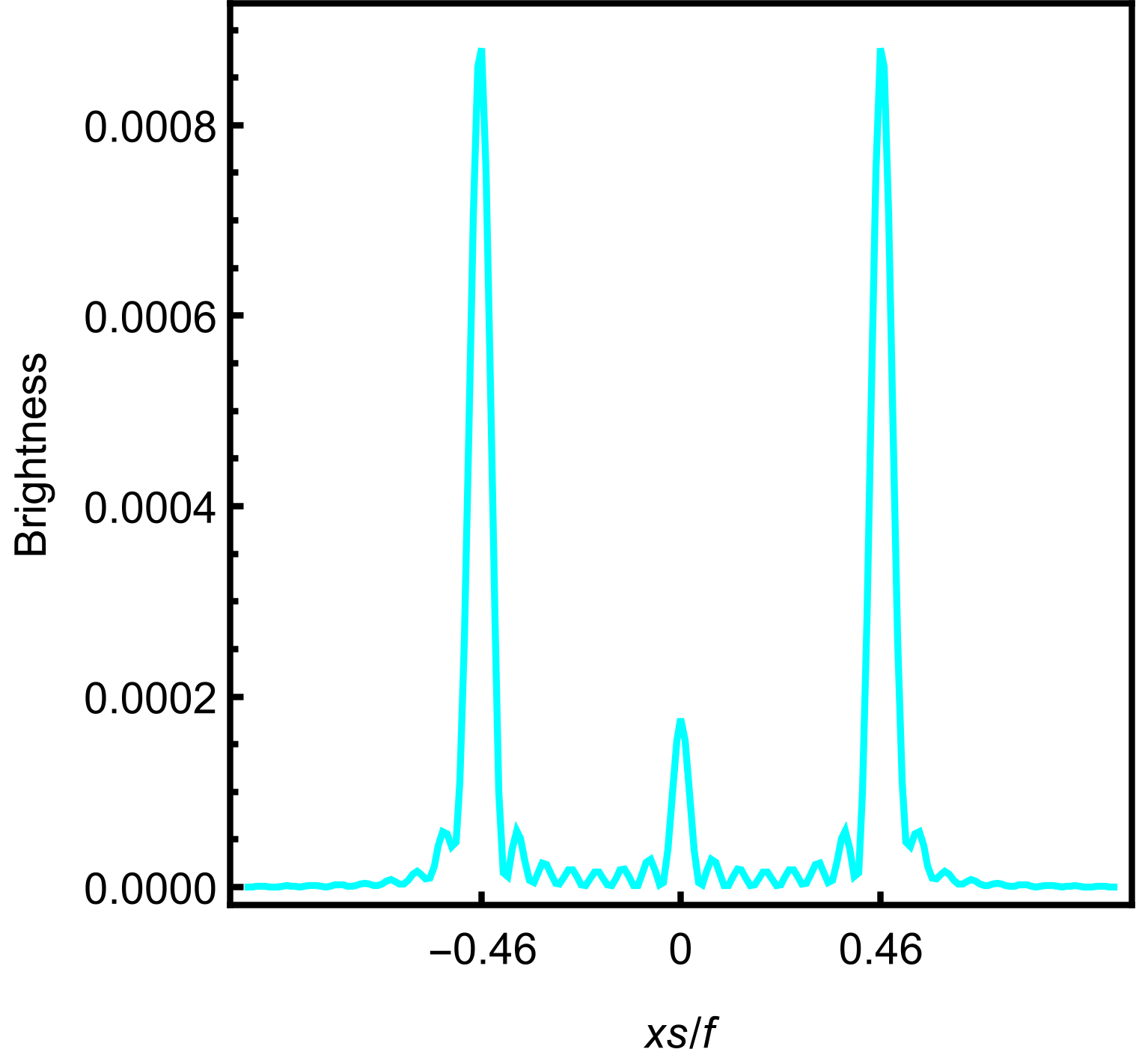}
    \caption{$y_{h}=5.5$($\mathit{T}=0.469$)}
  \end{subfigure}
  \hfill
  \begin{subfigure}[b]{0.45\columnwidth}
    \centering
    \includegraphics[width=\textwidth,height=0.9\textwidth]{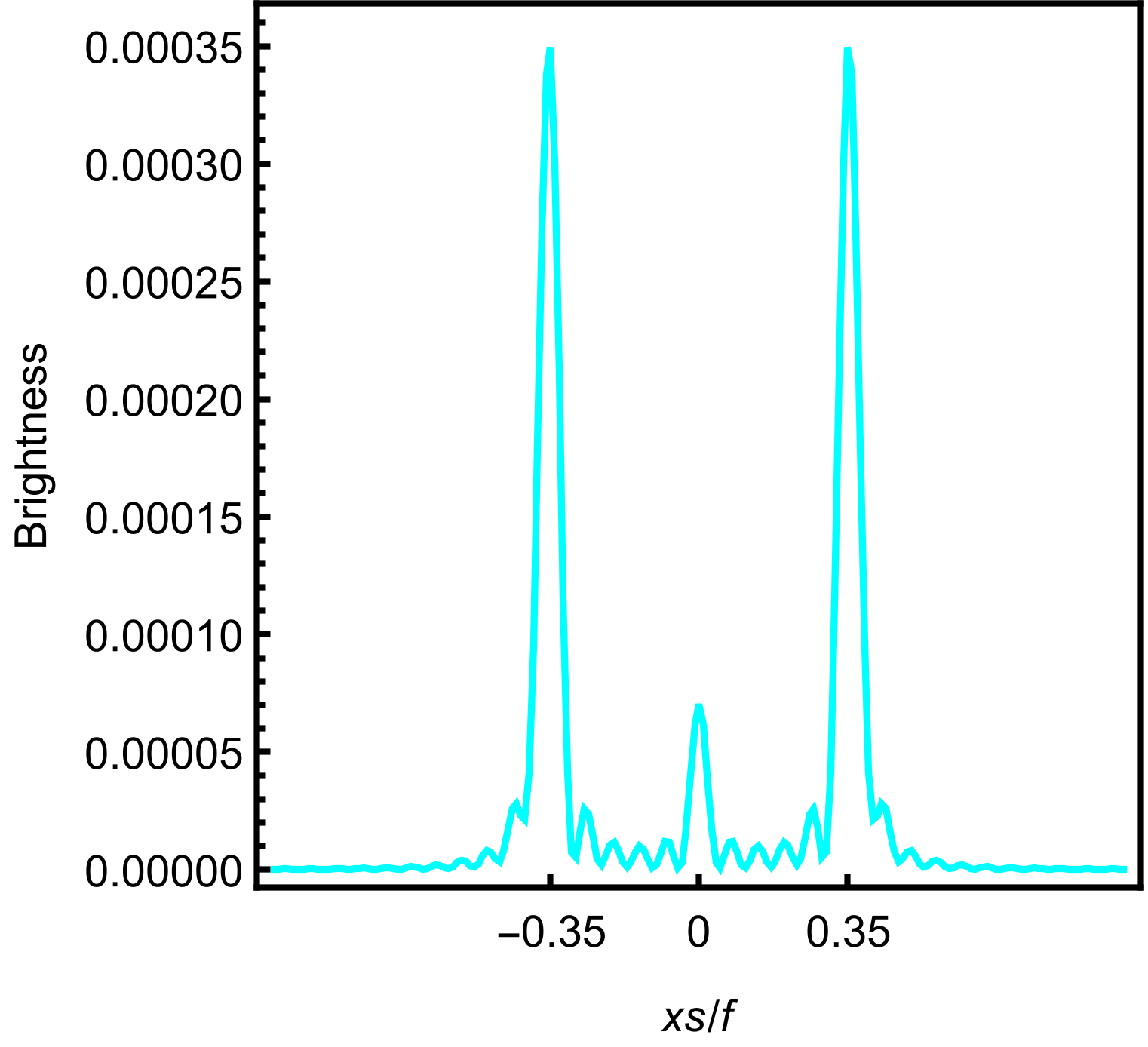}
    \caption{$y_{h}=8$($\mathit{T}=0.657$)}
  \end{subfigure}
  \caption{Effect of $\mathit{T}$ on the brightness, where $\theta _{obs}=0$, $\alpha=1.6$, $\omega=90$.}
  \label{12}%
\end{figure}

Similarly, Figure \ref{12} depicts the relationship between luminosity and the event horizon. When  $y_{h}=0.5$, $\mathit{xs/f}=1$, the luminosity is maximum. With increasing $y_{h}$, the value of $\mathit{xs/f}$ corresponding to the maximum luminosity decreases, indicating a decreases in the Einstein ring's radius.

\begin{figure}[htbp]
  \centering
  \begin{subfigure}[b]{0.49\columnwidth}
    \centering
    \includegraphics[width=\textwidth,height=0.8\textwidth]{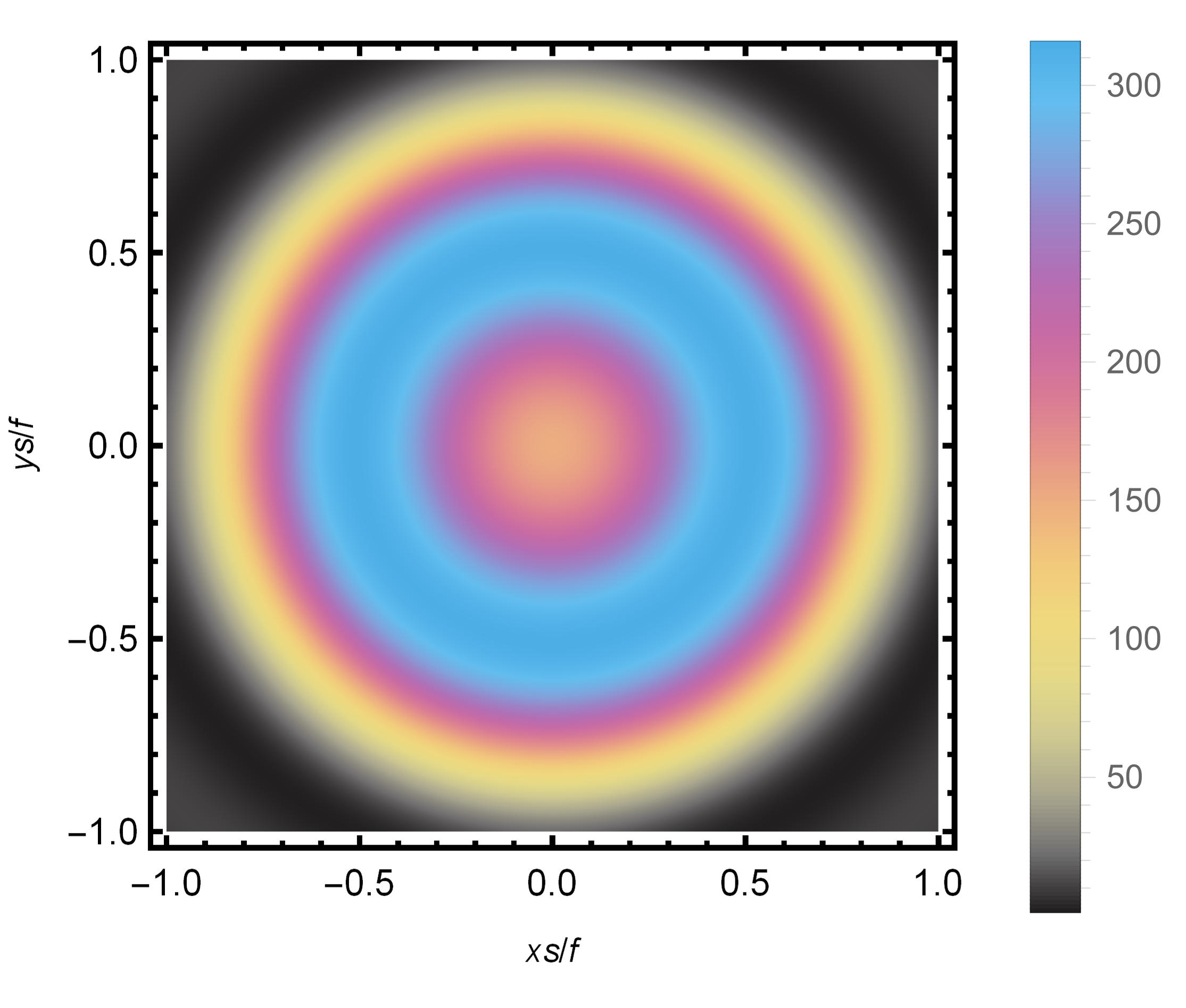}
    \caption{$\omega=10$}
  \end{subfigure}
  \hfill
  \begin{subfigure}[b]{0.49\columnwidth}
    \centering
    \includegraphics[width=\textwidth,height=0.8\textwidth]{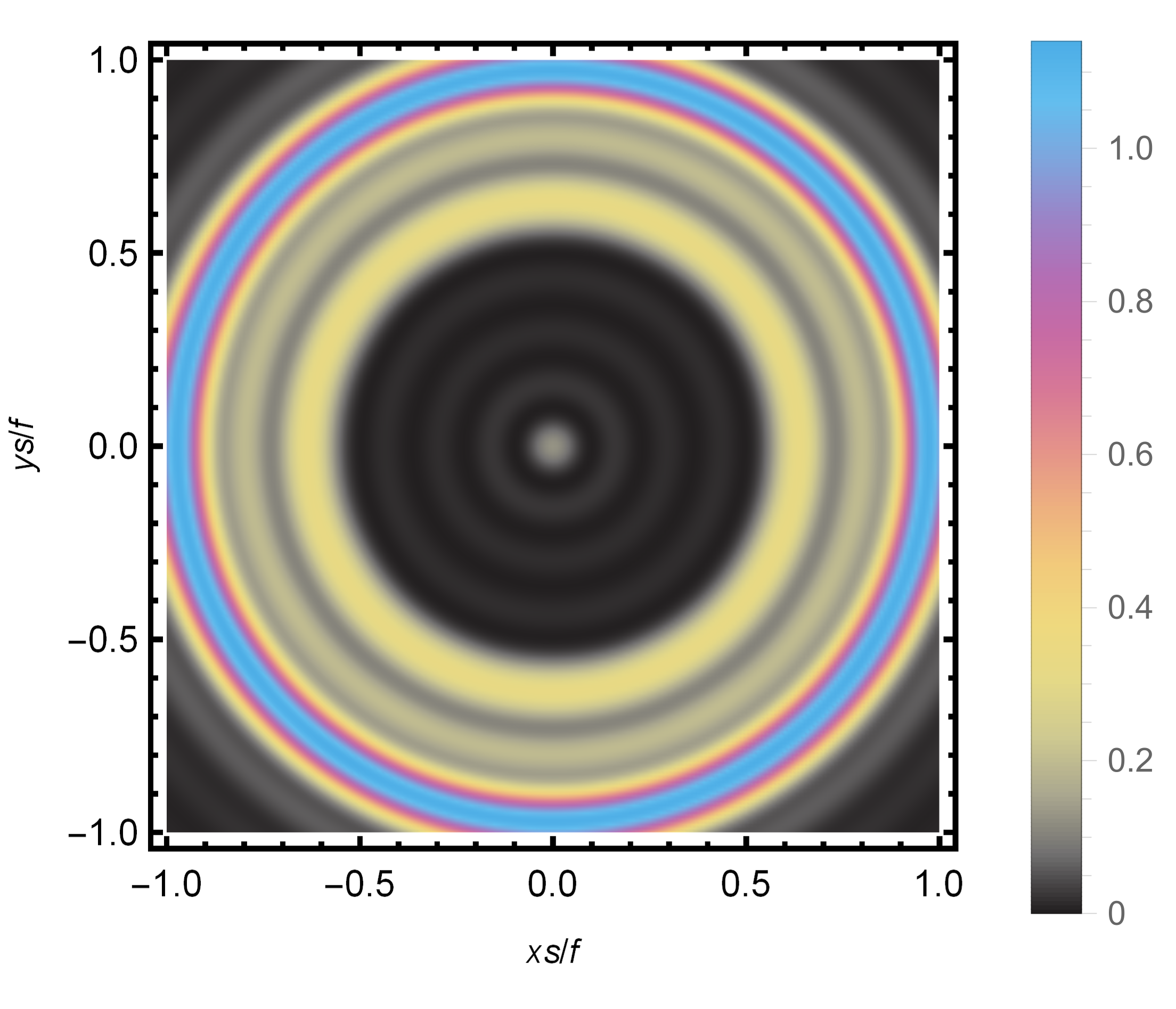}
    \caption{$\omega=40$}
  \end{subfigure}
\begin{subfigure}[b]{0.49\columnwidth}
    \centering
    \includegraphics[width=\textwidth,height=0.8\textwidth]{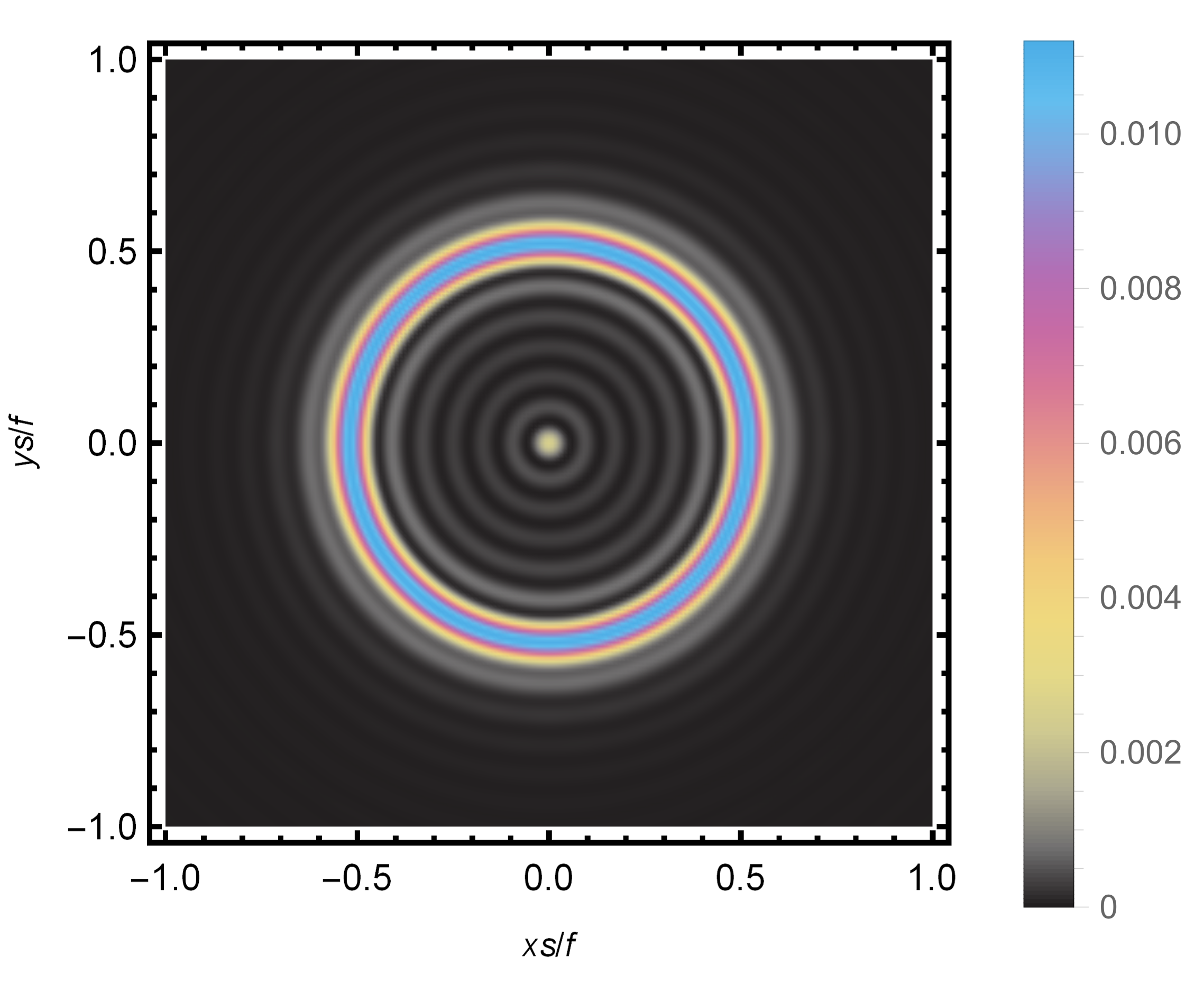}
    \caption{$\omega=70$}
  \end{subfigure}
  \hfill
  \begin{subfigure}[b]{0.49\columnwidth}
    \centering
    \includegraphics[width=\textwidth,height=0.8\textwidth]{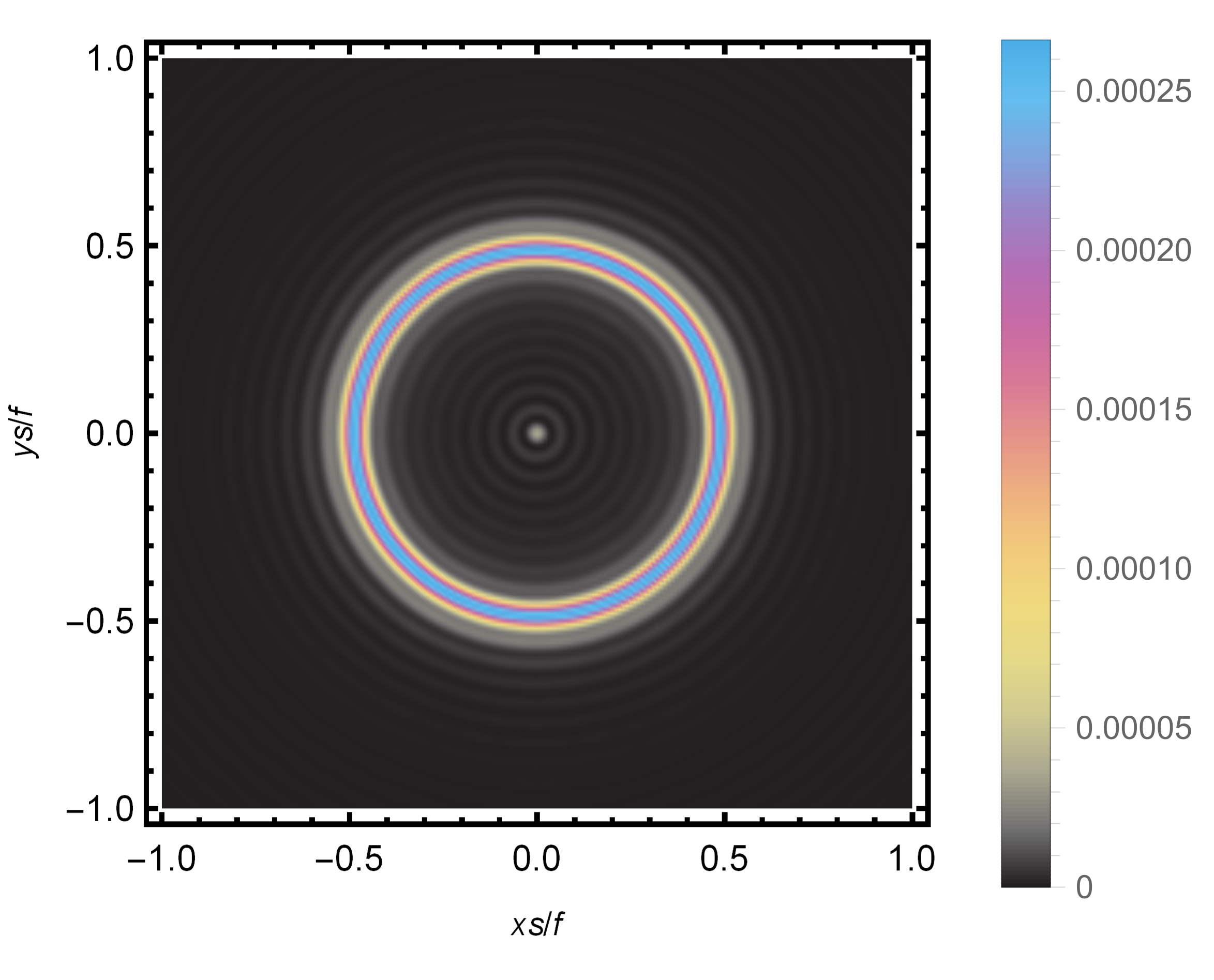}
    \caption{$\omega=100$}
  \end{subfigure}
  \caption{Effect of $\omega$ on the Einstein ring, where $\theta _{obs}=0$, $\alpha=1.6$, $y_{h}=5$.}
  \label{13}%
\end{figure}

\begin{figure}[htbp]
  \centering
  \begin{subfigure}[b]{0.45\columnwidth}
    \centering
    \includegraphics[width=\textwidth,height=0.9\textwidth]{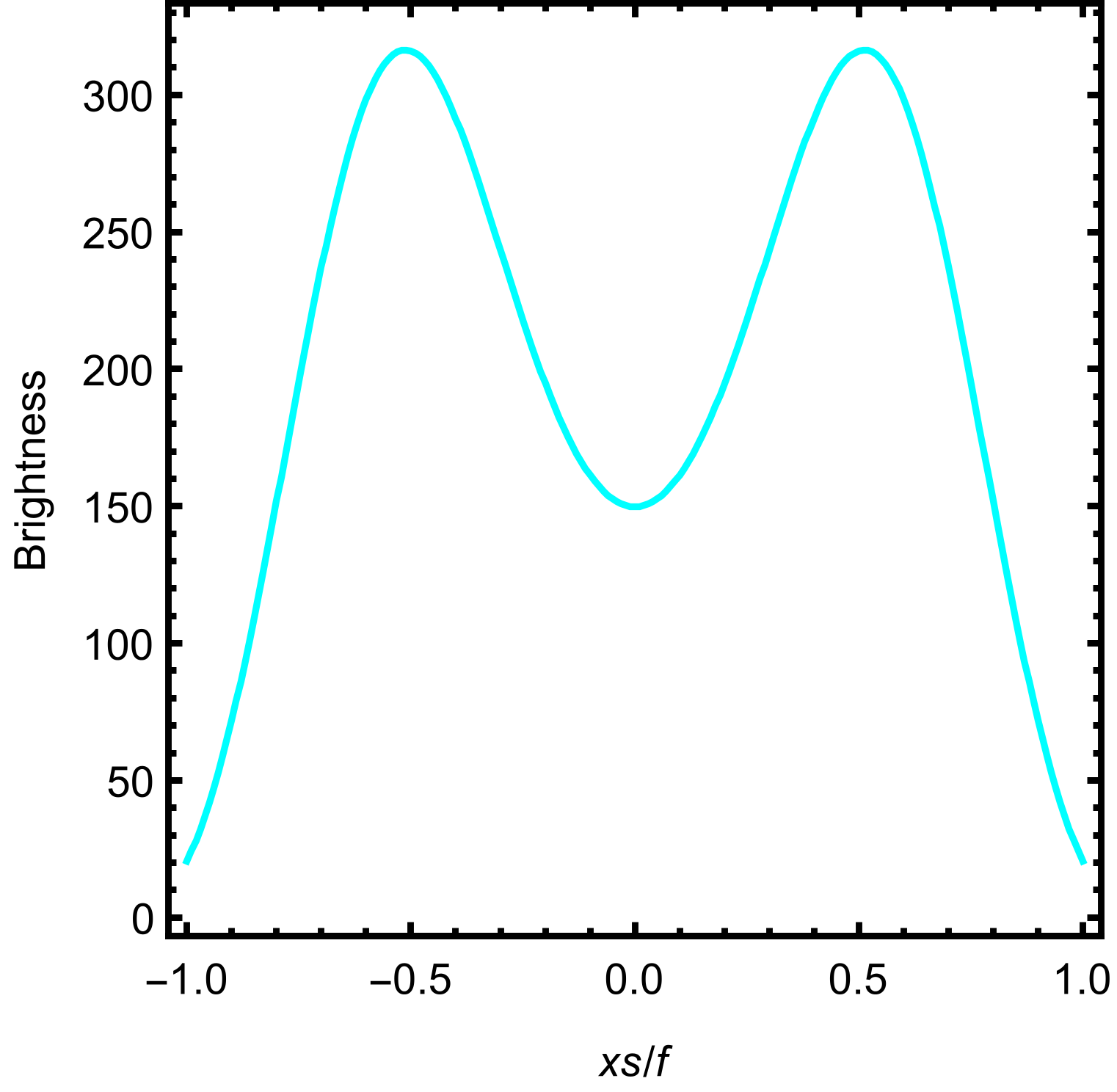}
    \caption{$\omega=10$}
  \end{subfigure}
  \hfill
  \begin{subfigure}[b]{0.45\columnwidth}
    \centering
    \includegraphics[width=\textwidth,height=0.9\textwidth]{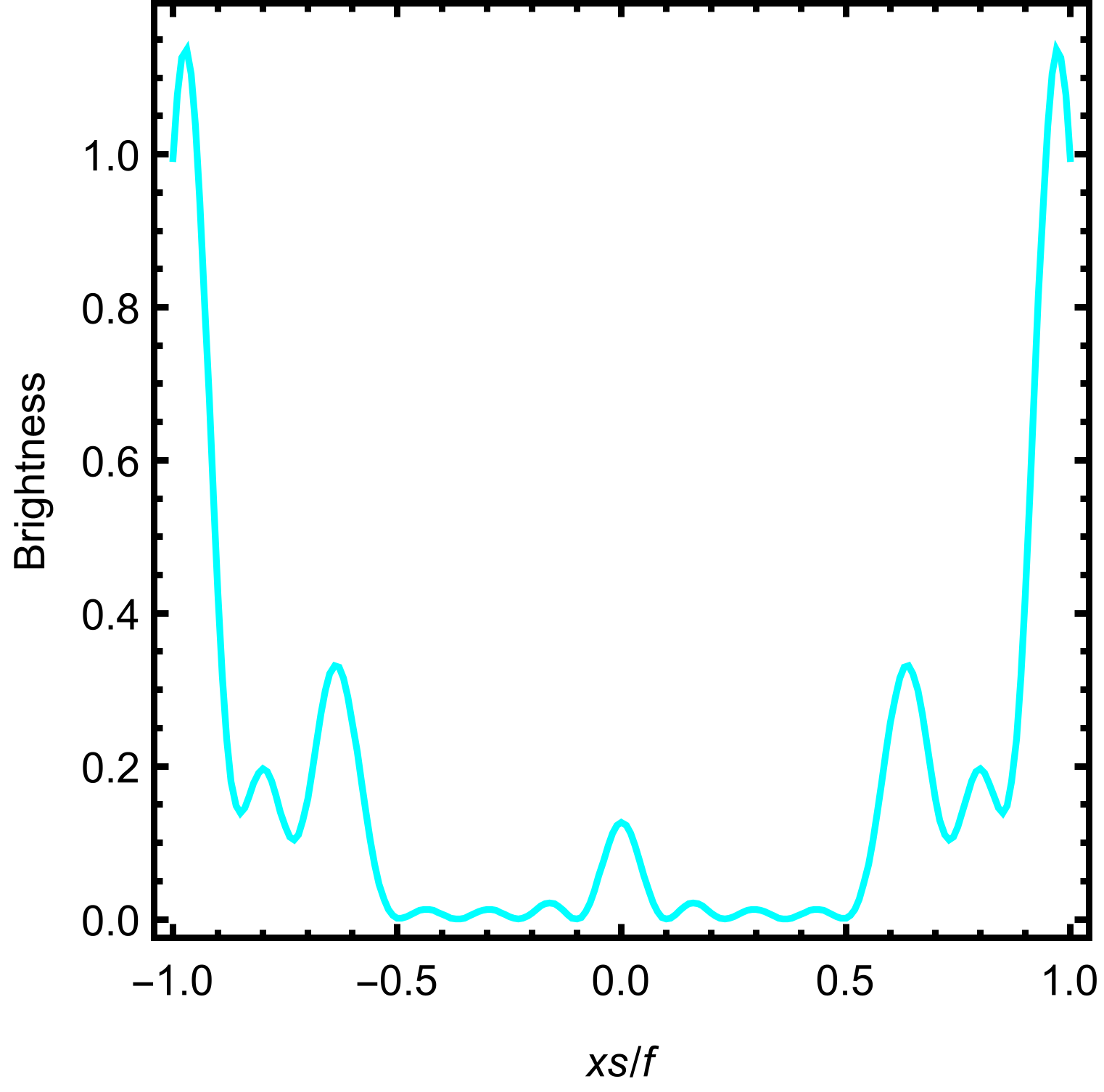}
    \caption{$\omega=40$}
  \end{subfigure}
\begin{subfigure}[b]{0.45\columnwidth}
    \centering
    \includegraphics[width=\textwidth,height=0.9\textwidth]{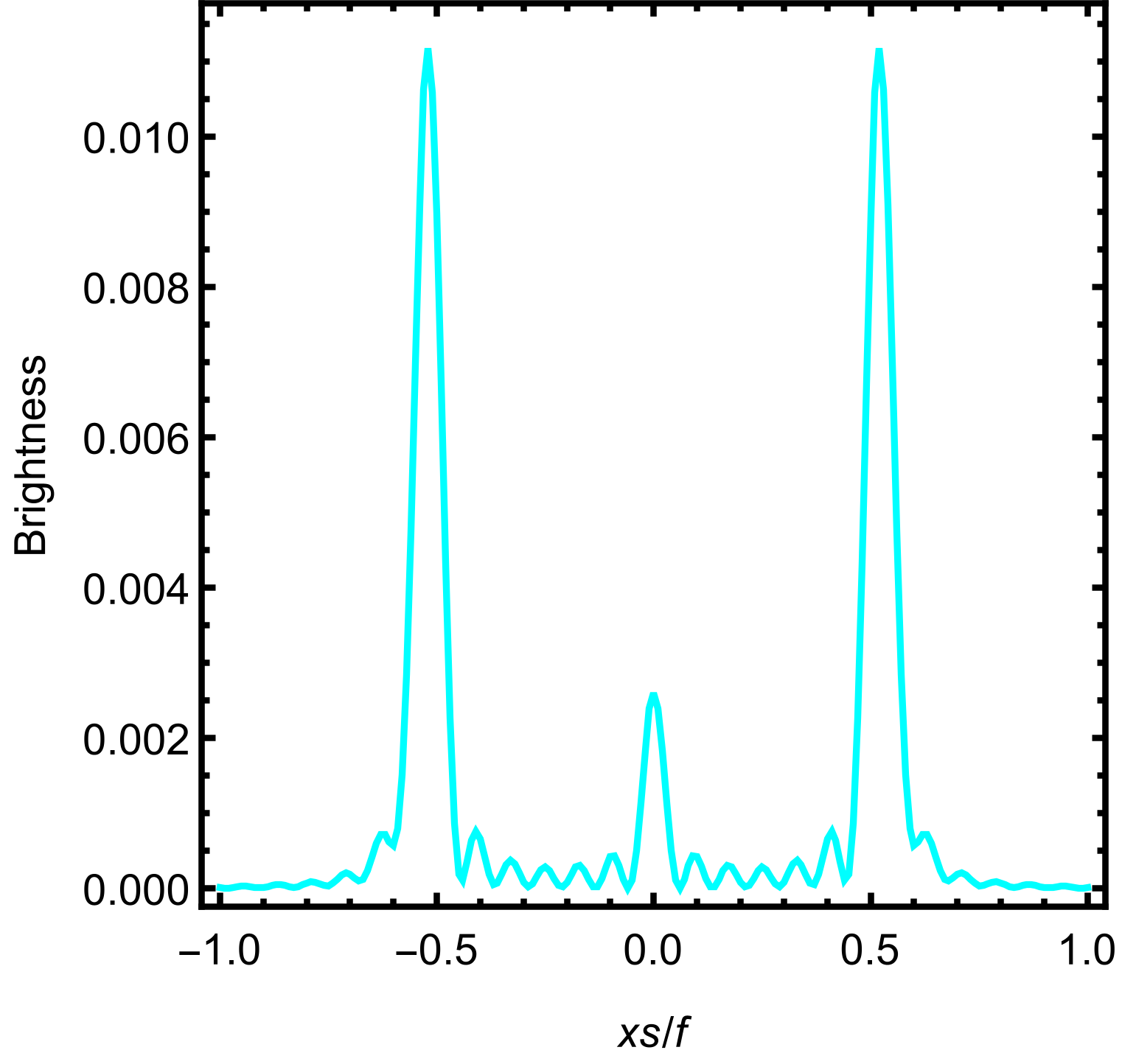}
    \caption{$\omega=70$}
  \end{subfigure}
  \hfill
  \begin{subfigure}[b]{0.45\columnwidth}
    \centering
    \includegraphics[width=\textwidth,height=0.9\textwidth]{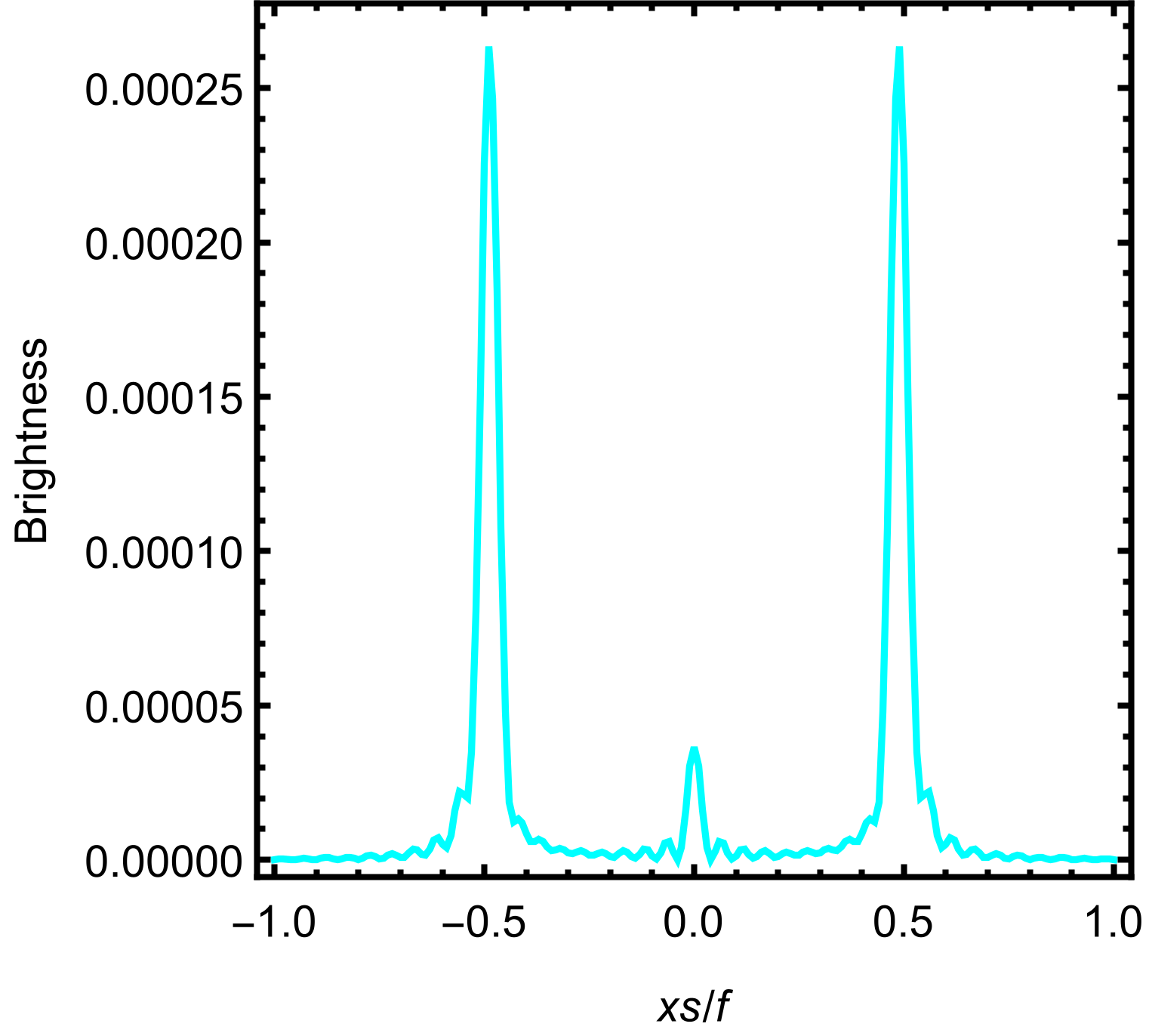}
    \caption{$\omega=100$}
  \end{subfigure}
  \caption{Effect of $\omega$ on the brightness, where $\theta _{obs}=0$, $\alpha=1.6$, $y_{h}=5$.}
  \label{14}%
\end{figure}

The impact of wave sources, such as the wave source frequency $\omega$, on the Einstein ring can also be analyzed. As demonstrated in Figure \ref{13}, a discernible trend emerges where the resolution of the Einstein ring improves as the wave source frequency increases. At $\omega=40$, multiple diffraction rings are observable, indicating significant interference effects. However, as the frequency rises to $\omega=100$, these additional diffraction rings gradually fade, leaving only the primary Einstein ring distinct. This transition can be attributed to the increasing dominance of geometric optics at higher frequencies. To complement our understanding of these findings, Figure \ref{14} depicts the correlation between luminosity and angular frequency in Figure \ref{14}. It is evident that high frequencies are crucial for effectively observing and resolving the radius of the ring.

\subsection{The effect of control
parameter $\beta$ on the images}
\begin{figure*}[htbp]
\centering
\begin{subfigure}[b]{0.24\textwidth}
  \centering
  \includegraphics[width=\textwidth]{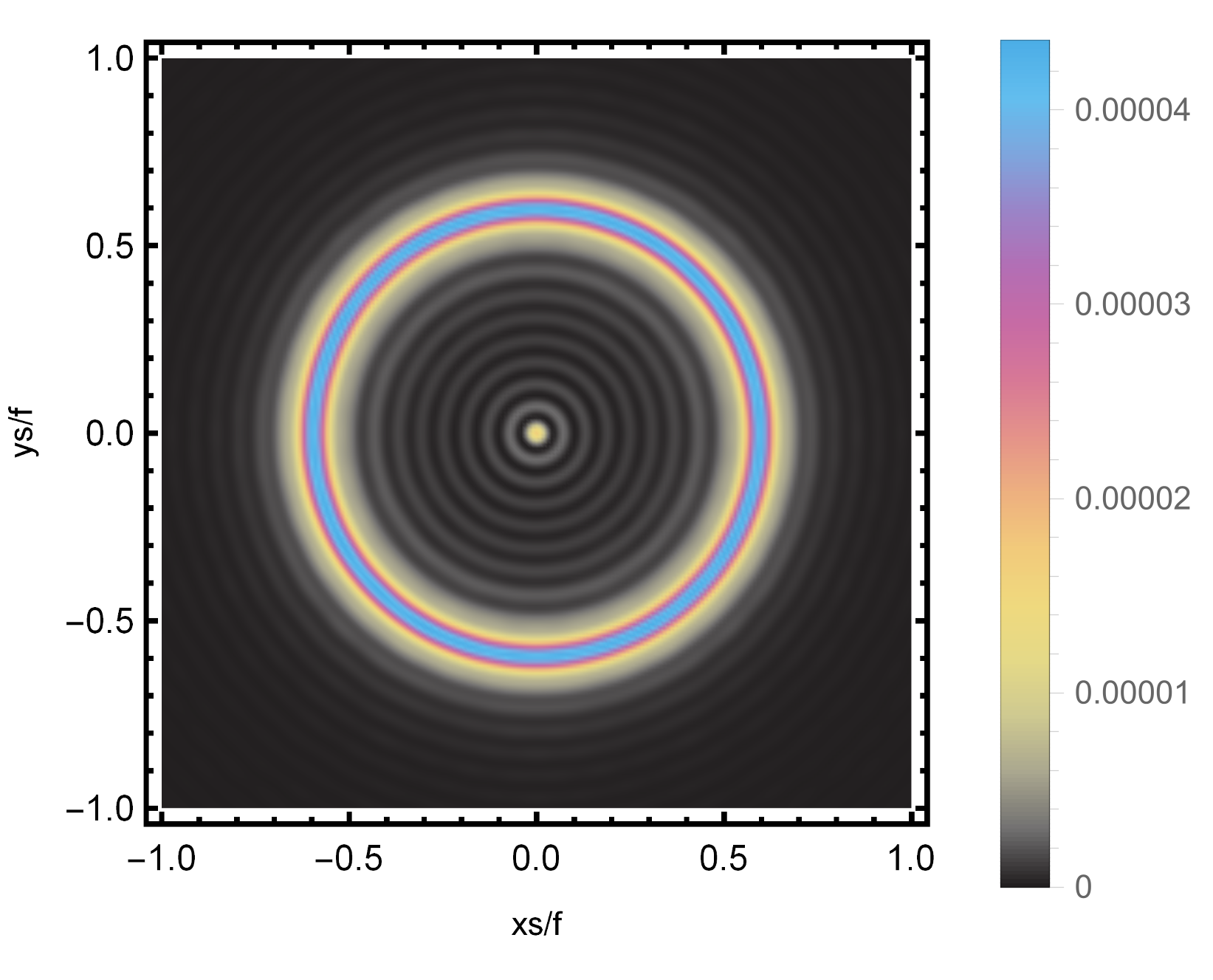} \caption{$\beta=0.3$,$\theta _{obs}=0$}
\end{subfigure}
\hfill
\begin{subfigure}[b]{0.24\textwidth}
  \centering
  \includegraphics[width=\textwidth]{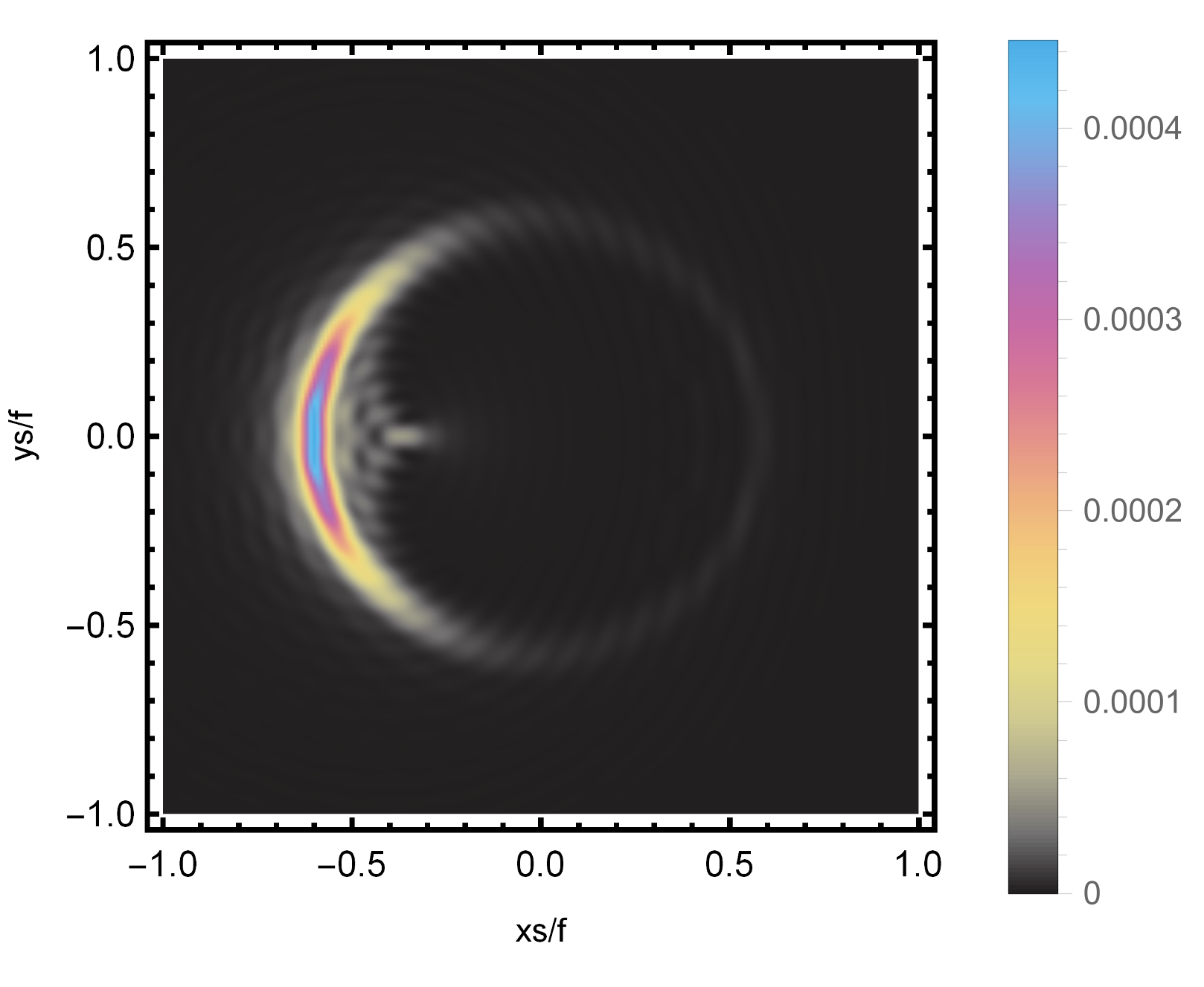}
  \caption{$\beta=0.3$,$\theta _{obs}=\pi/6$}
\end{subfigure}
\hfill
\begin{subfigure}[b]{0.24\textwidth}
  \centering
  \includegraphics[width=\textwidth]{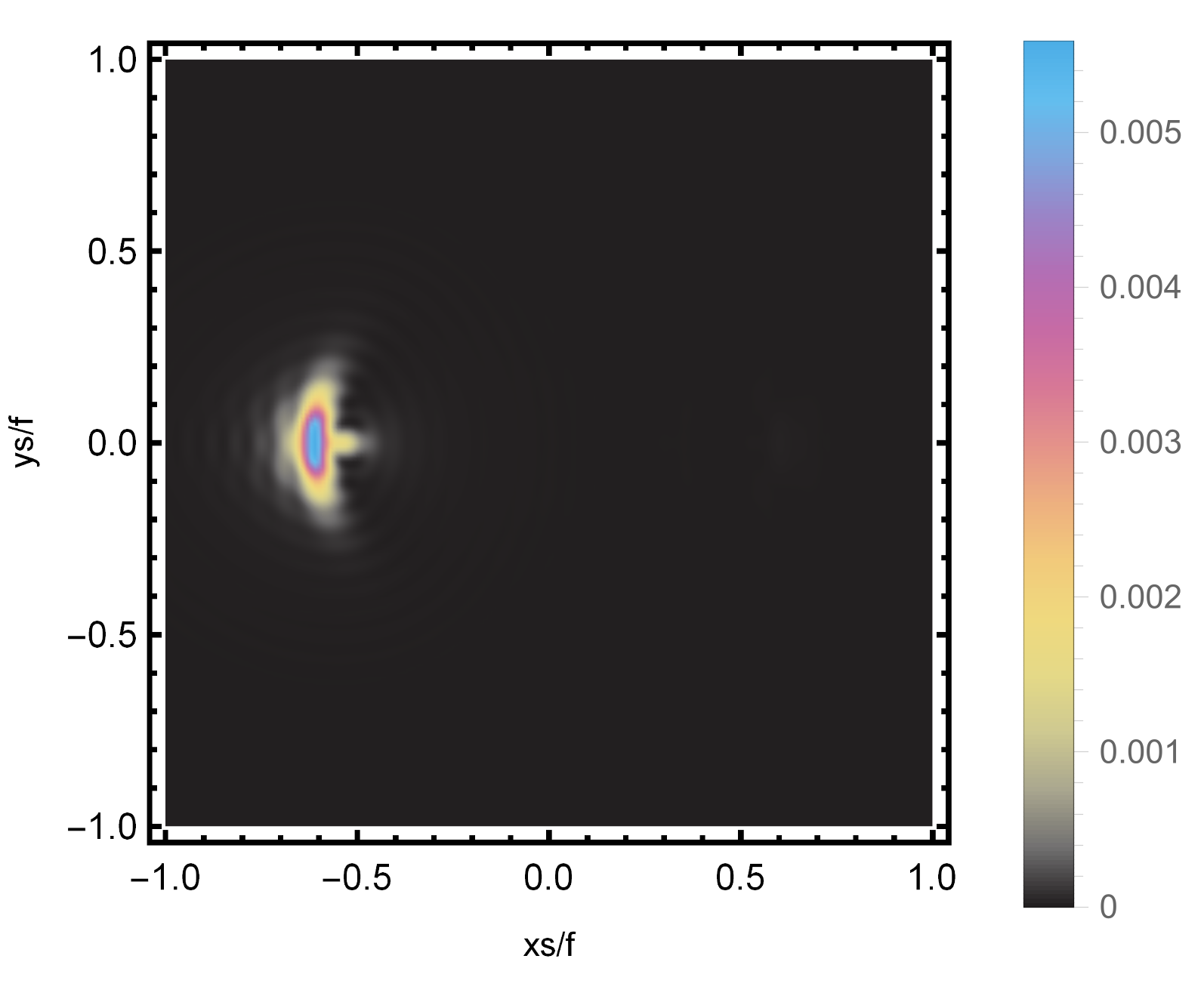}
  \caption{$\beta=0.3$,$\theta _{obs}=\pi/3$}
\end{subfigure}
\hfill
\begin{subfigure}[b]{0.24\textwidth}
  \centering
  \includegraphics[width=\textwidth]{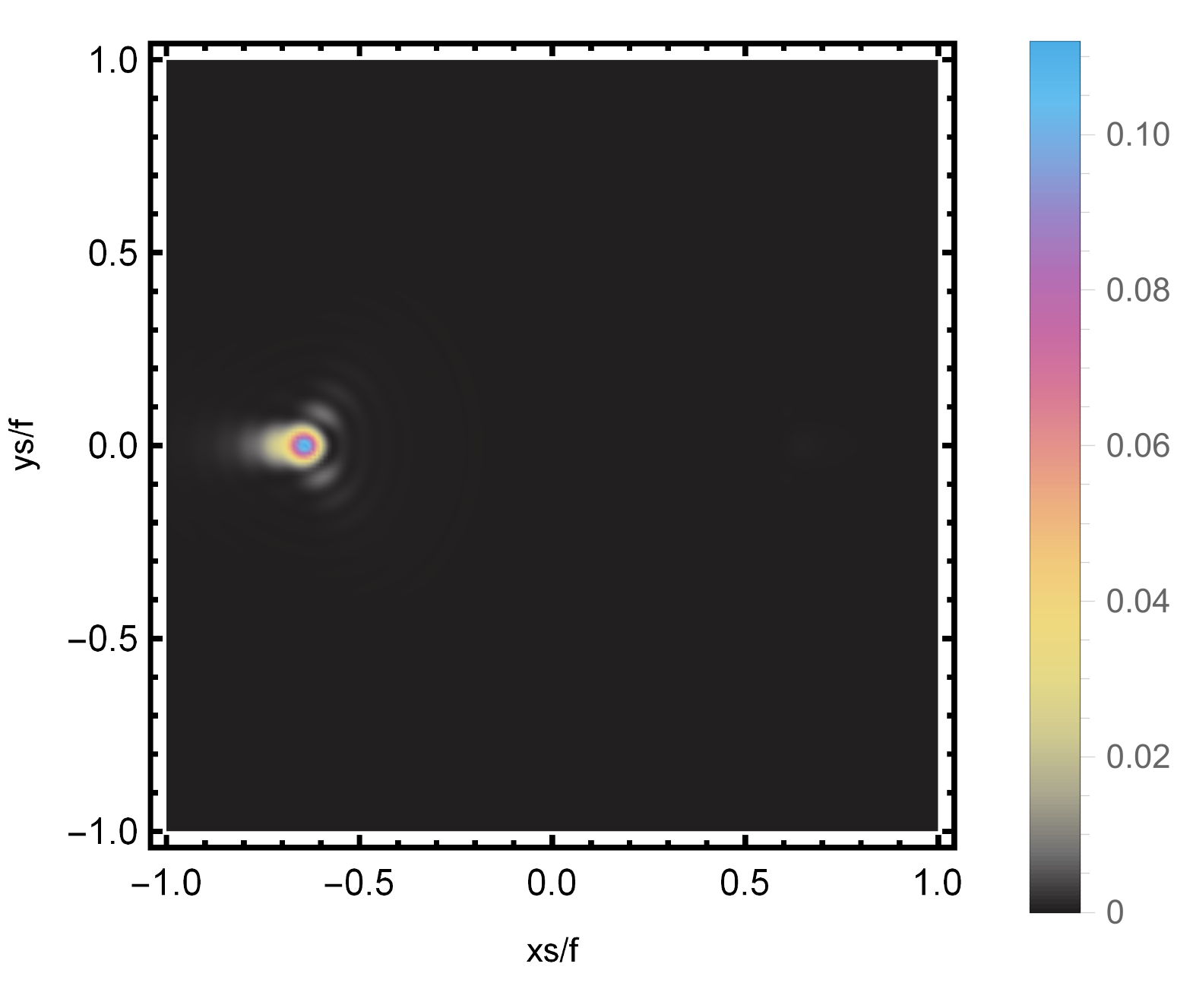}
  \caption{$\beta=0.3$,$\theta _{obs}=\pi/2$}
\end{subfigure}
\hfill
\begin{subfigure}[b]{0.24\textwidth}
  \centering
\includegraphics[width=\textwidth]{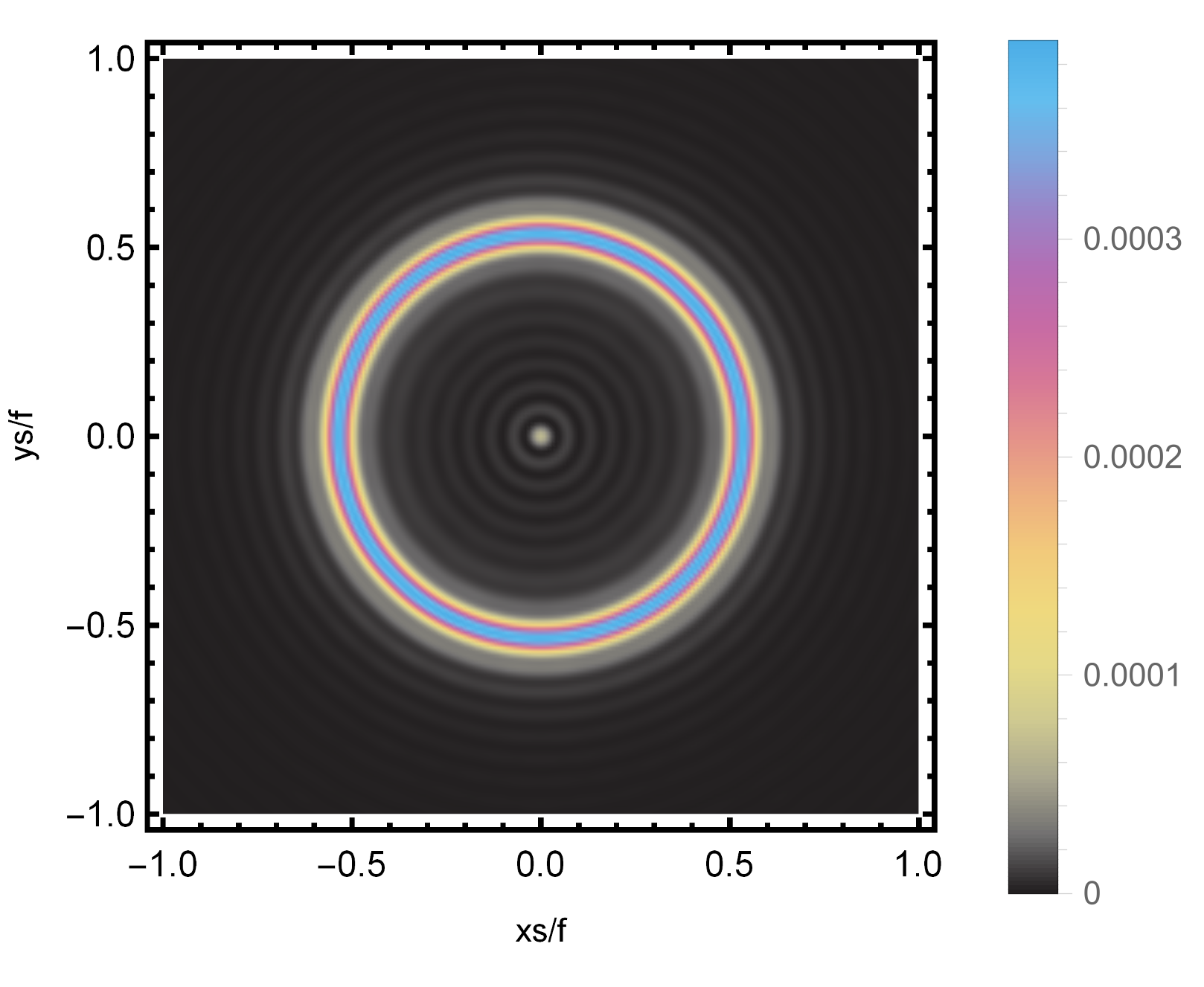}
  \caption{$\beta=0.45$,$\theta _{obs}=0$}
\end{subfigure}
\hfill
\begin{subfigure}[b]{0.24\textwidth}
  \centering
  \includegraphics[width=\textwidth]{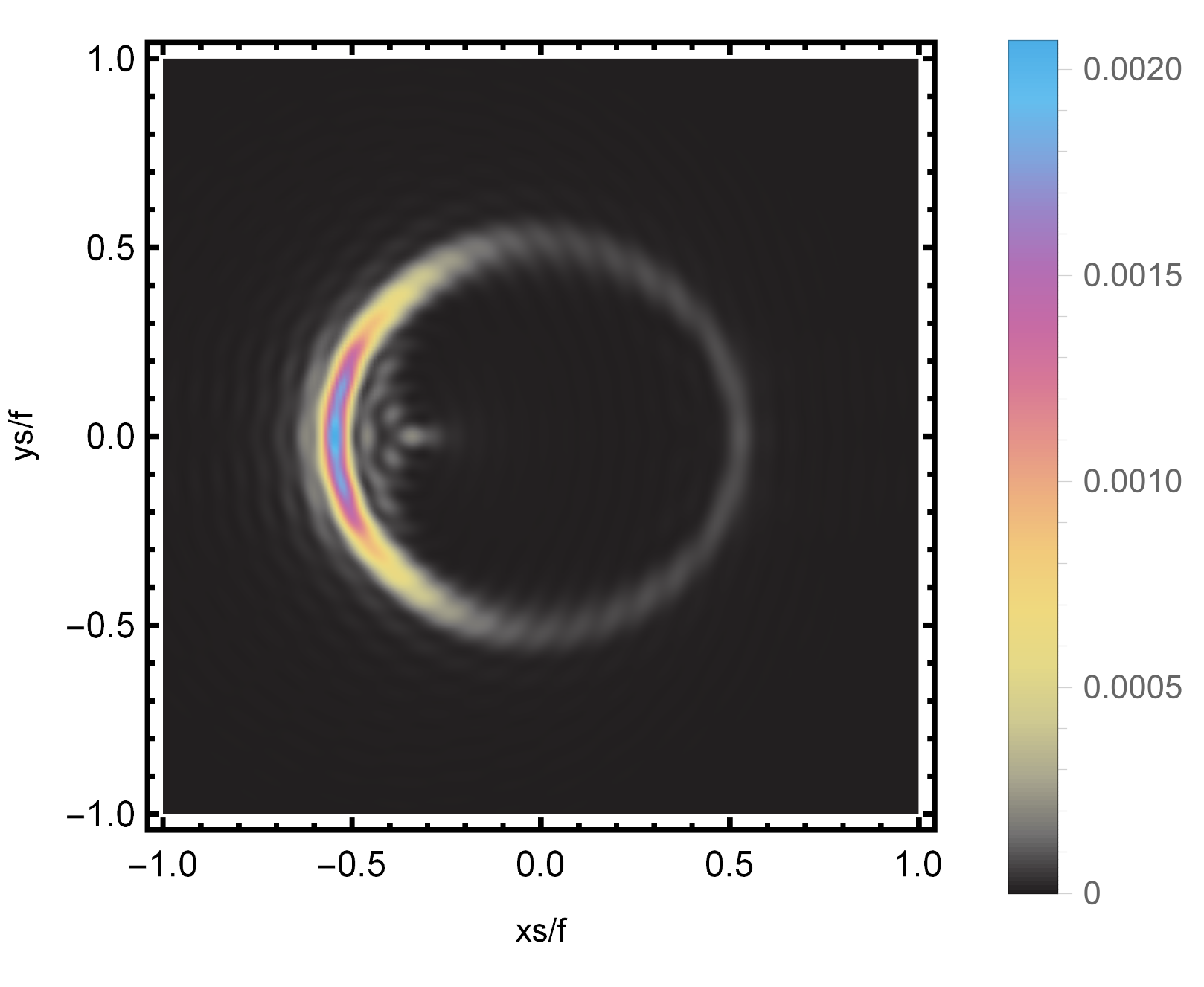}
  \caption{$\beta=0.45$,$\theta _{obs}=\pi/6$}
\end{subfigure}
\hfill
\begin{subfigure}[b]{0.24\textwidth}
  \centering
  \includegraphics[width=\textwidth]{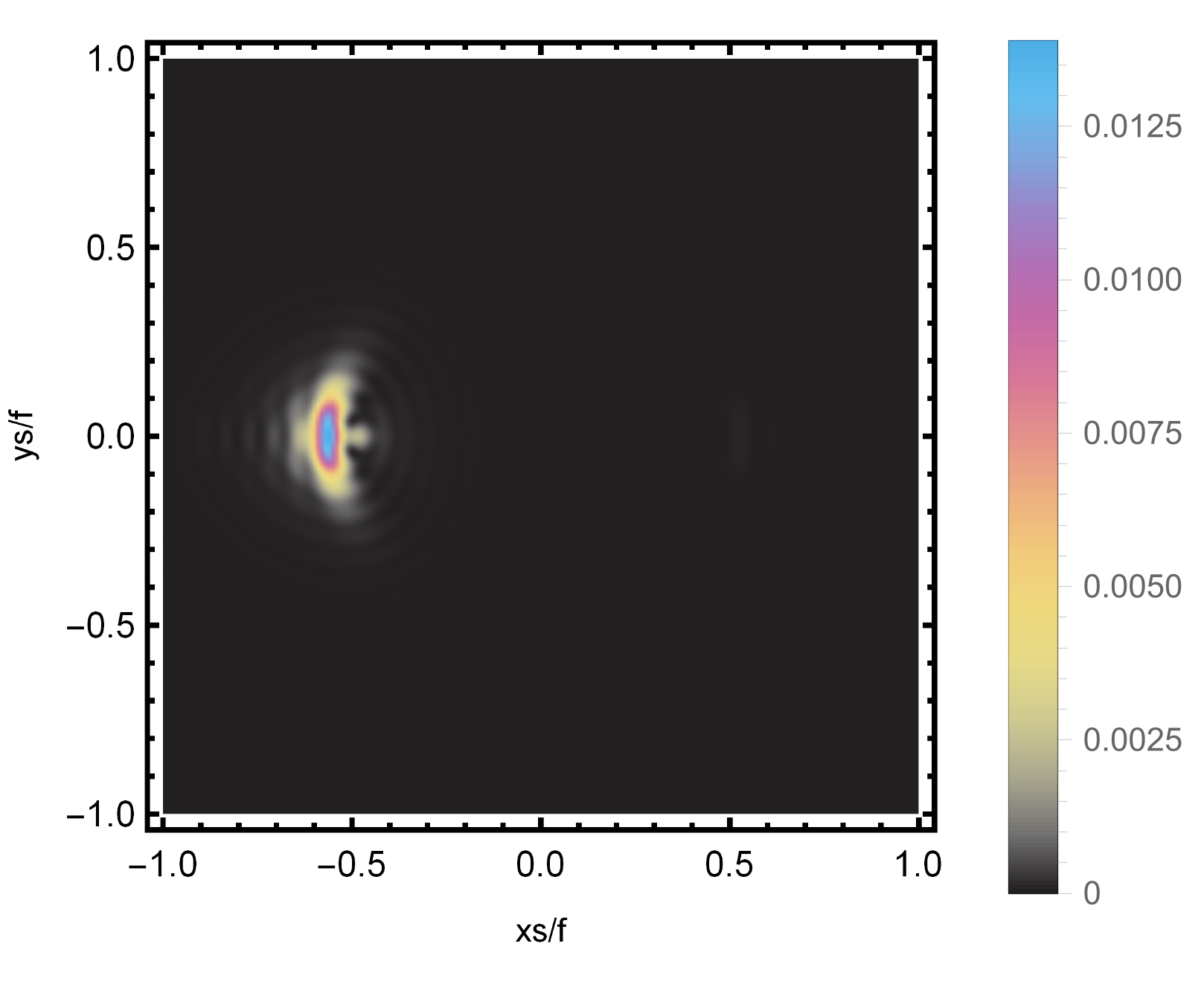}
  \caption{$\beta=0.45$,$\theta _{obs}=\pi/3$}
\end{subfigure}
\hfill
\begin{subfigure}[b]{0.24\textwidth}
  \centering
  \includegraphics[width=\textwidth]{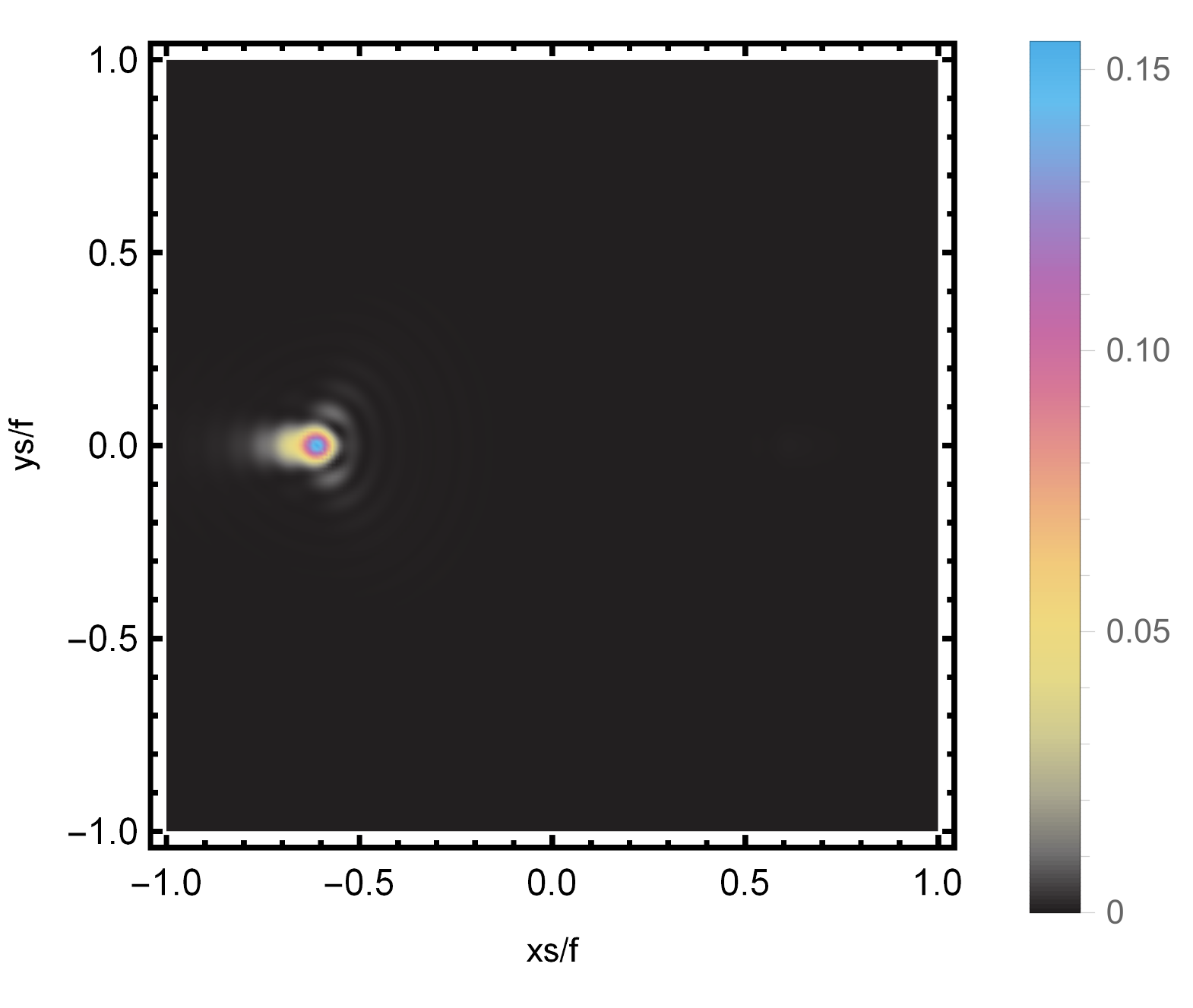}
  \caption{$\beta=0.45$,$\theta _{obs}=\pi/2$}
\end{subfigure}
\hfill
\begin{subfigure}[b]{0.24\textwidth}
  \centering
\includegraphics[width=\textwidth]{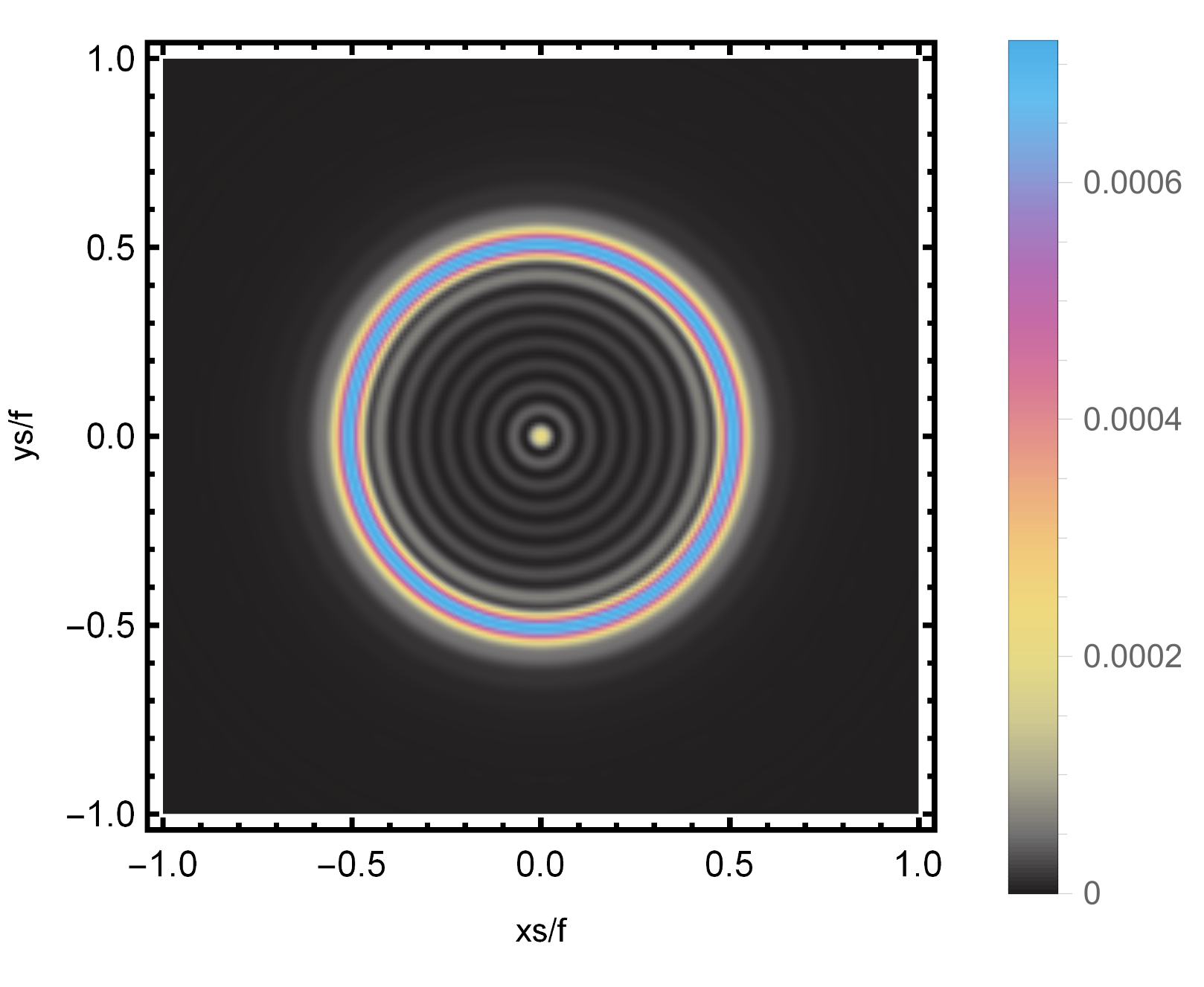} \caption{$\beta=0.6$,$\theta _{obs}=0$}
\end{subfigure}
\hfill
\begin{subfigure}[b]{0.24\textwidth}
  \centering
  \includegraphics[width=\textwidth]{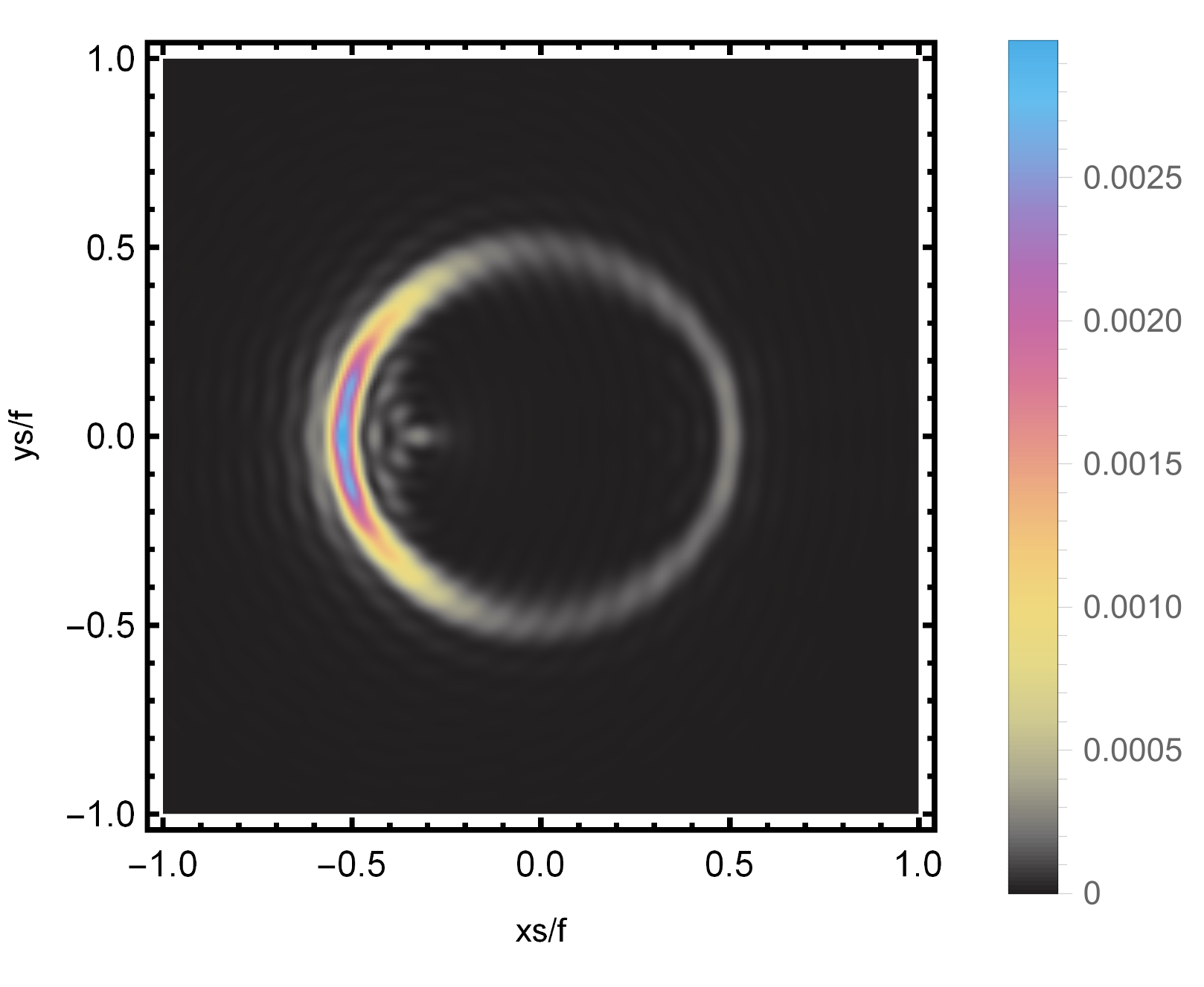}
  \caption{$\beta=0.6$,$\theta _{obs}=\pi/6$}
\end{subfigure}
\hfill
\begin{subfigure}[b]{0.24\textwidth}
  \centering
  \includegraphics[width=\textwidth]{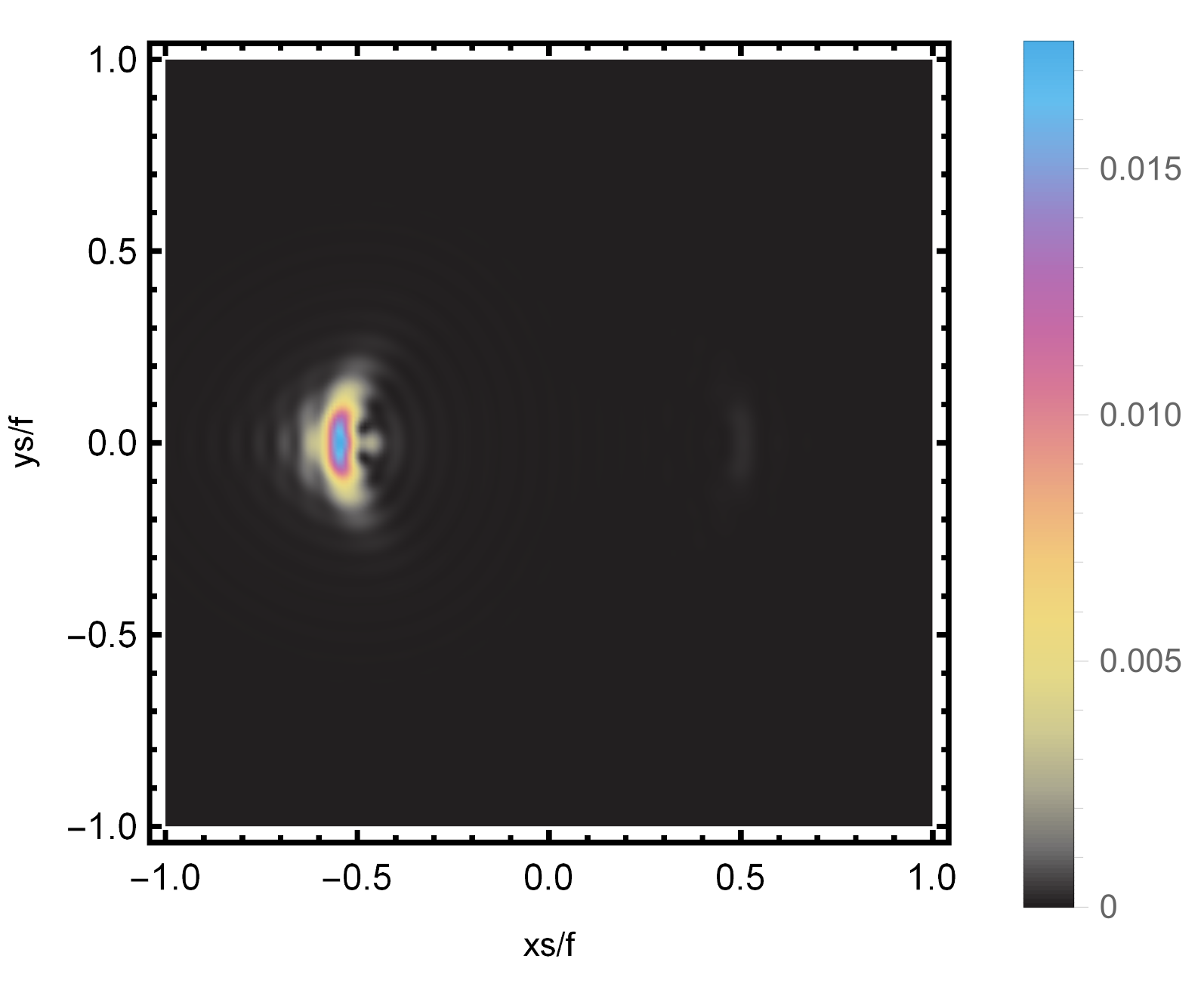}
  \caption{$\beta=0.6$,$\theta _{obs}=\pi/3$}
\end{subfigure}
\hfill
\begin{subfigure}[b]{0.24\textwidth}
  \centering
  \includegraphics[width=\textwidth]{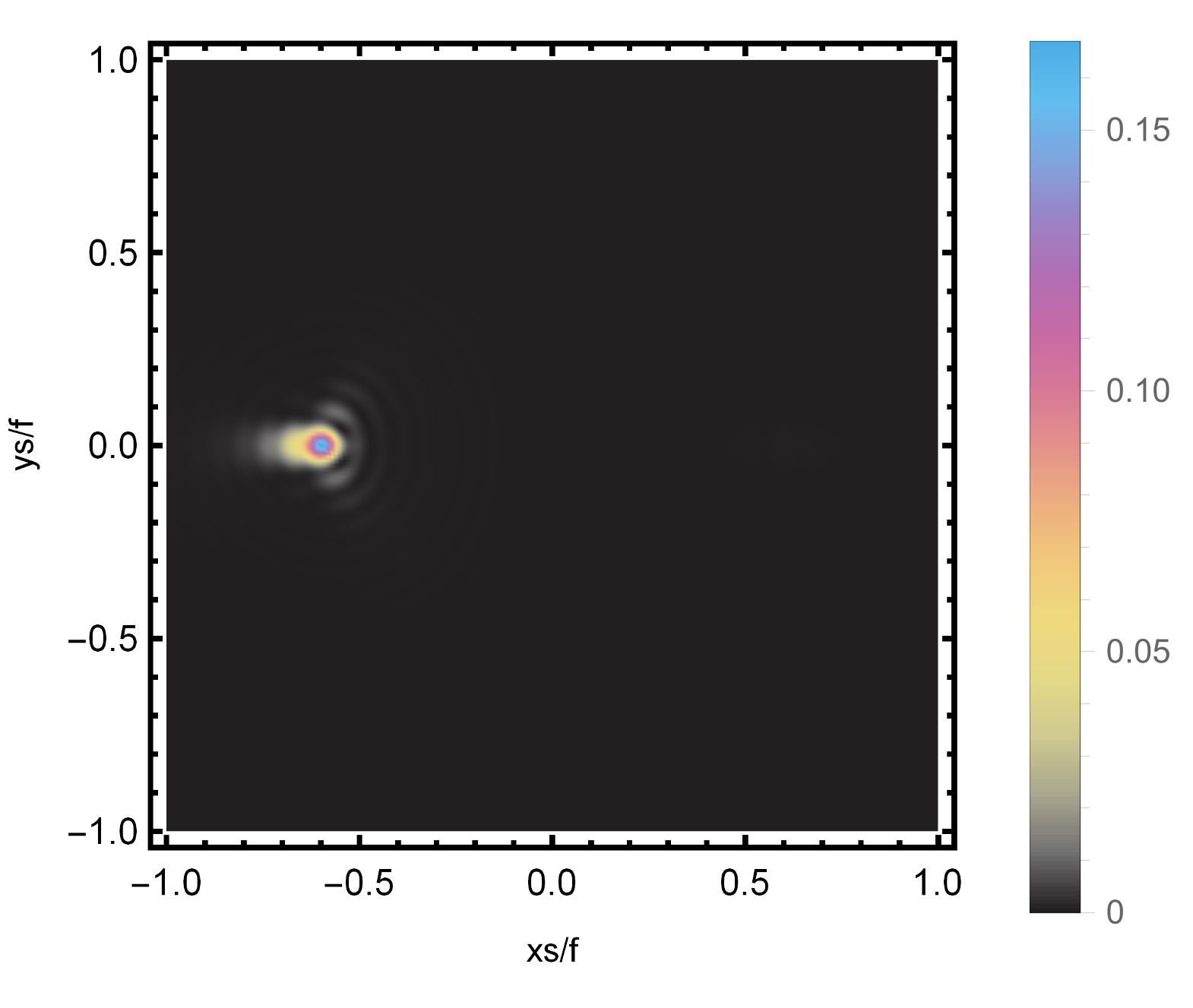}
  \caption{$\beta=0.6$,$\theta _{obs}=\pi/2$}
\end{subfigure}
\hfill
\begin{subfigure}[b]{0.24\textwidth}
  \centering
\includegraphics[width=\textwidth]{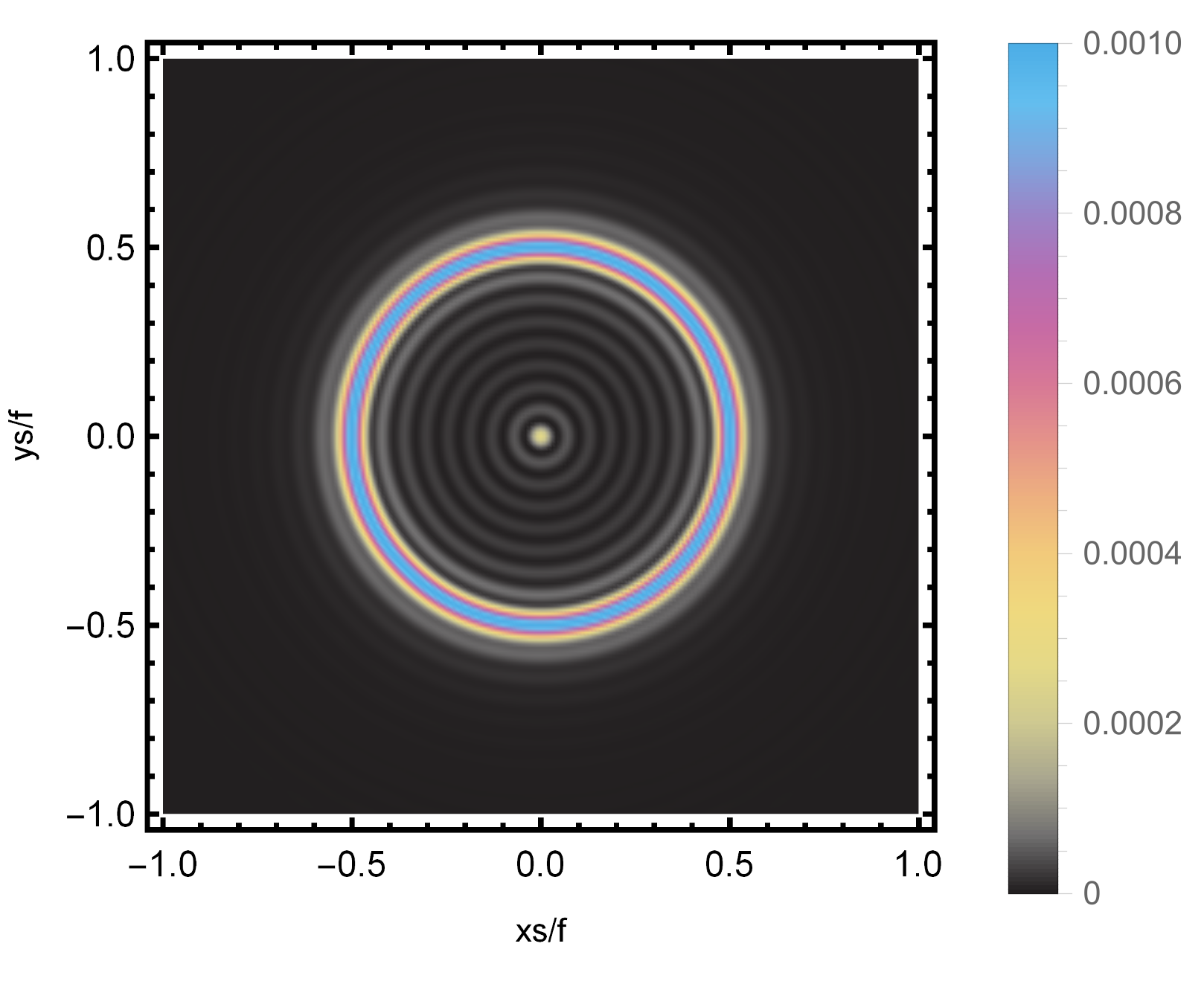}
  \caption{$\beta=0.75$,$\theta _{obs}=0$}
\end{subfigure}
\hfill
\begin{subfigure}[b]{0.24\textwidth}
  \centering
  \includegraphics[width=\textwidth]{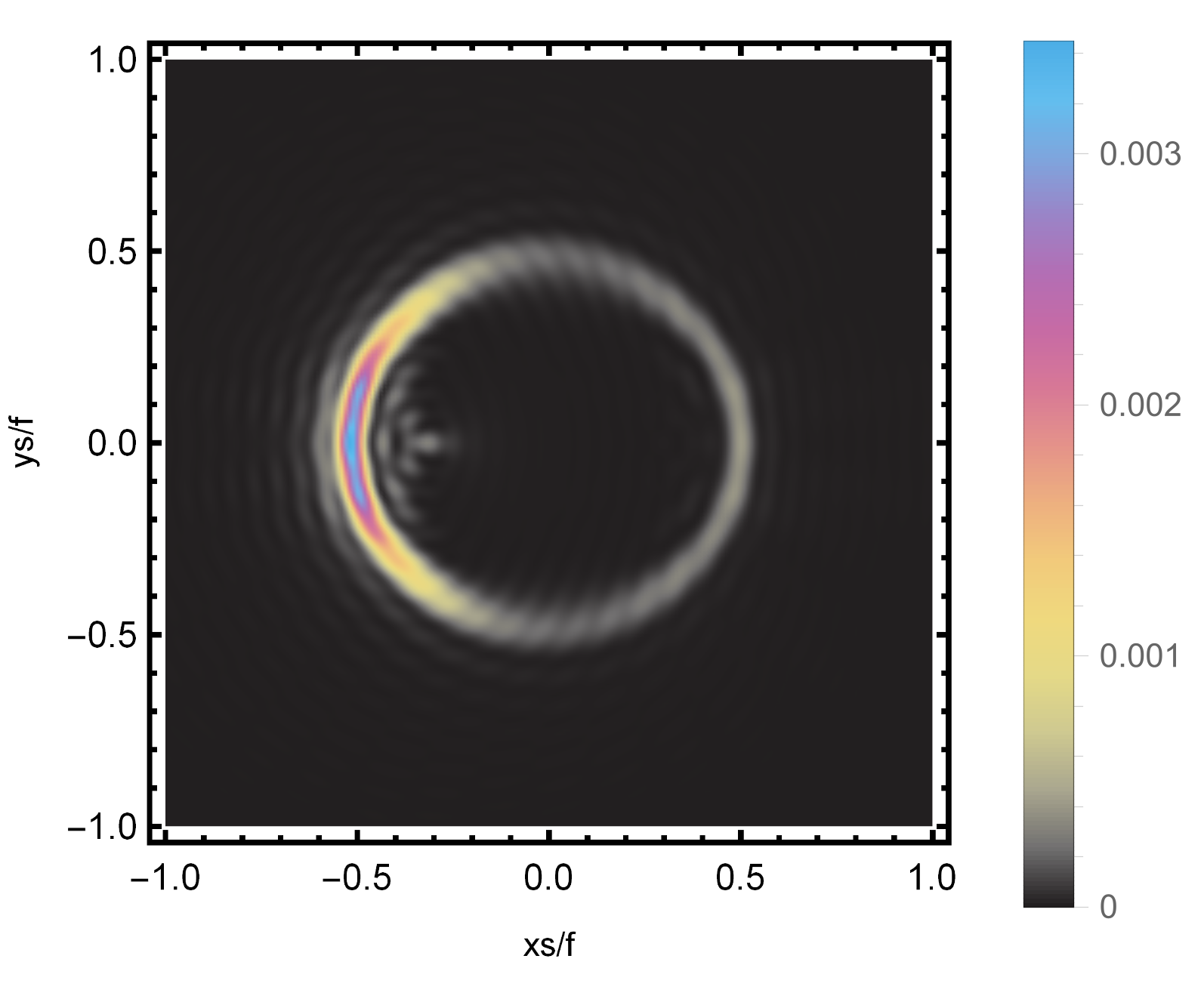}
  \caption{$\beta=0.75$,$\theta _{obs}=\pi/6$}
\end{subfigure}
\hfill
\begin{subfigure}[b]{0.24\textwidth}
  \centering
  \includegraphics[width=\textwidth]{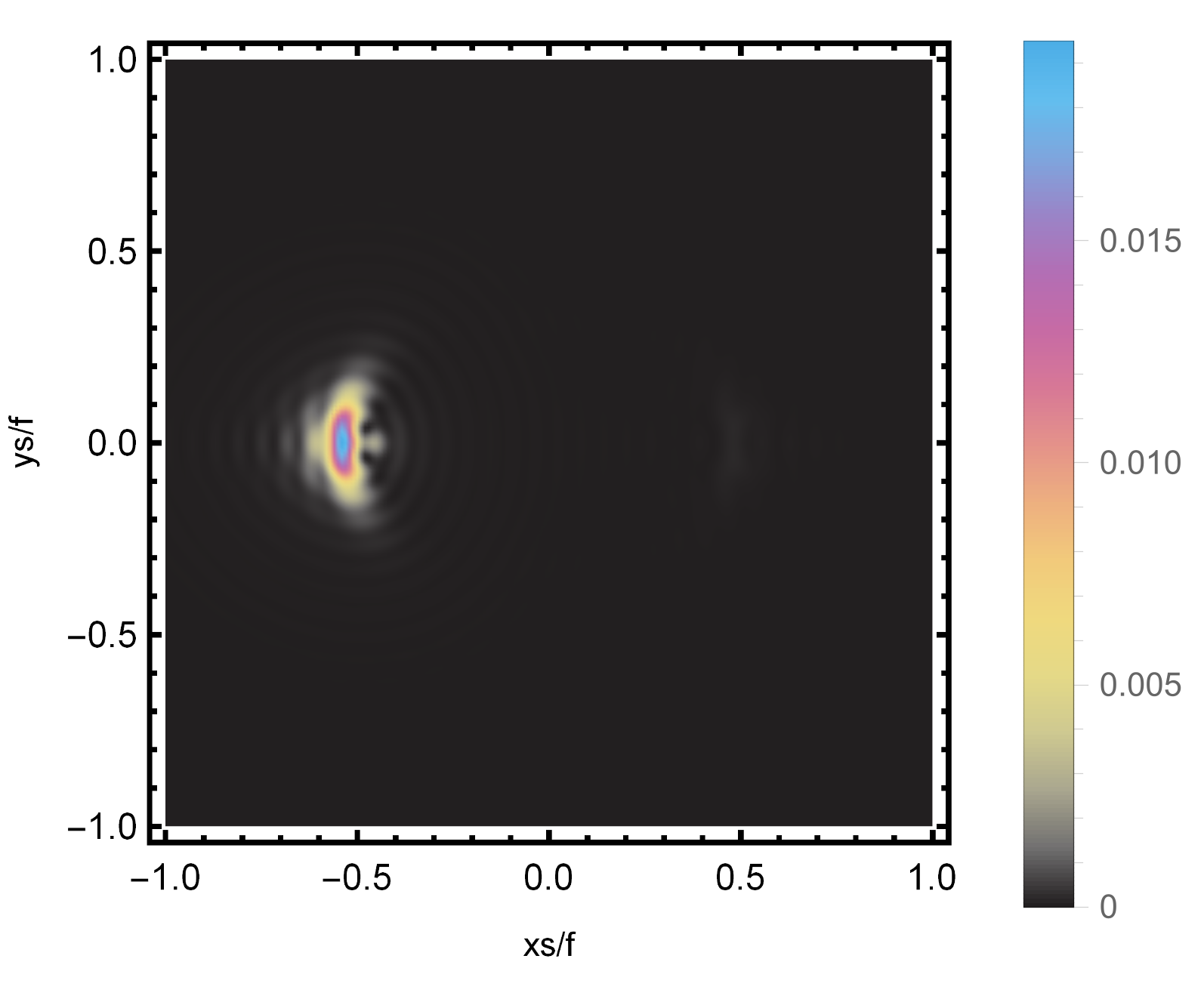}
  \caption{$\beta=0.75$,$\theta _{obs}=\pi/3$}
\end{subfigure}
\hfill
\begin{subfigure}[b]{0.24\textwidth}
  \centering
  \includegraphics[width=\textwidth]{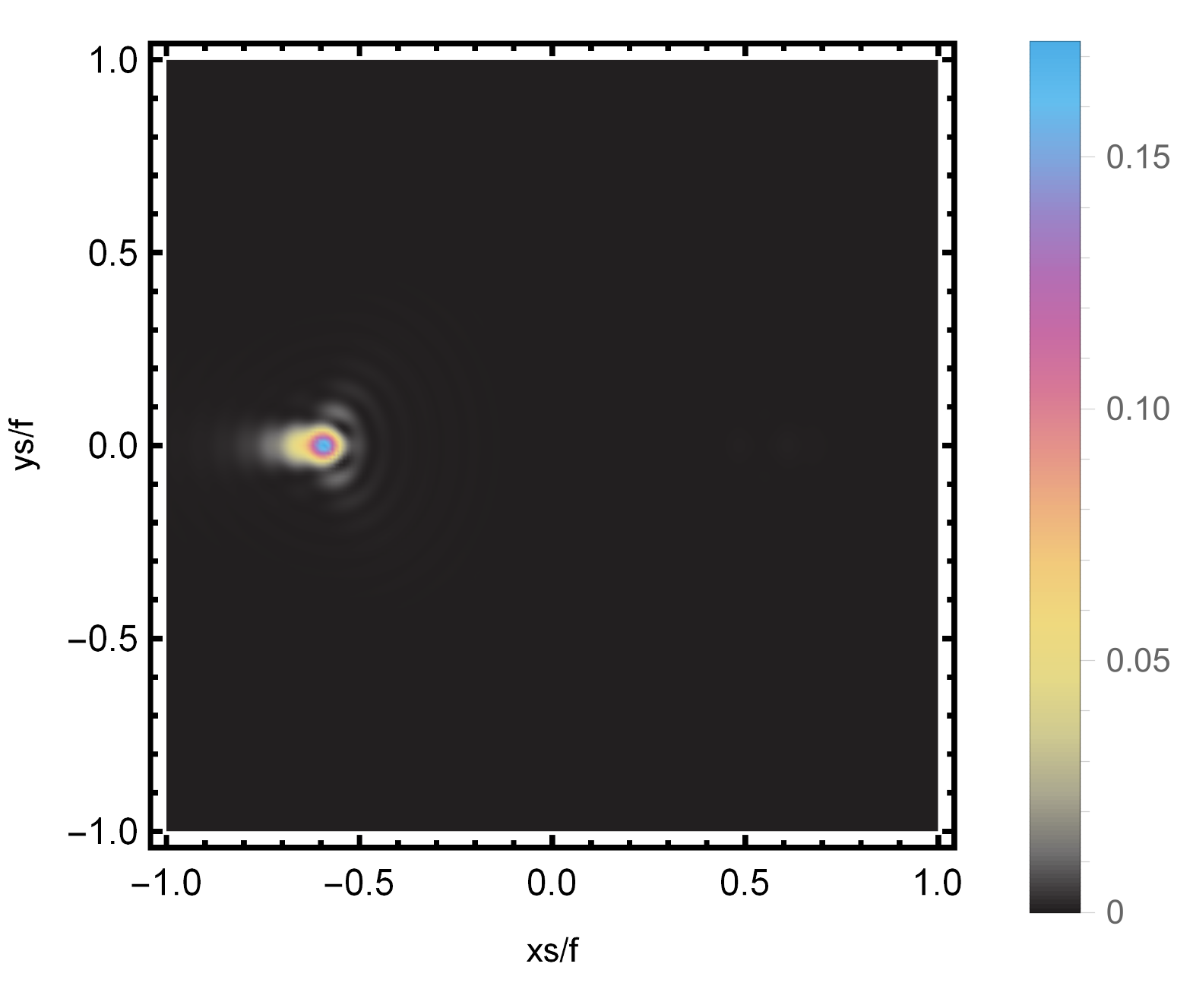}
  \caption{$\beta=0.75$,$\theta _{obs}=\pi/2$}
\end{subfigure}
\hfill
\begin{subfigure}[b]{0.24\textwidth}
  \centering
\end{subfigure}
\caption{Observational appearance of the response on the screen for different $\beta$ at various observation angles, where $y_{h}=5$, $\omega =90$.}
\label{15}%
\end{figure*}

\begin{figure}[htbp]
  \centering
  \begin{subfigure}[b]{0.45\columnwidth}
    \centering
    
    \includegraphics[width=\textwidth,height=0.9\textwidth]{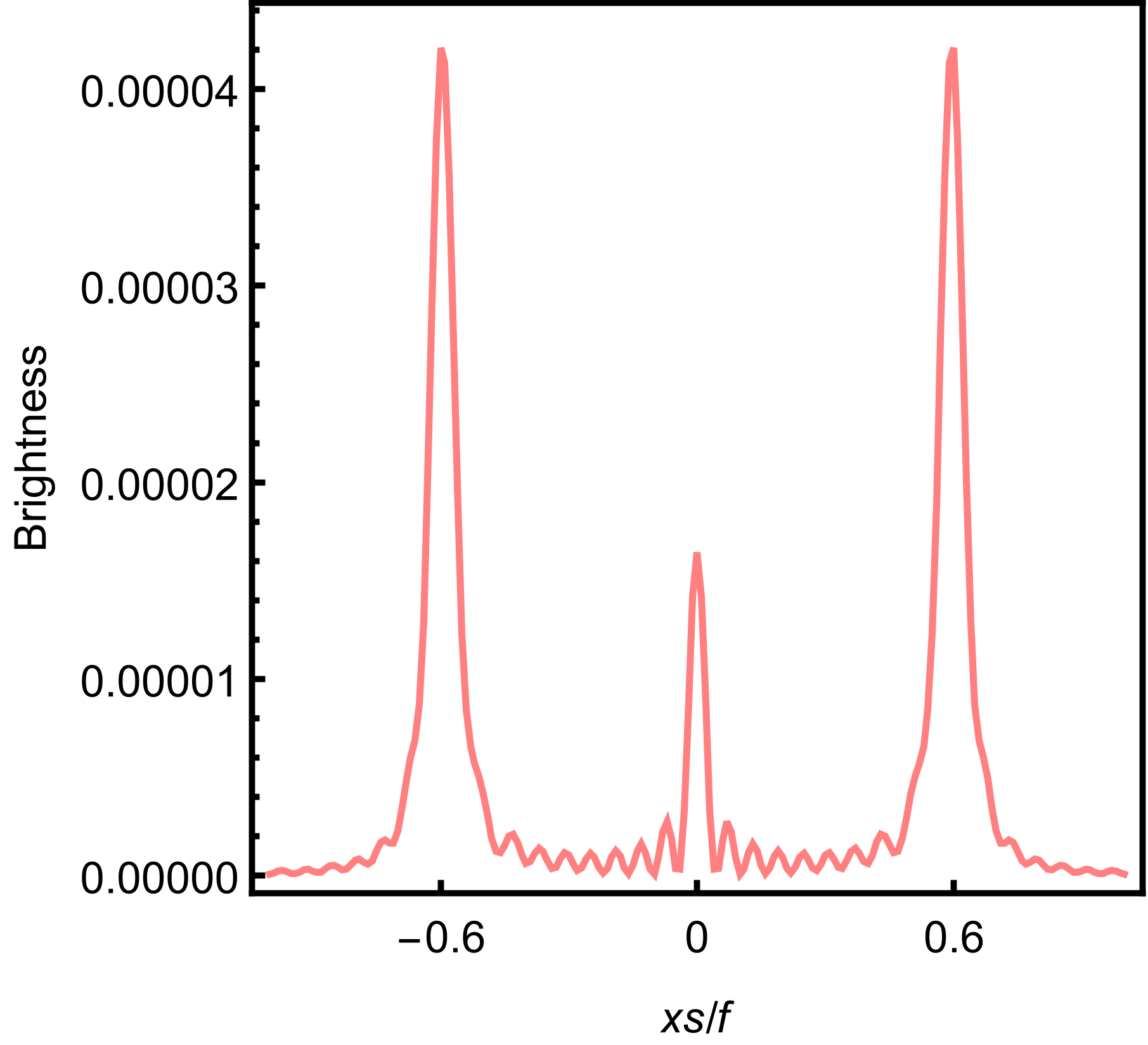}
    \caption{$\beta=0.3$}
  \end{subfigure}
  \hfill
  \begin{subfigure}[b]{0.45\columnwidth}
    \centering
    \includegraphics[width=\textwidth,height=0.9\textwidth]{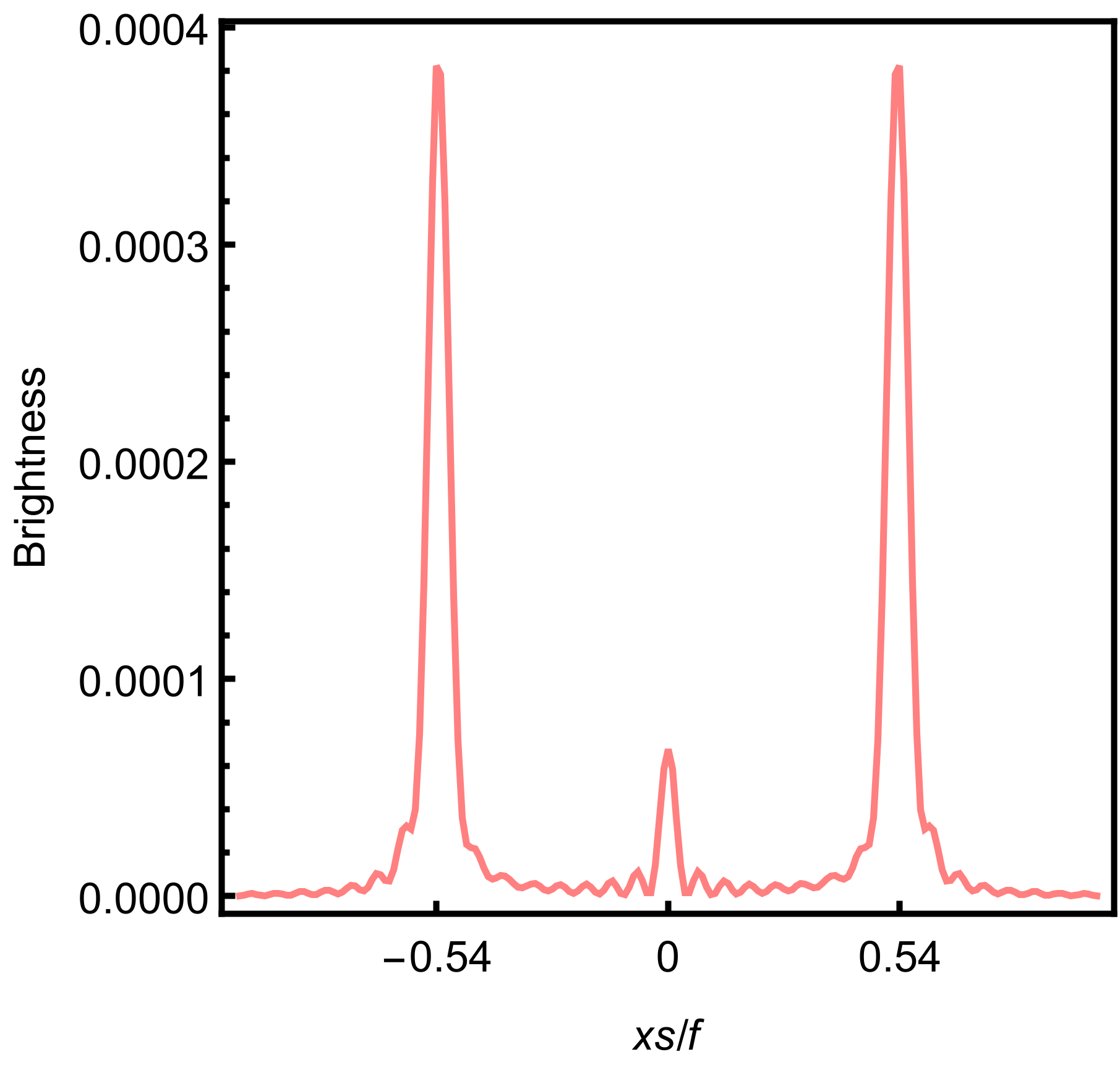}
    \caption{$\beta=0.45$}
  \end{subfigure}

\begin{subfigure}[b]{0.45\columnwidth}
    \centering
    \includegraphics[width=\textwidth,height=0.9\textwidth]{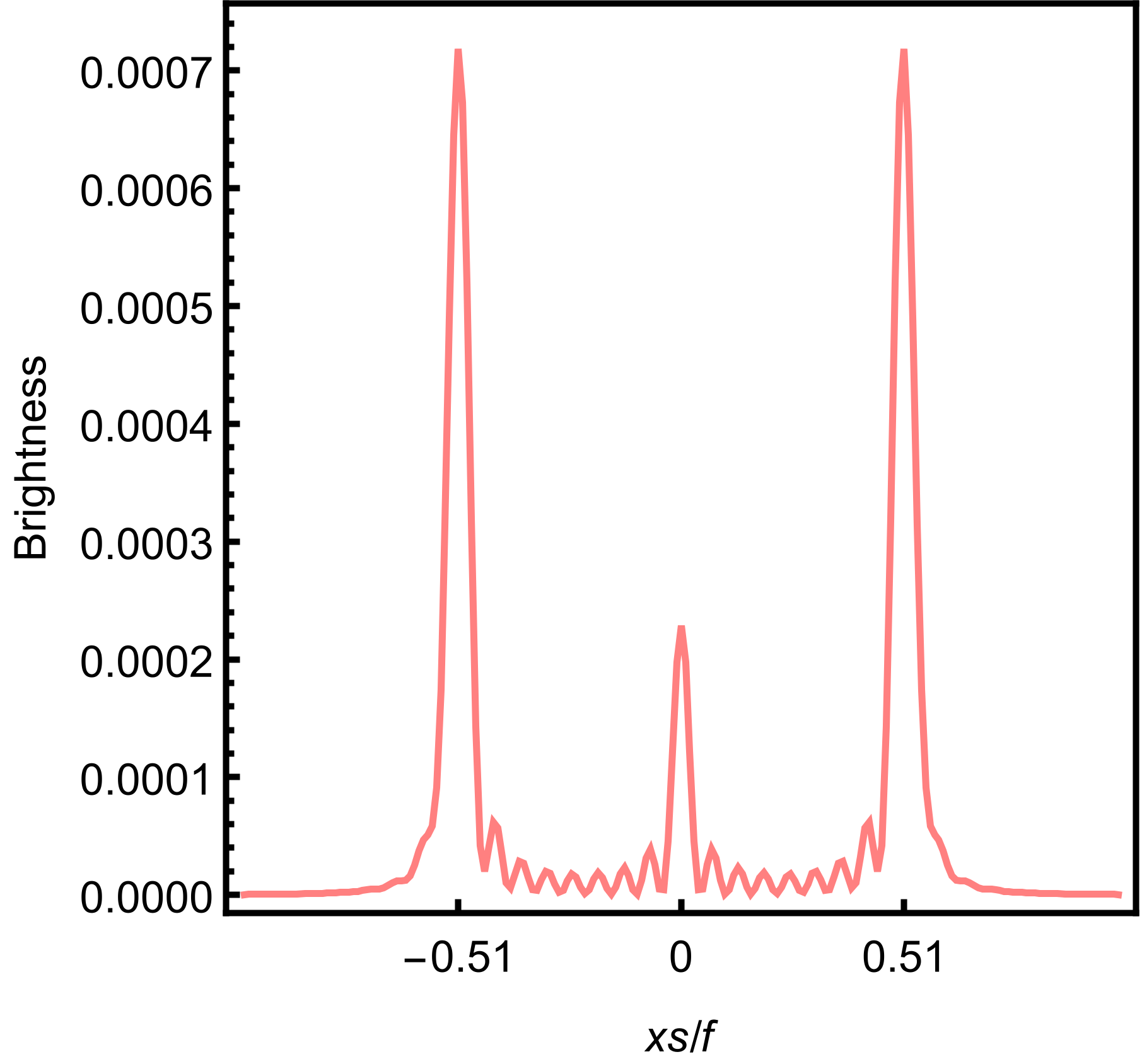}
    \caption{$\beta=0.6$}
  \end{subfigure}
  \hfill
  \begin{subfigure}[b]{0.45\columnwidth}
    \centering
    \includegraphics[height=0.9\textwidth,width=\textwidth]{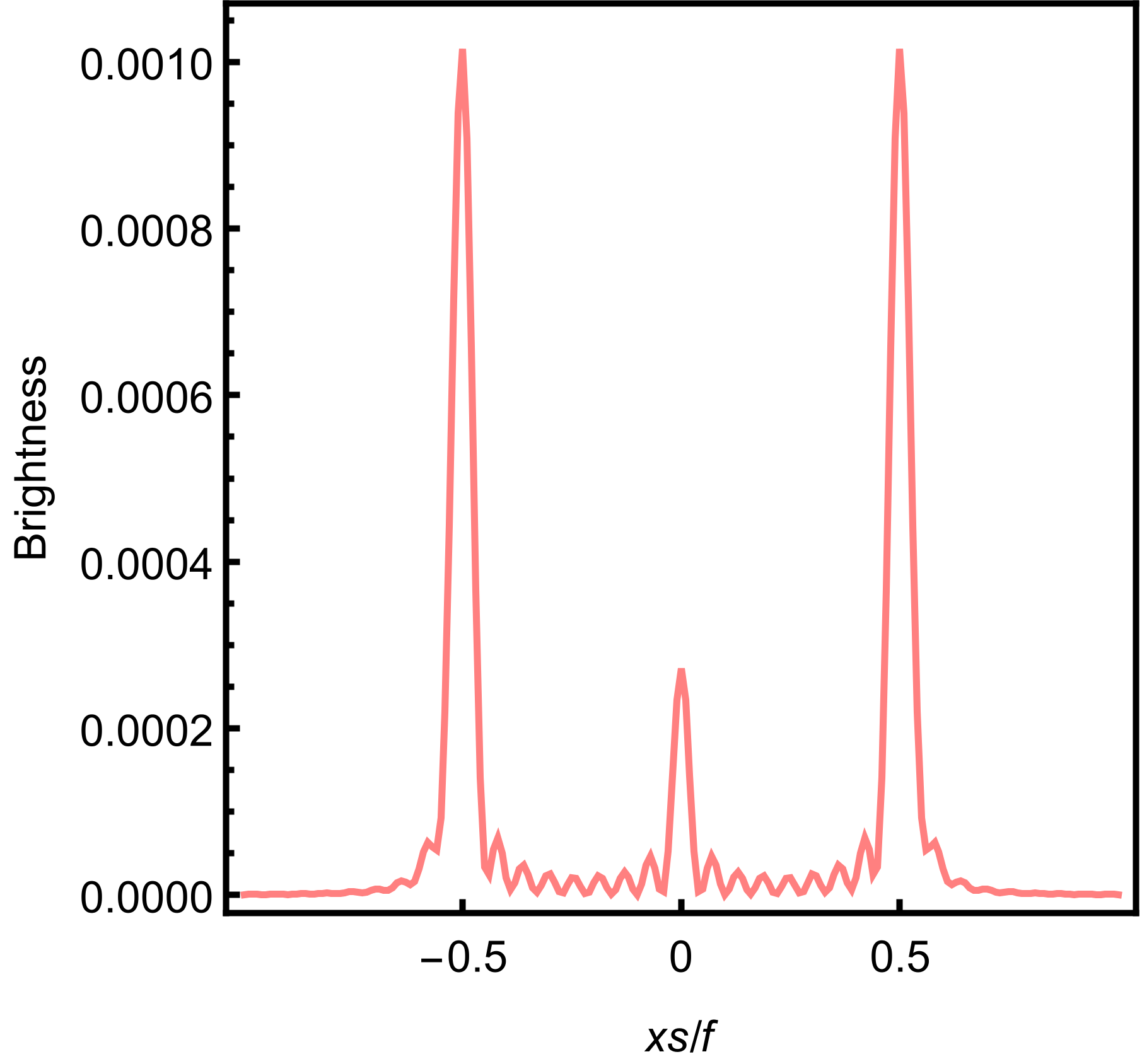}
    \caption{$\beta=0.75$}
  \end{subfigure}
  
  \caption{Effect of $\beta$ on the brightness, where $\theta _{obs}=0$, $y_{h}=5$,$\omega=90$. }
  \label{16}%
\end{figure}

Similarly, as the control parameter $\beta$ changes, a series of concentric rings can be observed in the leftmost column of Figure \ref{15}. As the observer's position $\theta _{obs}$ varies from $\theta _{obs}=0$ to $\theta _{obs}=\pi/2$, the Einstein ring transitions from strict spherical symmetry to axial symmetry, accompanied by a decrease in partial brightness. Subsequently, the ring gradually diminishes, ultimately leaving only a single light spot. Figure \ref{16} indicates that the radius of the Einstein ring gradually decreases as $\beta$ increases. Specifically, when $\beta=0.3$, the radius of the ring is 0.6, and when $\beta=0.75$, the radius of the ring decreases to  0.5. Furthermore, as $\beta$ increases, the luminosity gradually intensifies. 

\begin{figure}[htbp]
  \centering
  \begin{subfigure}[b]{0.49\columnwidth}
    \centering
    \includegraphics[width=\textwidth,height=0.8\textwidth]{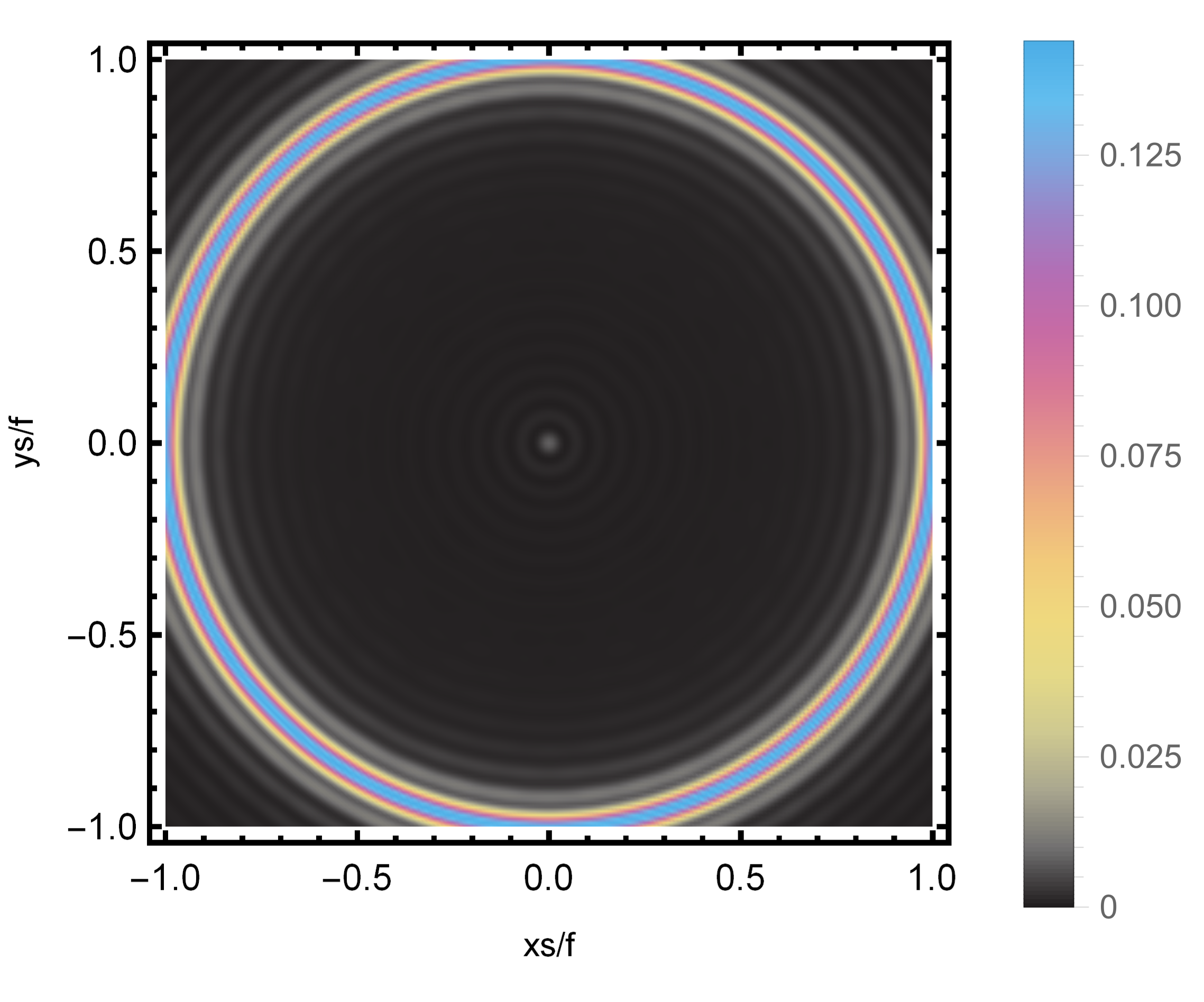}
    \caption{$y_{h}=0.5$,$\mathit{T}=0.515$}
  \end{subfigure}
  \hfill
  \begin{subfigure}[b]{0.49\columnwidth}
    \centering
    \includegraphics[width=\textwidth,height=0.8\textwidth]{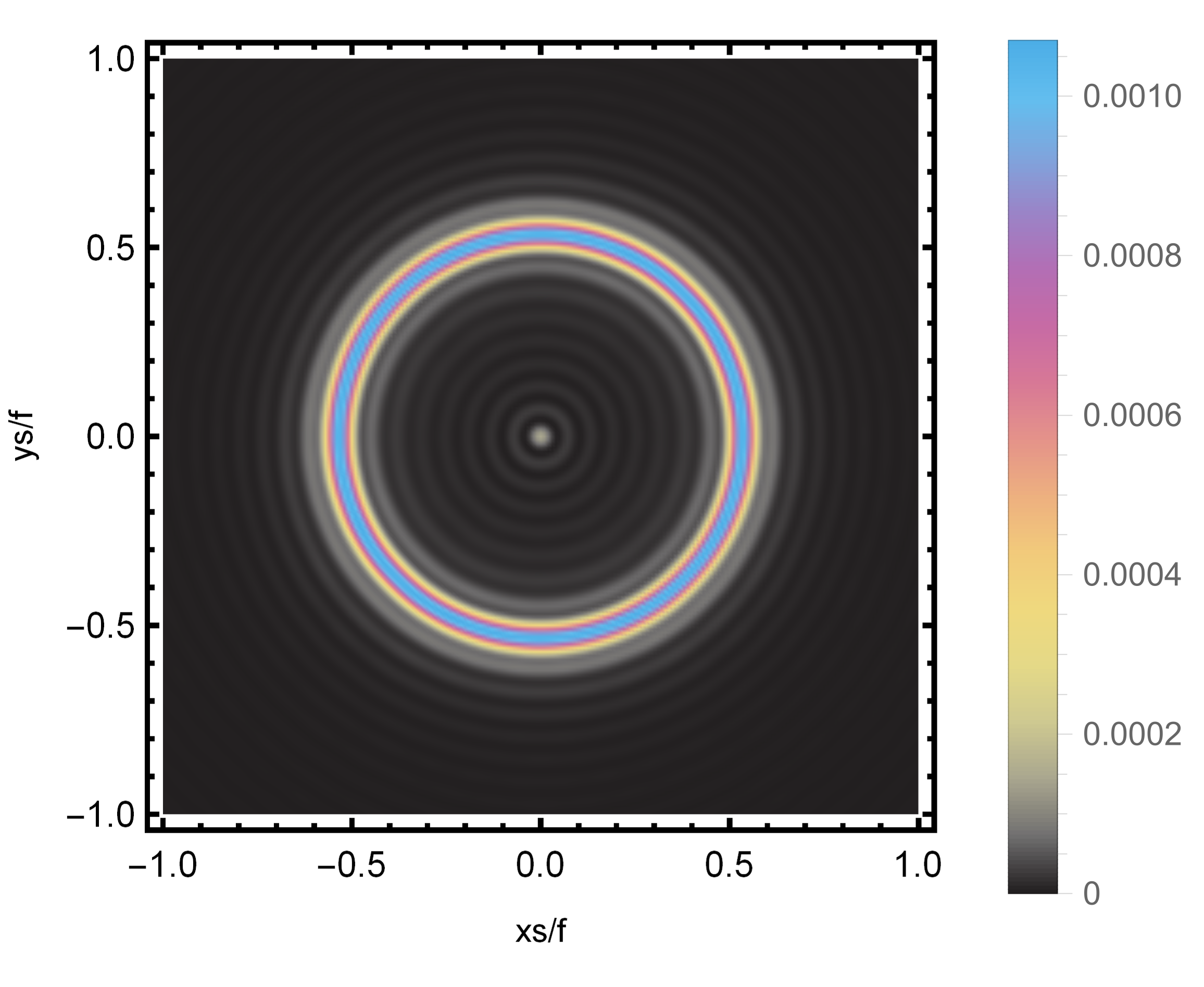}
    \caption{$y_{h}=4.5$,$\mathit{T}=0.403$}
  \end{subfigure}

\begin{subfigure}[b]{0.49\columnwidth}
    \centering
    \includegraphics[width=\textwidth,height=0.8\textwidth]{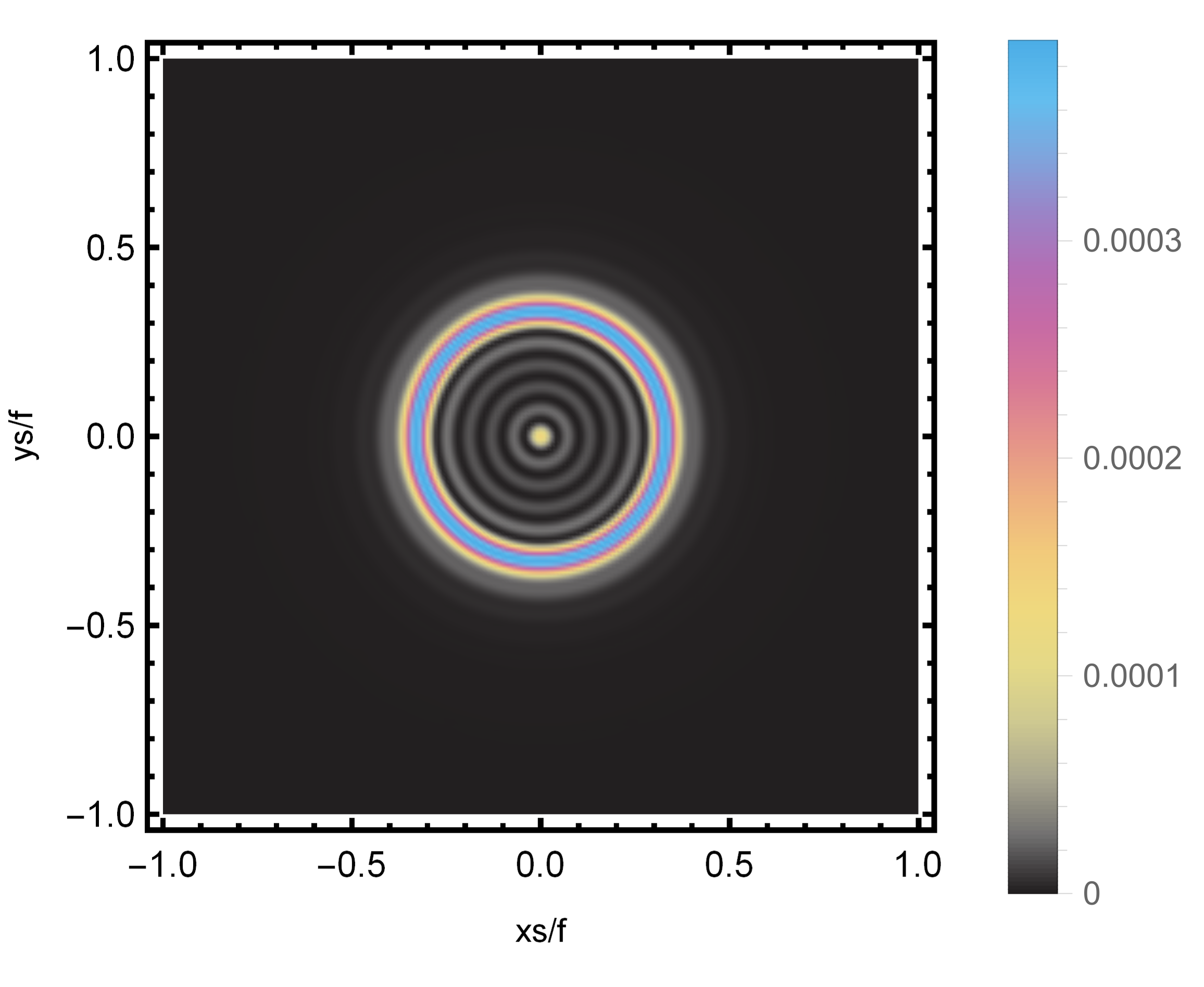}
    \caption{$y_{h}=8.5$,$\mathit{T}=0.698$}
  \end{subfigure}
  \hfill
  \begin{subfigure}[b]{0.49\columnwidth}
    \centering
    \includegraphics[width=\textwidth,height=0.8\textwidth]{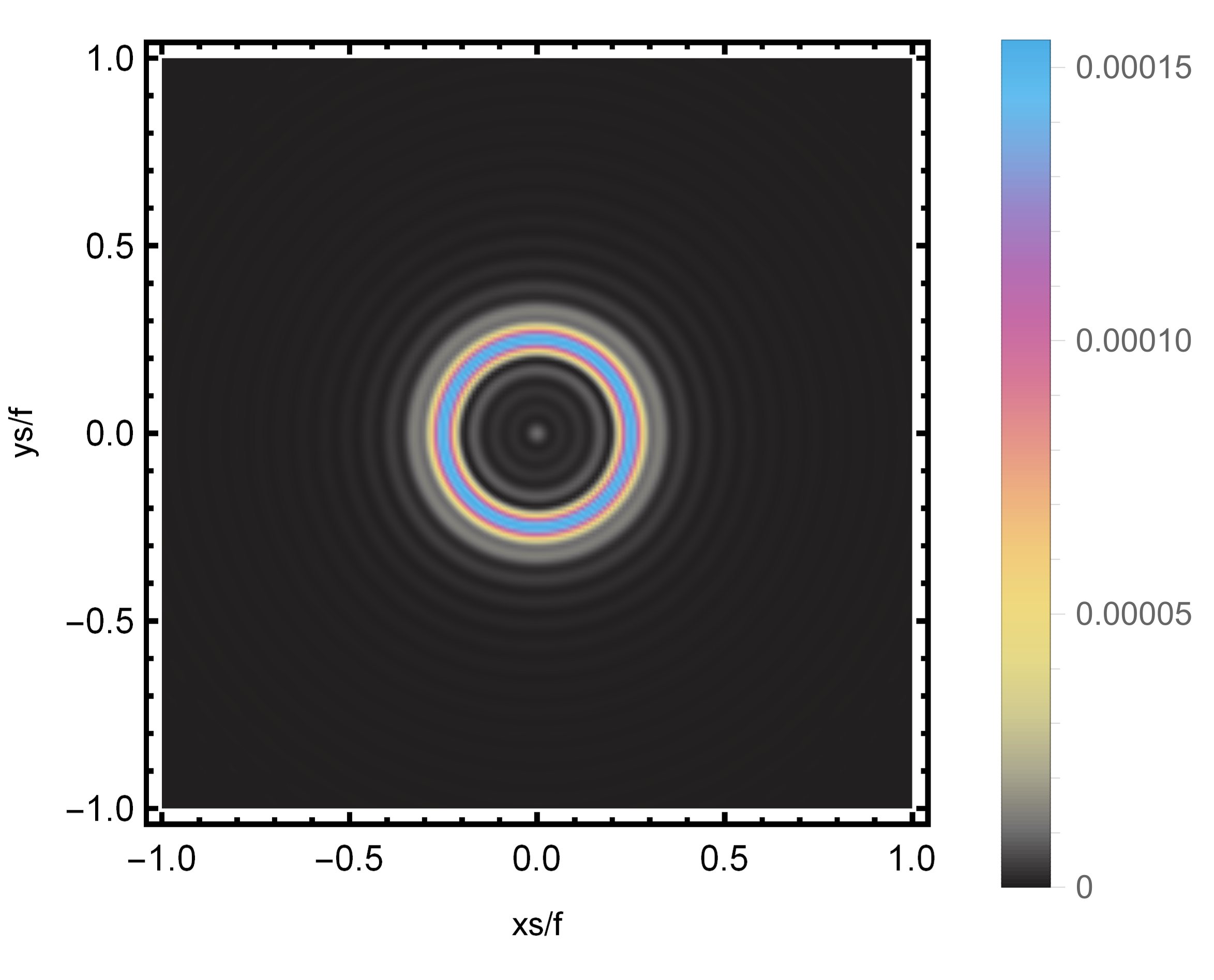}
    \caption{$y_{h}=12.5$,$\mathit{T}=1.009$}
  \end{subfigure}
  
  \caption{Effect of $\mathit{T}$ on the brightness, where $\theta_{obs}=0$, $\beta=1$, $\omega=90$. }
  \label{17}%
\end{figure}

\begin{figure}[htbp]
  \centering
  \begin{subfigure}[b]{0.45\columnwidth}
    \centering
    \includegraphics[width=\textwidth,height=0.9\textwidth]{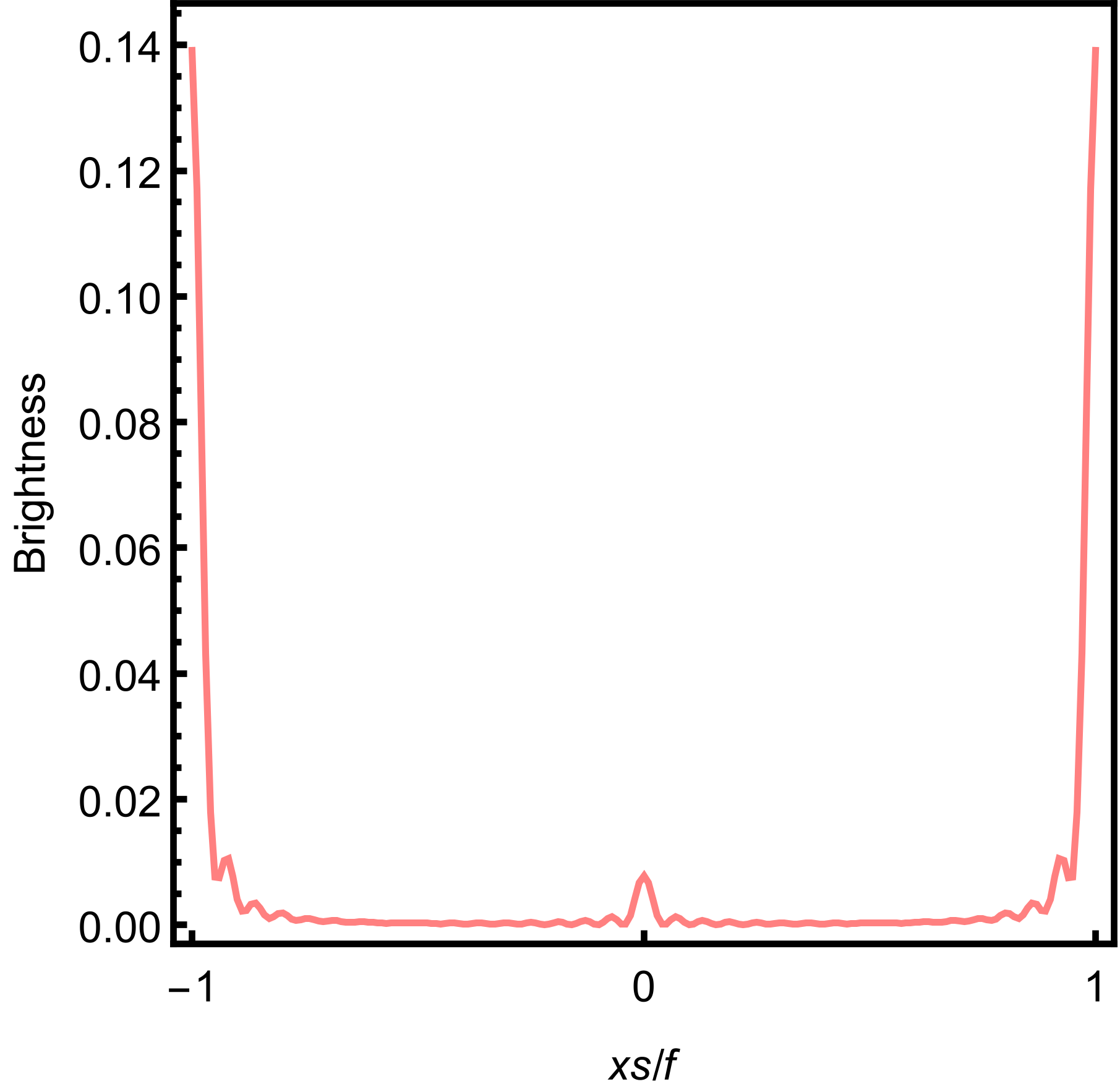}
    \caption{$y_{h}=0.5$,$\mathit{T}=0.515$}
  \end{subfigure}
  \hfill
  \begin{subfigure}[b]{0.45\columnwidth}
    \centering
    \includegraphics[width=\textwidth,height=0.9\textwidth]{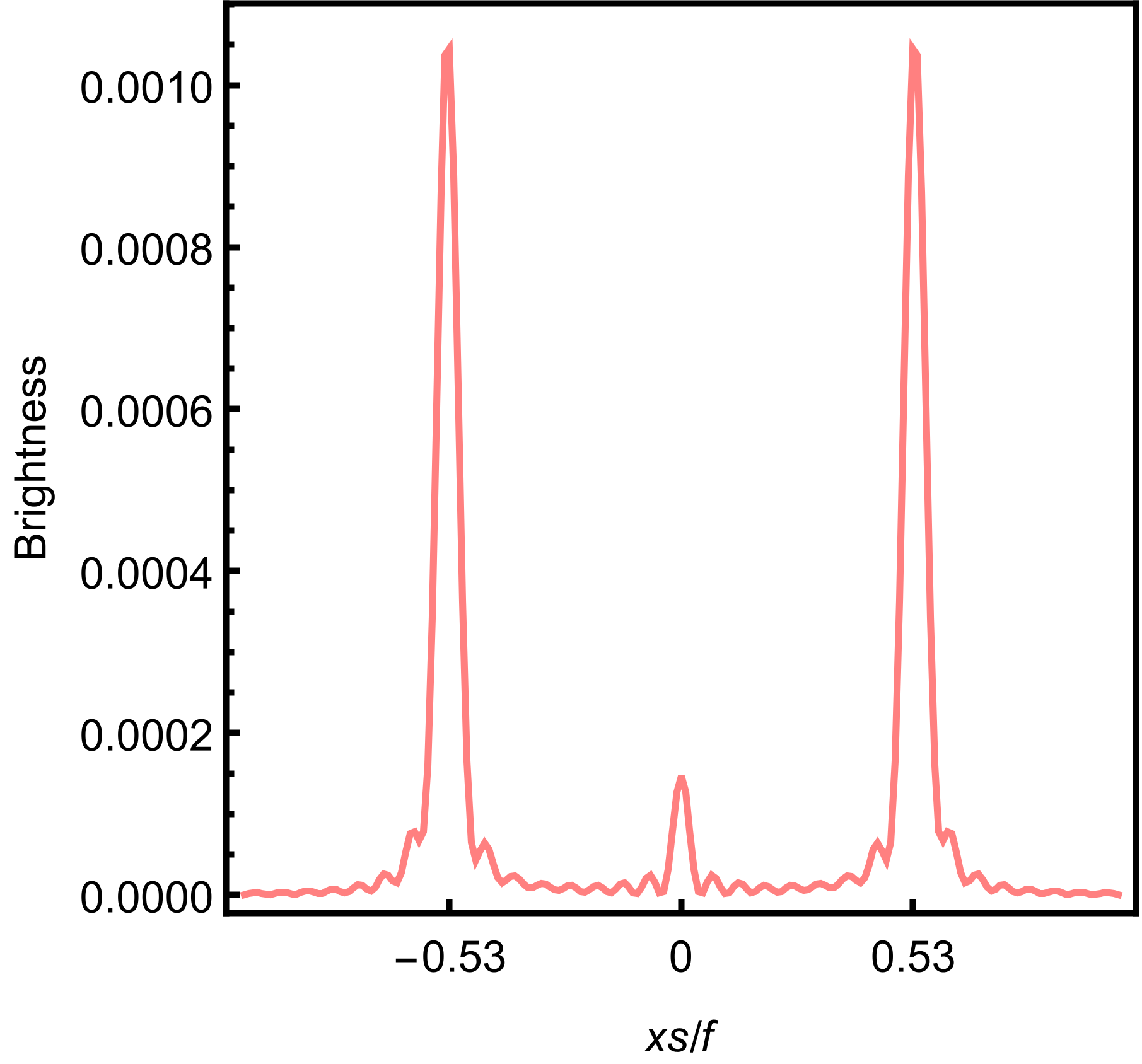}
    \caption{$y_{h}=4.5$,$\mathit{T}=0.403$}
  \end{subfigure}

\begin{subfigure}[b]{0.45\columnwidth}
    \centering
    \includegraphics[width=\textwidth,height=0.9\textwidth]{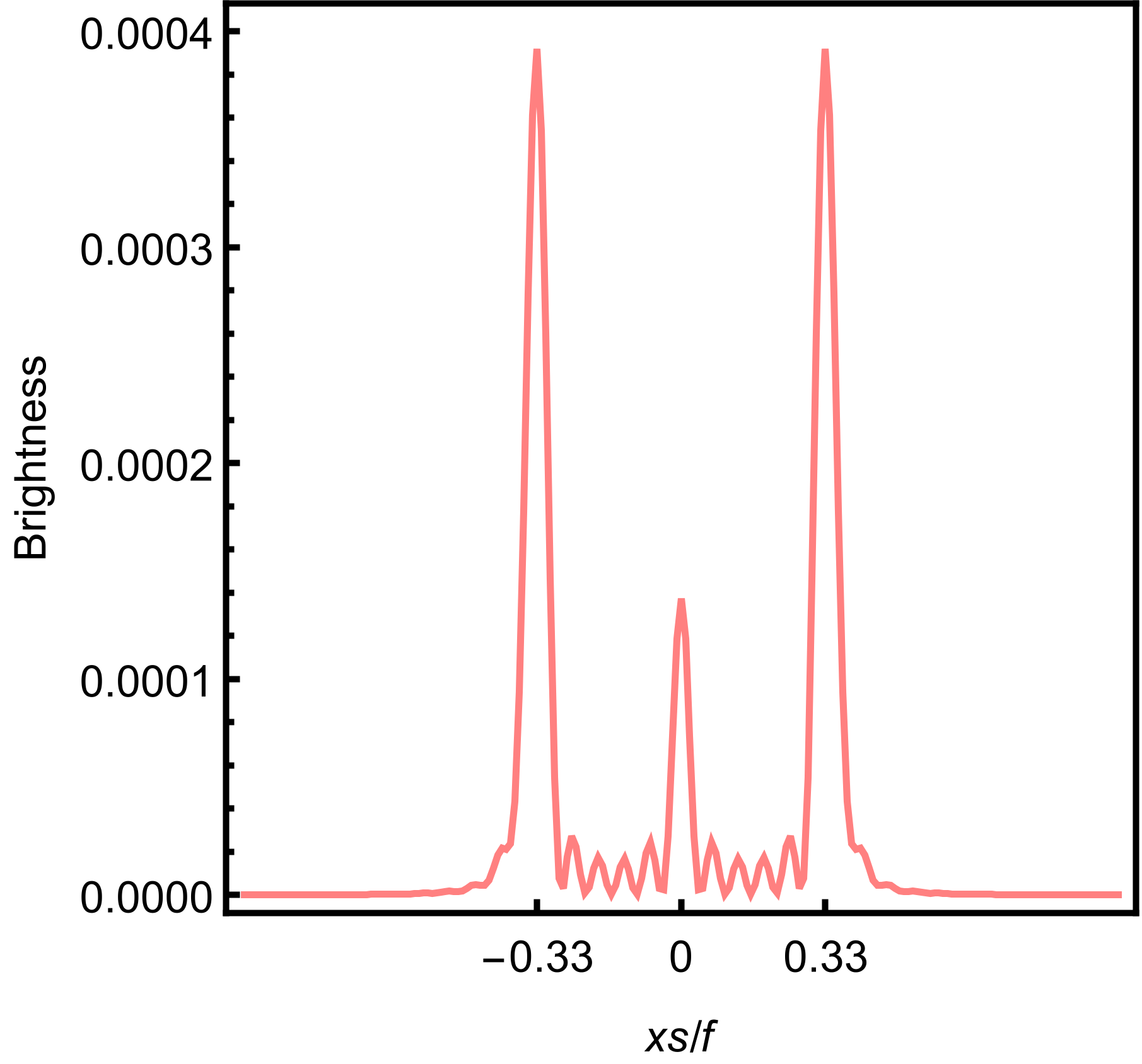}
    \caption{$y_{h}=8.5$,$\mathit{T}=0.698$}
  \end{subfigure}
  \hfill
  \begin{subfigure}[b]{0.45\columnwidth}
    \centering
    \includegraphics[width=\textwidth,height=0.9\textwidth]{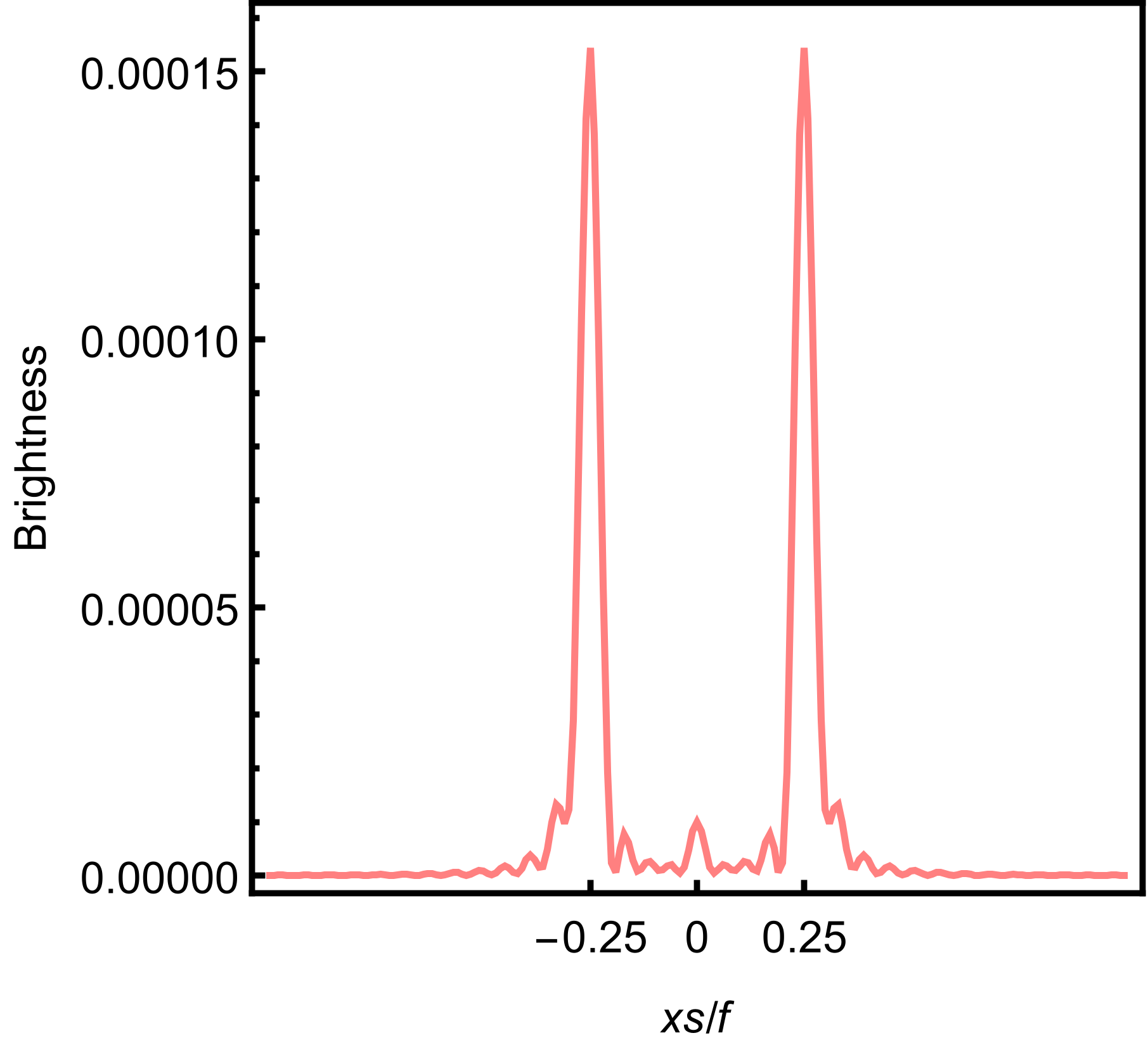}
    \caption{$y_{h}=12.5$,$\mathit{T}=1.009$}
  \end{subfigure}
  
  \caption{Effect of $\mathit{T}$ on the brightness, where $\theta _{obs}=0$, $\beta=1$, $\omega=90$.}
  \label{18}%
\end{figure}

In Figure \ref{17}, we analyze the impact of the event horizon on the Einstein ring of a BH. The results reveal a reciprocal relationship as $y_{h}$ increases, the radius of the ring decreases correspondingly. Figure \ref{18} illustrates the relationship between luminosity and the radius of the event horizon. As $y_{h}$ gradually increases, the horizontal coordinate corresponding to the maximum luminosity gradually decreases, indicating that the radius of the Einstein ring  diminishes with the increase of $y_{h}$.

\subsection{Images from the viewpoint of geometric optics  }
In this section, we will discuss Einstein rings from the perspective of geometric optics.  Our primary focus is on studying the incidence angle of photons, based on the spherically symmetric spacetime background described in Eq.(\ref{jie}) and Eq.(\ref{jie1}). We define conserved quantities within the metric framework as $\omega =G(r)\partial t/\partial \lambda $ and $\tilde{L} =r^{2} \partial \varphi /\partial t $, where $\omega$ represents the energy of the photon, $\tilde{L}$ denotes the angular momentum of the photon and $\lambda$ is the affine parameter. Due to the spherical symmetry of spacetime, it is only necessary to consider the case where photon orbits are located at the equator. $\theta \equiv \pi/2 $. The four-velocity $\upsilon ^{\zeta } =(\mathrm{d}/\mathrm{d}\lambda )^{\zeta } $ is satisfied \begin{eqnarray}
-G(r)(\mathrm{d}t/ \mathrm{d}\lambda)^{2} +G(r)^{-1} (\mathrm{d}r/ \mathrm{d}\lambda)^{2}\nonumber\\+r^{2} \sin \theta
(\mathrm{d}\varphi / \mathrm{d}\lambda)^{2} =0,
\end{eqnarray} and \begin{equation}
\dot{r}^{2}  =\omega -\tilde{L} z(r).
\end{equation} here $z(r)=G(r)/r^{2}$ and $\dot{r} =\partial r /\partial \lambda$. The incident angle $n^{\zeta } \equiv \partial / \partial r^{\zeta }$ with boundary $\theta _{in}$ as normal vector is defined as\cite{Liu:2022cev,Zeng:2023tjb} \begin{equation}
\cos \theta _{in}=\frac{g_{\alpha \beta } \upsilon ^{\alpha } n^{\beta } }{\left | \upsilon  \right | \left | n \right | }\mid _{r=\infty }
=\sqrt{\frac{\dot{r}^{2} /G }{\dot{r}^{2} /G+\tilde{L} /r^{2} } } \mid _{r=\infty },
\end{equation} further simplification leads to \begin{equation}
\sin\theta _{in}^{2} =1-\cos \theta _{in}^{2} =\frac{\tilde{L} ^{2}z(r) }{{\dot{r}^{2}+\tilde{L} ^{2}z(r)} }\mid _{r=\infty }=\frac{\tilde{L} ^{2}}{\omega ^{2} }.
\end{equation}

\begin{figure}
	\centering 
\includegraphics[width=0.4\textwidth, angle=0]{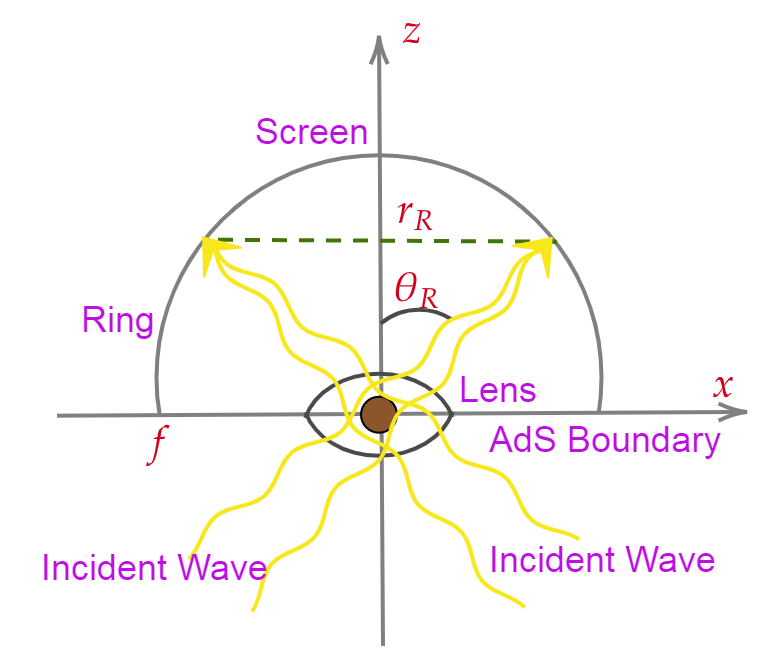}
	\caption{Relation between ring radius $r_{R}$ and ring angle $\theta _{R}$.} 
	\label{19}%
\end{figure}

When the two endpoints of a geodesic are aligned with the center of the BH, due to axial symmetry\cite{Hashimoto:2018okj}, an observer will perceive a ring-shaped image with a radius that corresponds to the incident angle $\theta _{in}$. As a photon reaches the location of the photon sphere, it neither escapes nor falls into the BH, but instead begins to orbit around it. At this point, the angular momentum is denoted by $L$, and the equation of orbital motion for the photon at the photon ring is determined by the following conditions\cite{Hashimoto:2019jmw,Hashimoto:2018okj,Liu:2022cev} \begin{equation}
\dot{r}=0,\mathrm{d}z /\mathrm{d}r=0.
\end{equation} In the case, $\sin\theta _{in}=\mathit{L} /{\omega }$ is given, and from Figure \ref{19} $\sin \theta _{R}$ is observed to satisfy the relationship \begin{equation}
\sin \theta _{R} =r_{R} /f.
\end{equation}

The incident angle of the photon and the angle of the photon ring both describe the angle of the photon ring that can be observed by an observer, and they should be essentially equal, that is \begin{equation}
r_{R} /f=L/\omega.
\label{27}
\end{equation}

\begin{figure}[htbp]
  \centering
  \begin{subfigure}[b]{0.45\textwidth}
    \centering
    \includegraphics[width=0.65\textwidth,height=0.65\textwidth]{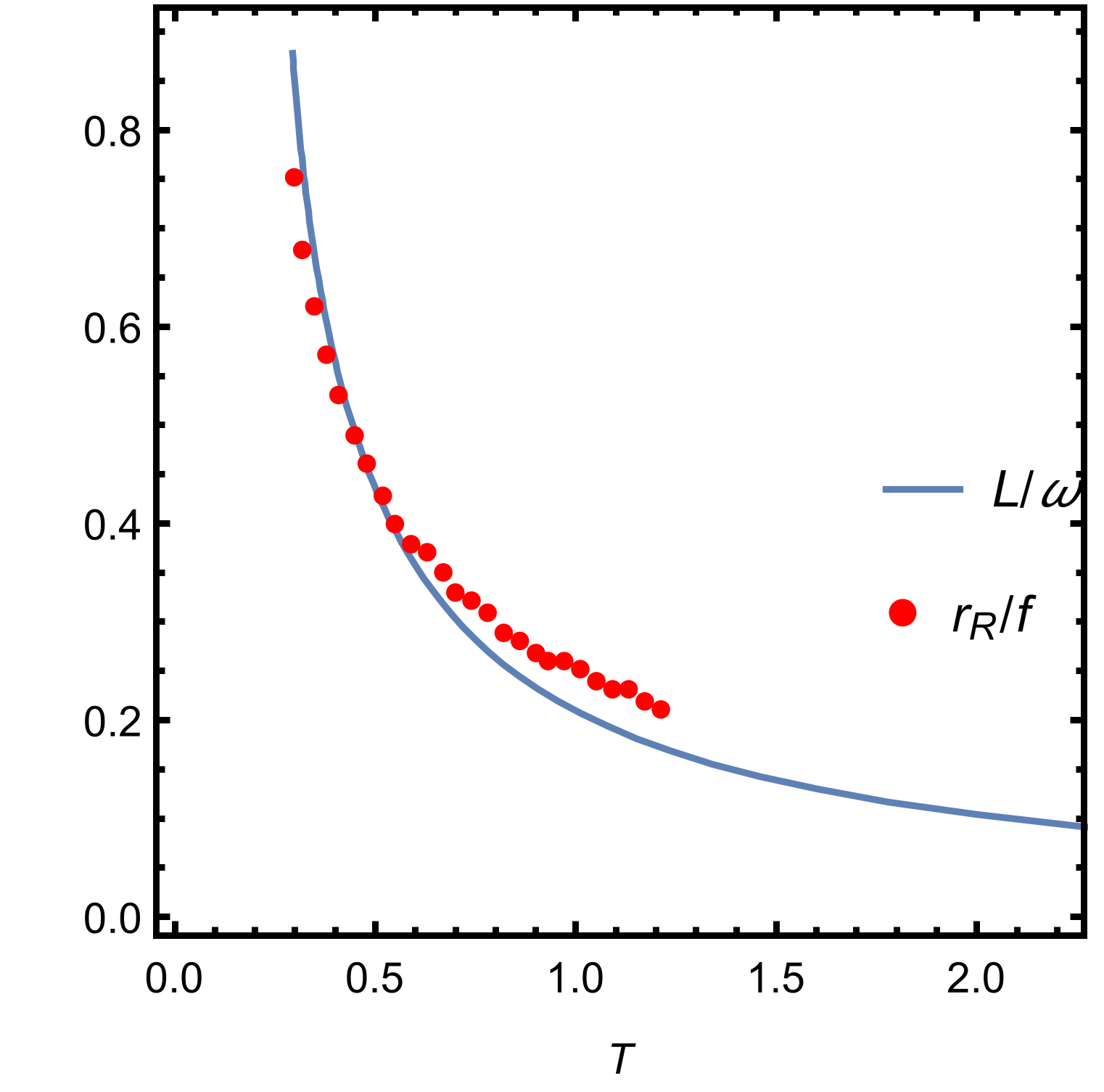}
    \caption{$\alpha=0.1$}
  \end{subfigure}
  \hfill
  \begin{subfigure}[b]{0.45\textwidth}
    \centering
    \includegraphics[width=0.65\textwidth,height=0.65\textwidth]{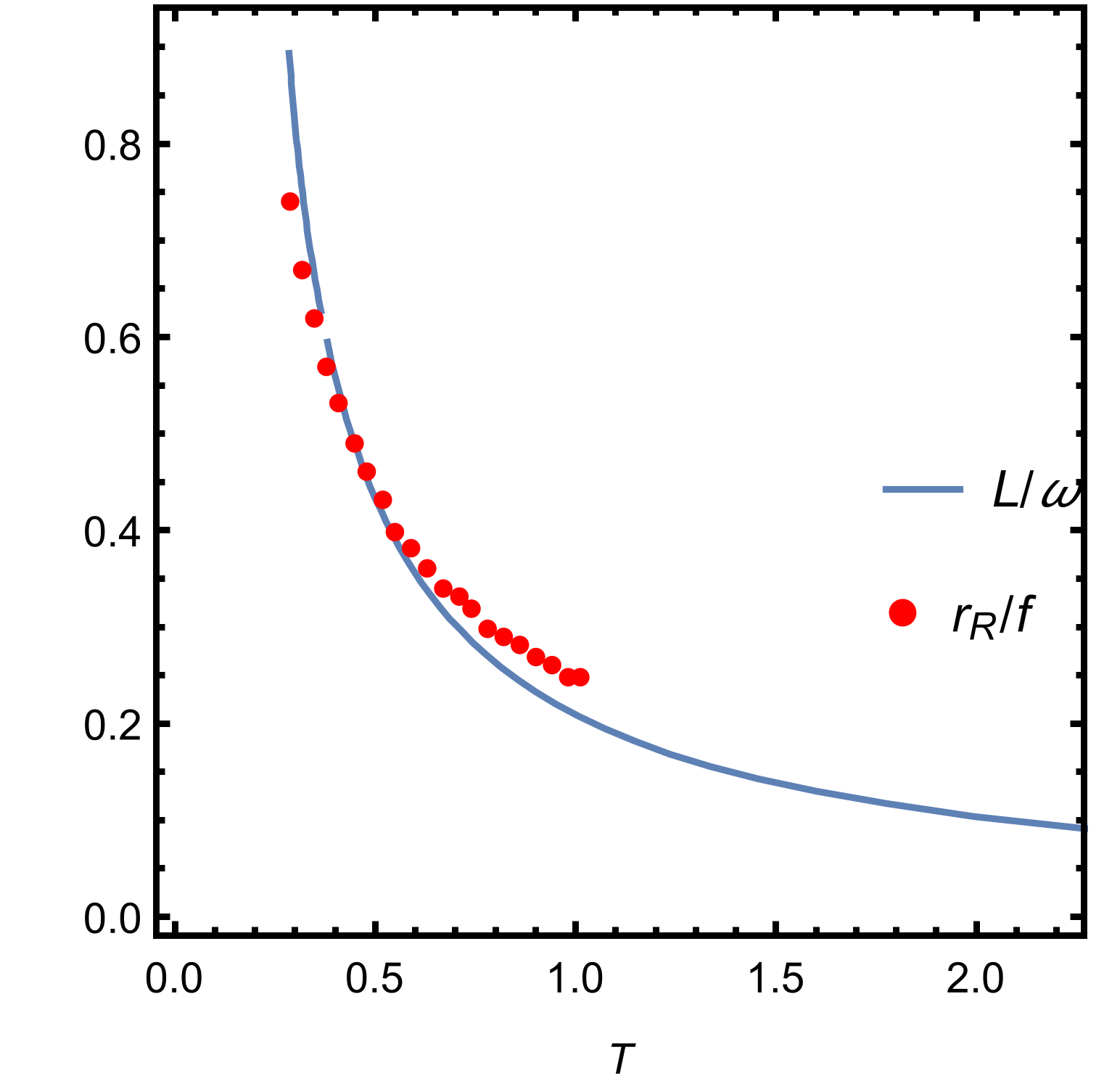}
    \caption{$\alpha=1.1$}
  \end{subfigure}

\begin{subfigure}[b]{0.45\textwidth}
    \centering
    \includegraphics[width=0.65\textwidth,height=0.65\textwidth]{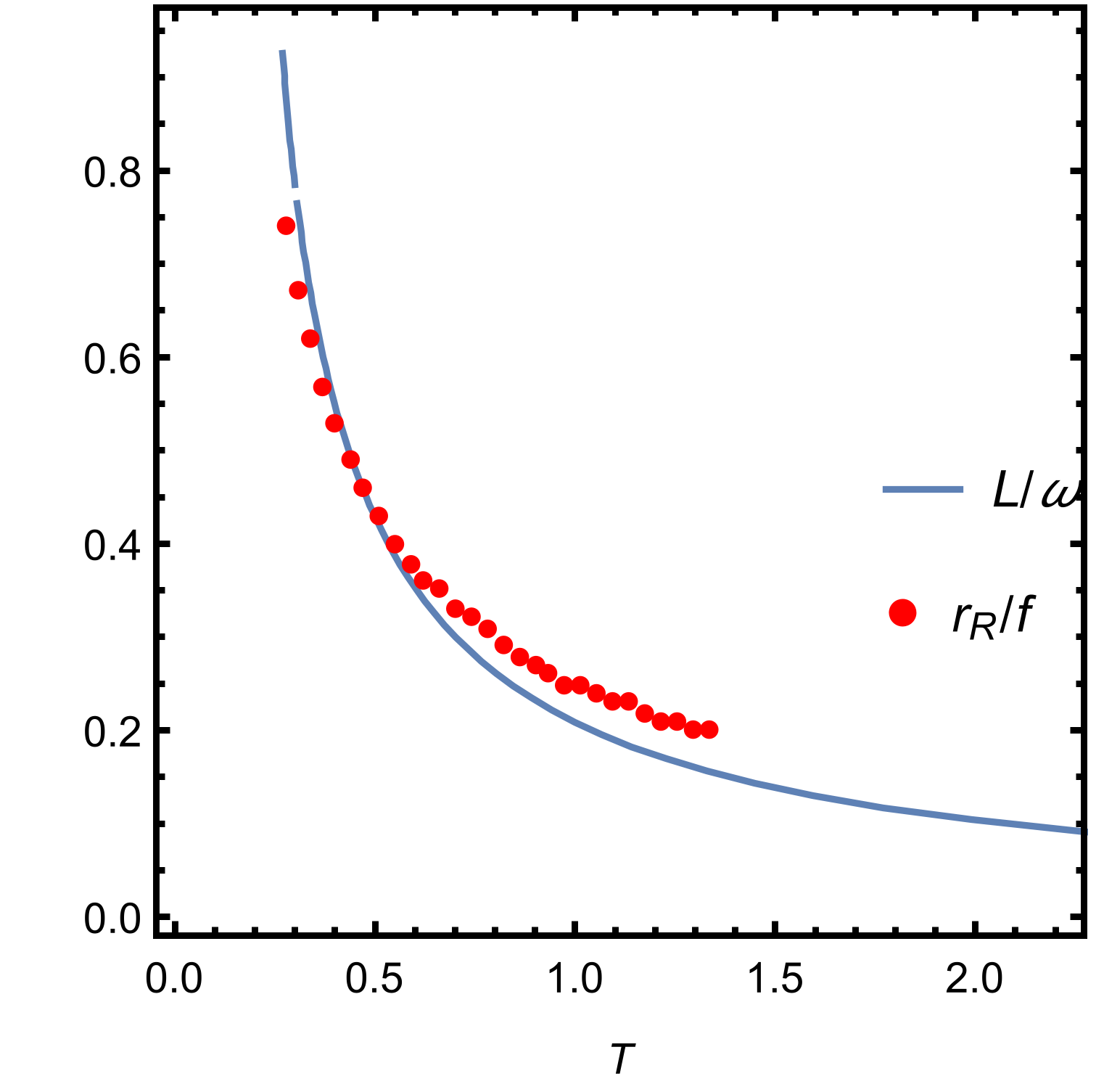}
    \caption{$\alpha=3$}
  \end{subfigure}
  \hfill
  
  \caption{Comparing the Einstein ring radius between the geometric optics and wave optics for different values of $\alpha$ with $\omega =90$. Where the discrete red points represent the Einstein ring radius derived from wave optics, while the blue curve depicts the variation in the radius of the circular orbit as the temperature changes, which is based on geometric optics.}
  \label{22}%
\end{figure}

\begin{figure}[htbp]
  \centering
  \begin{subfigure}[b]{0.45\textwidth}
    \centering
    \includegraphics[width=0.65\textwidth,height=0.65\textwidth]{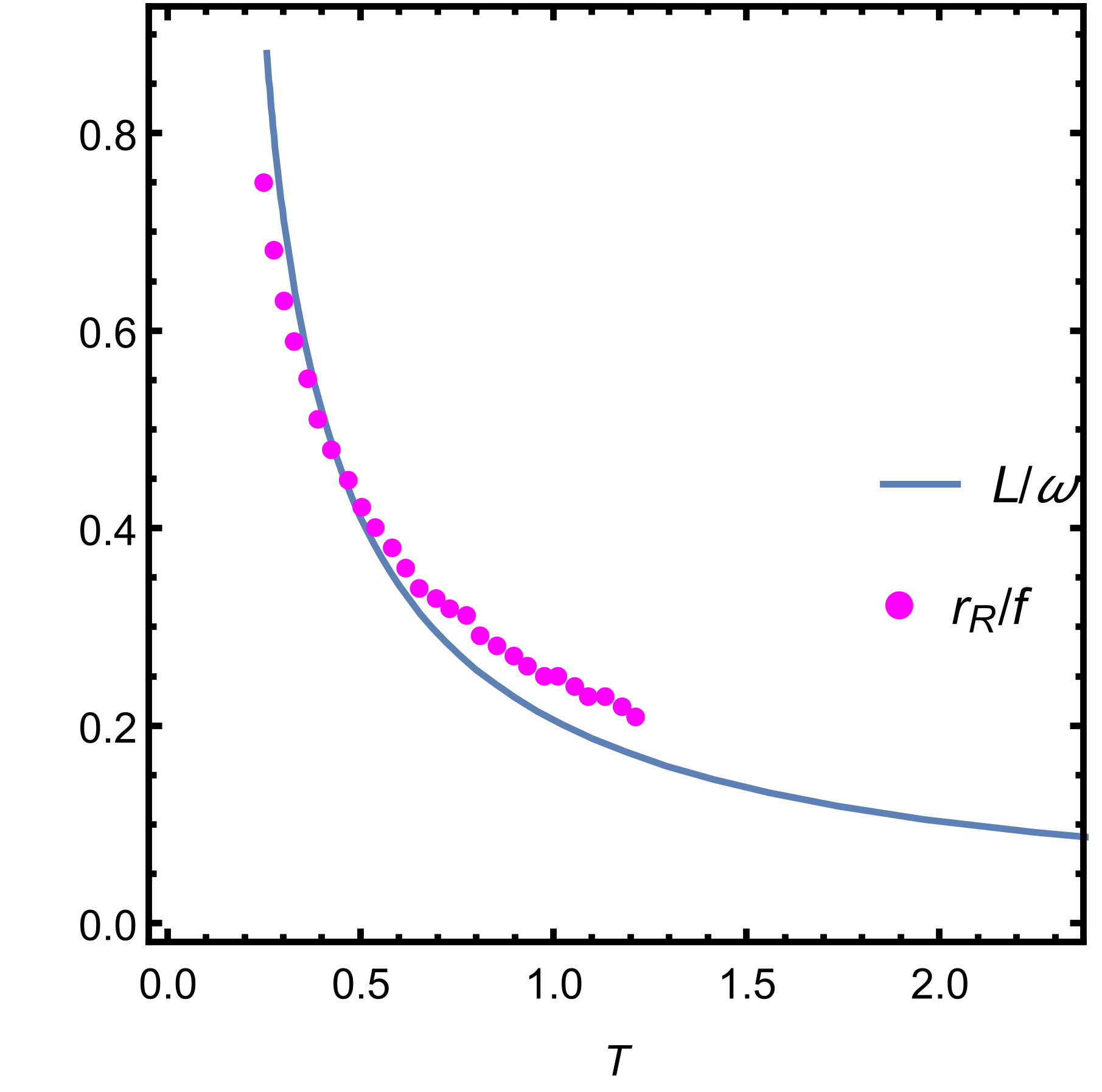}
    \caption{$\beta=0.5$}
  \end{subfigure}
  \hfill
  \begin{subfigure}[b]{0.45\textwidth}
    \centering
    \includegraphics[width=0.65\textwidth,height=0.65\textwidth]{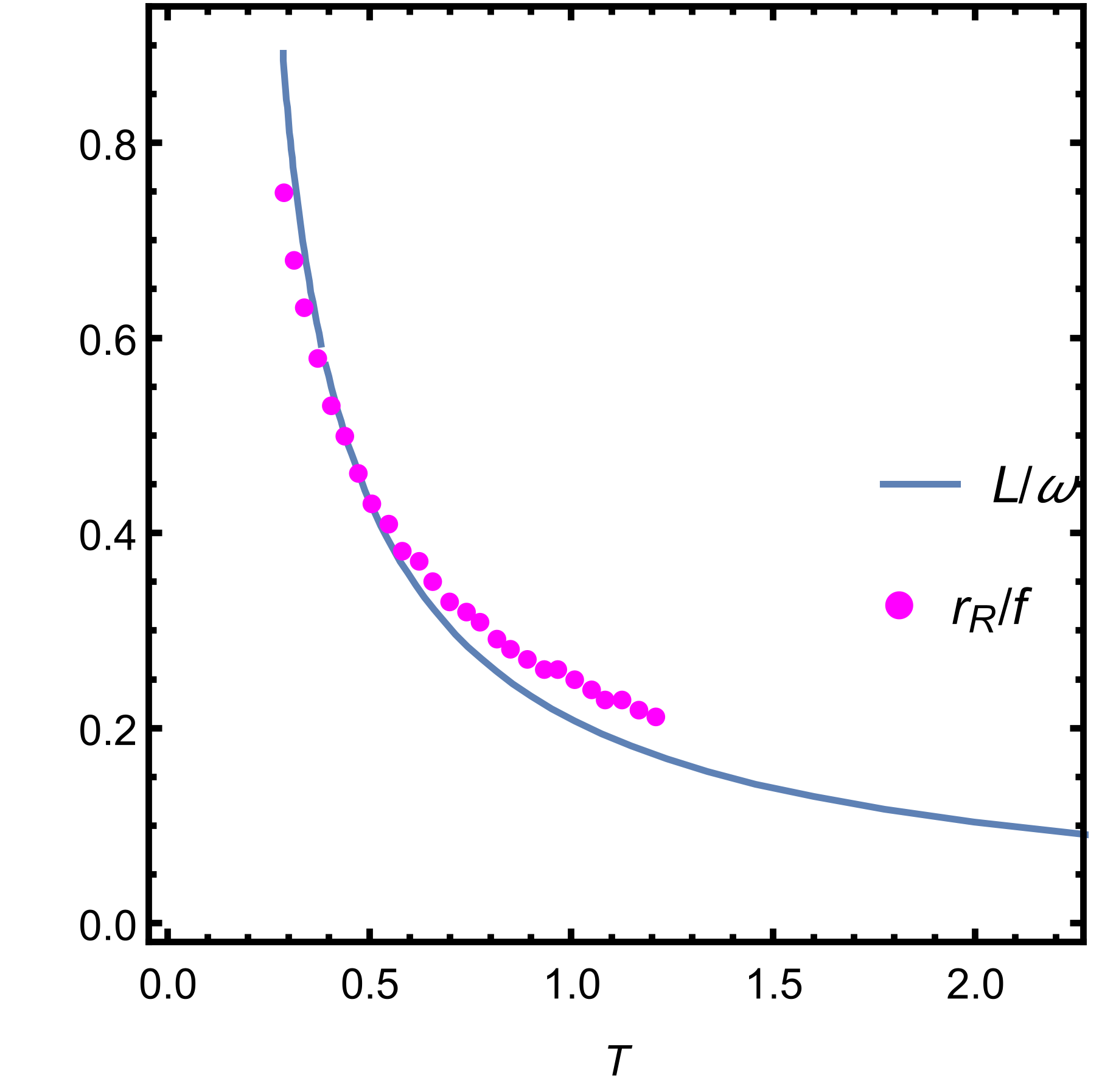}
    \caption{$\beta=1$}
  \end{subfigure}

\begin{subfigure}[b]{0.45\textwidth}
    \centering
    \includegraphics[width=0.65\textwidth,height=0.65\textwidth]{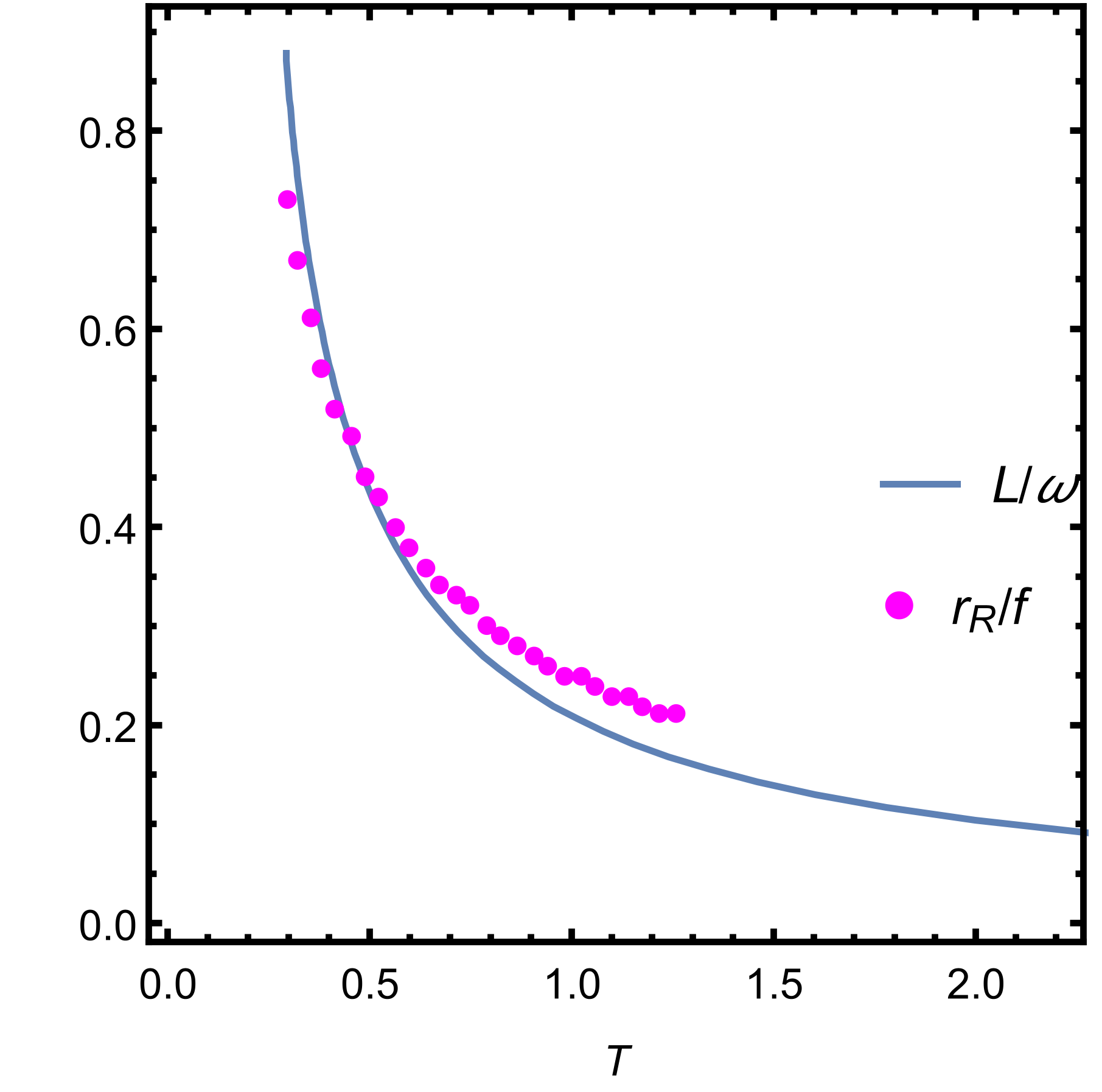}
    \caption{$\beta=2$}
  \end{subfigure}
  \hfill
  
  \caption{Comparing the Einstein ring radius between the geometric optics and wave optics for different values of $\beta$ with $\omega =90$. Where the discrete magenta points represent the Einstein ring radius derived from wave optics, while the blue curve depicts the variation in the radius of the circular orbit as the temperature changes, which is based on geometric optics.}
  \label{23}%
\end{figure}

When relevant physical parameters undergo changes, numerical methods are employed to verify the results obtained from Eq.(\ref{27}). Figure \ref{22} illustrates the radius of the Einstein ring for  various values of parameter $\alpha$, while Figure \ref{23} depicts the radius of the Einstein ring for different values of parameter $\beta$. it is evident that the observed angle of the Einstein ring, obtained by wave optics, closely approximates the incident angle of the photon ring, which is derived from geometric optics. Notably, this conclusion holds irrespective of the deformation parameter $\alpha$ and the control parameter $\beta$, despite their influence on the fitting accuracy. In other words, the Einstein ring can be constructed holographically via wave optics.

%\section{Discussion}
%%\label{}

\section{Summary and conclusions}
%%\label{}

Using wave optics, the Einstein rings of the deformed AdS-Schwarzschild BH were investigated, based on AdS/CFT correspondence. By solving the asymptotic behavior of the wave function near the boundary, one can obtain the response of a Gaussian wave source at the south pole, measured at the north pole. The results indicate that the response function is the diffraction pattern formed by the scalar waves emitted from the wave source after being scattered by the BH. When $\beta=1$, the amplitude of the response function increases with the increase of the deformation parameter $\alpha$, but the response function decreases with the increase of the wave source frequency $\omega$, and varies significantly with temperature. Similarly, when $\alpha=1$, the amplitude of the response function decreases with the increase of both the control parameter $\beta$ and the wave source frequency $\omega$. Moreover, the amplitude of the response function also varies with temperature.

An optical system featuring a convex lens was employed for observing the corresponding holographic images. The result showed that when the control parameter $\beta$ is fixed, the radius of the holographic ring gradually increases with the increase of the deformation parameter $\alpha$. Additionally, as the event horizon increases, the ring radius decreases. Moreover, as the wave source frequency increases, the image becomes clearer. On the contrary, the deformation parameter $\alpha$ is fixed, the radius of the Einstein ring gradually decreases with the increase of control parameter $\beta$. Similarly, as the horizon radius increases, the ring radius decreases. Also, as the wave source frequency increases, the ring image becomes clearer. 

Under the framework of geometric optics, a further study was conducted on the incident angle of the photon ring. Theoretical analysis and numerical simulation revealed that this result is consistent with the observation angle of the Einstein ring obtained through wave optics. This conclusion does not depend on deformation parameter $\alpha$ and control parameter $\beta$, although the $\alpha$ and $\beta$ affect the fitting accuracy.

\section*{Acknowledgements}
This work is supported by the National Natural Science Foundation of China (Grants No.
11675140, No. 11705005, and No. 12375043), and Innovation and Development Joint Foundation
of Chongqing Natural Science Foundation (Grant No. CSTB2022NSCQ-LZX0021) and Basic
Research Project of Science and Technology Committee of Chongqing (Grant No. CSTB2023NSCQMSX0324), and the Sichuan Science and Technology Program (Grant No. 2023ZYD0023, 2024NSFSC1999).

\appendix

%`\section{Appendix title 1}

%`\section{Appendix title 2}


\end{document}